\documentclass[12pt,psfig,a4]{article}
\makeatletter
\def\@seccntformat#1{\@ifundefined{#1@cntformat}%
   {\csname the#1\endcsname\quad}  
   {\csname #1@cntformat\endcsname}
}
\let\oldappendix\appendix 
\renewcommand\appendix{%
    \oldappendix
    \newcommand{\section@cntformat}{\appendixname~\thesection:\,\,}
}
\makeatother

\usepackage[T1]{fontenc}
\usepackage{amsfonts}
\usepackage[top=2.3cm,margin=1in]{geometry}
\usepackage{graphics}
\usepackage{lscape}
\usepackage{subfigure}
\usepackage{hyperref}
\usepackage{graphicx}
\usepackage{setspace}
\usepackage{latexsym}
\usepackage{mathtools}
\usepackage{enumitem}
\usepackage{amsmath}
\usepackage{amsthm}
\usepackage{color,soul}
\usepackage{multirow}
\usepackage{hhline}
\usepackage[table]{xcolor}
\usepackage{booktabs}
\usepackage{rotating}
\usepackage[hang,small,bf]{caption}
\usepackage{caption}

\usepackage{epsfig}
\usepackage{dsfont}
\usepackage{imakeidx}
\usepackage[mathscr]{euscript}
\usepackage[overload]{empheq}

\usepackage[ruled,lined,boxed]{algorithm2e}
\usepackage{epstopdf}

\usepackage[round]{natbib}

\DeclareMathOperator*{\argmin}{arg\,min}
\DeclareMathOperator*{\argmax}{arg\,max}

\PassOptionsToPackage{
        natbib=true,
        style=authoryear-comp,
        hyperref=true,
        backend=biber,
        maxbibnames=99,
        firstinits=true,
        uniquename=init,
        maxcitenames=2,
        parentracker=true,
        url=false,
        doi=false,
        isbn=false,
        eprint=false,
        backref=true,
            }   {biblatex}

\newcommand{\Lyx}{L\kern-.1667em\lower.25em\hbox{y}\kern-.125emX\spacefactor1000}
\input{tcilatex}

\begin{document}

\title{\textbf{Mean-Variance-VaR portfolios: MIQP formulation and performance analysis}}
\author{Francesco Cesarone$^1$, Manuel L Martino$^1$, Fabio Tardella$^2$ \\
{\small $^1$\emph{ Roma Tre University - Department of Business Studies}}\\
{\small $^2$\emph{ University of Florence - Department of Information Engineering (DINFO)}}\\
{\footnotesize francesco.cesarone@uniroma3.it, manuelluis.martino@uniroma3.it, fabio.tardella@unifi.it}\\
}
\date{\today}
\maketitle

\begin{abstract}
Value-at-Risk is one of the most popular risk management tools in the financial industry.
Over the past 20 years several attempts to include VaR in the portfolio selection process have been proposed.
However, using VaR as a risk measure in
portfolio optimization models leads to problems that are computationally hard to solve.
In view of this, few practical applications of VaR in portfolio selection have appeared in the literature up to now.

\noindent
In this paper,
we propose to add the VaR criterion to the classical Mean-Variance approach in order to better address
the typical regulatory constraints of the financial industry.
We thus obtain a portfolio selection model characterized by three criteria:
expected return, variance, and VaR at a specified confidence level.
The resulting optimization problem consists in minimizing variance
with parametric constraints on the levels of expected return and VaR.
This model can be formulated as a Mixed-Integer Quadratic Programming (MIQP) problem.
An extensive empirical analysis on seven real-world datasets demonstrates the practical applicability
of the proposed approach.
Furthermore,
the out-of-sample performance of the optimal Mean-Variance-VaR portfolios seems to be generally
better than that of the optimal Mean-Variance
and Mean-VaR portfolios.

\medskip
\noindent
\textbf{Keywords}: Portfolio Optimization, Asset Allocation, Value-at-Risk, MIQP, Multi-objective Optimization.

\end{abstract}

\section{Introduction}

The milestone of modern finance theory for portfolio selection
is undoubtedly the seminal work of ~\cite{10.2307/2975974,markowitz1959portfolio}
on the gain-risk analysis.
Indeed, his famous Mean-Variance model is still widely used by both academics and practitioners to support portfolio selection decisions.
The success of this bi-objective optimization problem has inevitably drawn many criticisms and proposals of alternative or more refined models \citep{kolm201460}.
In fact, many refinements of the Markowitz model have been provided in the literature, such as the inclusion of three or more objectives for selecting a portfolio.
Indeed, to better control the shape and characteristics of the portfolio return distribution, many scholars have attempted to extend the MV model to higher order moments, such as the portfolio skewness and kurtosis.
This is justified by the empirical evidences that the asset returns can show asymmetric and/or fat tail distributions
\citep[see, e.g.,][]{mandelbrot1972certain,pagan1996econometrics,cont2001empirical}.
For instance, \cite{konno1995mean} have presented the Mean-Variance-Skewness (MVS) model,
a natural extension of the classical MV model
where the mean and the skewness of the portfolio return are maximized, while its variance is minimized.
Furthermore, the authors have compared some characteristics of the optimal portfolios obtained by the MVS model
with those achieved by the Mean-Absolute Deviation-Skewness (MADS) model, previously proposed in \cite{konno1993mean}, where risk is measured by means of the Absolute Deviation of the portfolio return.
\cite{araciouglu2011mean} have proposed a Polynomial Goal programming model,
where in addition to the first three moments of portfolio returns they have also included kurtosis which is minimized.

\noindent
We point out that a large number of  portfolio selection models have been formulated as multi-objective optimization problems.
For interested readers, we mention the work of \cite{steuer2003multiple} who have presented an extensive survey on  Multiple Criteria Decision Making applied to several important topics in finance.
%
\cite{chow1995portfolio} has proposed a tri-objective optimization model based on the portfolio expected return, variance and the tracking error w.r.t. a benchmark index, where the tracking error is defined as the standard deviation of the difference between the portfolio and benchmark returns.
\cite{lo2006s} have introduced liquidity as an additional criterion. More precisely, the authors have analysed how the inclusion of liquidity into the investment process affects the selected portfolios and the risk-return tradeoff.
\cite{anagnostopoulos2010portfolio} have provided a tri-objective optimization problem, where the portfolio expected return is maximized, while the portfolio variance and cardinality (i.e., the number of the selected assets) are minimized.
\cite{utz2014tri} have considered portfolio sustainability,
measured by the weighted sum of the assets' ESG scores,
as a third criterion.
The authors have analyzed how this linear objective can influence the levels of the portfolio risk and return.
To force diversification \cite{bera2008optimal} have added the maximization of the portfolio entropy to the MV model, which could otherwise generate portfolios highly concentrated on few assets.
For regulatory and reporting purposes in the portfolio selection process, \cite{roman2007mean} have emphasized the important role of risk measures focused on high portfolio losses (or equivalently, on low portfolio returns).
They have then proposed a tri-objective optimization problem,
where the portfolio expected return is maximized, while variance and Conditional-Value-at-Risk (CVaR) are minimized.
In the case of discrete random variables, the authors have reformulated this model as single-objective quadratic optimization problem.
However, \cite{cont2010robustness} claim that using CVaR instead of VaR as a risk measure could lead to a less robust risk management model.
Indeed, although CVaR has the advantage of being a coherent risk measure \citep{acerbi2002coherence},
the authors observe that it is much more sensitive to outliers than VaR.

\noindent
The use of VaR in portfolio optimization can be found in \cite{wang2000mean},
where the author first presents
a two-stage optimization approach based on the Mean-Variance and Mean-VaR analysis.
Then, he proposes a Mean-Variance-VaR model, but without providing any explicit formulation of this problem and
any empirical analysis on real markets.
\cite{basak2001value} have proposed a dynamic portfolio optimization problem that maximizes the expected utility of the portfolio wealth including a VaR constraint for regulatory requirements.
\cite{alexander2002economic} have described a Mean-VaR model assuming that the assets' returns
follow a multivariate normal distribution. Furthermore, the authors have also stretched their analysis
to the case of multivariate student-t returns.
\cite{gaivoronski2005value} have tackled the problem of VaR minimization by considering its approximation,
the Smoothed Value-at-Risk (SVaR), which requires less computational effort than the original problem.
Another significant theoretical and practical contribution to this topic has been provided by \cite{benati2007mixed}.
They have first proved that the Mean-VaR problem is NP-hard, and then they have proposed a Mixed-Integer Linear Programming (MILP) formulation for it.
\cite{pinar2013static} has presented a closed-form solution for the Mean-VaR model
applied to a market with a risk-free asset and $n$ normally distributed risky assets,
where short sales are allowed.
\cite{lotfi2017adjusted} have suggested a robust version of the Mean-VaR problem, where VaR is estimated with a parametric approach, a risk-free asset is considered, and short-sellings are allowed.

\noindent
In this paper, we propose to add the VaR criterion to the classical Mean-Variance approach in order to better address
the typical regulatory constraints of the financial industry.
We thus obtain a portfolio selection model characterized by three criteria:
expected return, variance, and VaR at a specified confidence level.
The resulting optimization
problem consists in minimizing variance with parametric constraints on the levels of expected return and VaR.
This model can be formulated as a Mixed-Integer Quadratic Programming (MIQP) problem.
An extensive empirical analysis on seven real-world datasets demonstrates the practical applicability of the proposed approach.
Furthermore,
the out-of-sample performance of the optimal Mean-Variance-VaR portfolios seems to be generally
better than that of the optimal Mean-Variance
and Mean-VaR portfolios.

\noindent
The paper is organized as follows.
In Section \ref{sec:Mean-Variance-VaRmodel}, we introduce the Mean-Variance-VaR model and we show
how to formulate it as an MIQP problem.
In Section \ref{sec:FindingMean-Variance-VaR},
we discuss how to practically obtain the Mean-Variance-VaR efficient surface,
minimizing variance with parametric constraints on the levels of the portfolio expected return and of VaR.
In Section \ref{sec:Empiricalresults}, we first report the computational times required by Gurobi \citep{gurobi}, one of the best currently available MIQP solvers,
to solve our model when varying the number $n$ of assets in the investment universe,
the number $T$ of historical scenarios, and the confidence level $\varepsilon$.
Then, we provide an extensive out-of-sample performance analysis based on
six real-world datasets
with a choice of parameters
for which the models can be solved in reasonable time.
Finally, Section \ref{sec:ConclusiveRemarks} summarizes the main contributions of our work
and describes some directions for further developments.

\section{The Mean-Variance-VaR model}
\label{sec:Mean-Variance-VaRmodel}

We consider an investment universe of $n$ assets, whose linear returns are represented by the random variables $R_1, \ldots, R_n$.
In the case of full investment, a long-only portfolio is identified with a vector $x \in \Delta =\biggl\{x \in \mathbb{R}^n: \sum_{k=1}^n\,x_k\,=\,1, \, x_k \ge 0, k= 1, \dots, n \biggr\}$,
where $x_k$ is the fraction of capital invested in asset $k$.
Thus, the portfolio linear return
is given by $R_{P}(x)=\Sigma_{k=1}^n  x_k R_k$.
We assume that the random variables $R_1, \ldots, R_n$ are
defined on a discrete probability space $\{\Omega,\mathcal{F},P\}$,
with $\Omega\,=\,\{\omega_1,\dots,\omega_T\}$, $\mathcal{F}$ a $\sigma$-field and $P(\omega_t)=p_t$.
In this work, we use a
look-back approach where the possible realizations of the discrete random returns are obtained from historical data.
As it is customary in portfolio optimization,
the investment decision is made using $T$ equally likely historical
scenarios \citep[see, e.g.,][and references therein]{carleo2017approximating}.
Then, under scenario $t \in \{1,\dots,T\}$, we denote by $r_{kt}$ the return of asset $k \in \{1,\dots,n\}$
and by $R_{Pt}(x)=\sum_{k=1}^{n}  x_k r_{kt}$ the portfolio return.\\
The classical Mean-Variance (MV) portfolio optimization model \citep{10.2307/2975974,markowitz1959portfolio} aims at determining the vector of portfolio weights $x= \left( x_1, x_2, \cdots, x_n \right)$ that minimizes the portfolio variance $\sigma_{P}^{2}(x) = \sum_{k=1}^n \sum_{j=1}^n x_k\,x_j\,\sigma_{kj}$, while restricting the portfolio expected return
$\mu_{P}(x)= \sum_{k=1}^n\,\mu_k\,x_k$ to attain a specified target level $\eta$.
Here we denote by $\mu_k=\frac{1}{T}\sum_{t=1}^{T}  r_{kt}$ the expected return of asset $k$, and by $\sigma_{kj}= \frac{1}{T}\sum_{t=1}^{T}  (r_{kt} - \mu_k) (r_{jt} - \mu_j)$ the covariance between assets $k$ and $j$.
Thus, the MV model can be formulated as the following convex QP problem
\begin{equation}  \label{Markowitz}
	\left\lbrace{
		\begin{array}{lll}
			\min & \sum_{k=1}^n \sum_{j=1}^n x_k\,x_j\,\sigma_{kj} &  \\
			\mbox{s.t.} &  &  \\
			& \sum_{k=1}^n\,\mu_k\,x_k\,\ge\, \eta &  \\
			& \sum_{k=1}^n x_k = 1 &  \\ 
			& x_k \ge 0 & k=1,\ldots,n    
		\end{array}
	}\right.
\end{equation}
where the last two constraints represent the \emph{budget} and the \emph{no-short sellings} constraints, respectively.

\noindent
Our aim is to include Value-at-Risk (VaR) as a risk measure in the portfolio selection process in addition to the portfolio expected return and variance.
VaR is one of the most popular risk management tools in the financial industry
and it is commonly used to control risk \citep{morgan1996riskmetrics}.
As usual, $VaR_\varepsilon$ represents the maximum loss at a given confidence
level $(1-\varepsilon)$ related to a predefined time horizon $\{1,\dots,T\}$,
where typically $\varepsilon=0.01, 0.05, 0.10$.
Therefore, for a given portfolio $x$,
$VaR_\varepsilon(x)$ is the value such that the portfolio loss $L_P(x)\,=\,-R_P(x)=\,-\sum_{k=1}^{n}\, x_k R_k$ exceeds $VaR_\varepsilon(x)$ with a probability of $\varepsilon100\%$.
More formally,
$VaR_\varepsilon(x)$ of the random portfolio return $R_P(x)$ is the $\varepsilon$-quantile $Q_\varepsilon(R_P(x))$ of its distribution with negative sign
\[
\begin{aligned}
	& VaR_\varepsilon(x)=-Q_\varepsilon(R_P(x)).
\end{aligned}
\]

\noindent
For the Mean-Variance-VaR approach, a random portfolio return $R_P(x)$ is preferred to $R_P(y)$
if and only if $\mu_{P}(x) \ge \mu_{P}(y)$, $\sigma_{P}^{2}(x) \le \sigma_{P}^{2}(y)$ and $VaR_\varepsilon(x) \le VaR_\varepsilon(y)$,
where at least one inequality is strict.
Therefore, the efficient surface of the Mean-Variance-VaR model can be obtained by finding the non-dominated portfolios,
which are the Pareto-optimal solutions of the following tri-objective optimization problem
\begin{equation}  \label{eq:Mean-Variance-VaR}
	\left\lbrace{
		\begin{array}{lll}
			\min & \left(-\mu_{P}(x), \sigma_{P}^{2}(x), VaR_\varepsilon(x) \right) &  \\
			\mbox{s.t.} &  &  \\
			& x \in \Delta =\biggl\{x \in \mathbb{R}^n: \sum_{k=1}^n\,x_k\,=\,1, \, x_k \ge 0, k= 1, \dots, n \biggr\} \, . &
		\end{array}
	}\right.
\end{equation}
%
%
To practically solve Problem \eqref{eq:Mean-Variance-VaR},
we transform it into a single-objective optimization problem,
by applying the classical $\epsilon$-constraint method as follows
\begin{equation}  \label{eq:Mean-Variance-VaR-single}
	\left\lbrace{
		\begin{array}{lll}
			\min & \sigma_{P}^{2}(x) &  \\
			\mbox{s.t.} &  &  \\
			& \mu_{P}(x)\,\ge\, \eta &  \\
			& VaR_\varepsilon(x)\,\le\, z &  \\
			& \sum_{k=1}^n x_k = 1 &  \\ 
			& x_k \ge 0 & k=1,\ldots,n    
		\end{array}
	}\right.
\end{equation}
where $\eta$ and $z$ are the required target levels of the portfolio expected return and VaR, respectively.
Similar to \cite{benati2007mixed}, we can substitute $VaR_\varepsilon(x)=-r_\varepsilon$ by adding the constraints
$r_\varepsilon\le\sum_{k=1}^n r_{k t}\,x_k + M\,(1-y_t)$, $\forall t=1,\dots,T$ and
$\sum_{t=1}^T y_t\ge(1-\varepsilon)\,T$, where $r_\varepsilon$ is a real variable, $M$ is a sufficiently large positive number, and $y_t$, with $t=1,\dots,T$, are boolean variables.
Thus, Problem \eqref{eq:Mean-Variance-VaR-single} can be formulated as the following Mixed-Integer Quadratic (MIQP) problem
\begin{subequations}  \label{eq:Mean-Variance-VaR-single2}
\begin{align}[left = \empheqlbrace\,]
\min\limits_{(x, r_\varepsilon, y)} & \sum_{k=1}^n \sum_{j=1}^n x_k\,x_j\,\sigma_{kj} & &\nonumber \\
\mbox{s.t.} &  & &\nonumber \\
& \sum_{k=1}^n\,\mu_k\,x_k\,\ge\, \eta & &\nonumber \\
& -r_\varepsilon \,\le\, z & & \qquad \qquad(4a) \\
& r_\varepsilon \le \sum_{k=1}^n r_{k t}\,x_k + M\,(1-y_t) & t=1,\dots,T  &\qquad \qquad(4b) \\
& \sum_{t=1}^T y_t \geq (1-\varepsilon)\,T & & \qquad \qquad(4c)\\
& \sum_{k=1}^n x_k = 1 & & \nonumber \\
& x_k \ge 0 & k=1,\ldots,n  & \nonumber \\
& y_t\in\{0,1\} & t=1,\dots,T & \nonumber
\end{align}
\end{subequations}
Note that when the portfolio loss $-\sum_{k=1}^n r_{k t}\,x_k$ is above $-r_\varepsilon$ at time $t$,
then, for sufficiently large $M>0$, we must have $1-y_t=1$ in constraint (4b)
so that $y_t$ must be equal to 0.
On the other hand, constraint (4c) imposes that the number of the portfolio loss scenarios that exceed
$-r_\varepsilon$ is not greater than $\varepsilon T$.
Thus, $-r_\varepsilon$ actually represents the VaR of the portfolio $x$
and is bounded above by $z$ (i.e., by the required target level of the portfolio VaR)
through constraint (4a).
In the empirical analysis we set $\varepsilon = 0.01, 0.05$.

\noindent
The Mean-Variance-VaR Pareto-optimal portfolios can be obtained as solutions
of Problem \eqref{eq:Mean-Variance-VaR-single2}
by appropriately varying the target level of the portfolio expected return $\eta$ and
the target level of the portfolio VaR $z$, as shown in the next section.

\section{Finding the Mean-Variance-VaR efficient surface}
\label{sec:FindingMean-Variance-VaR}

In this section
we discuss how to practically obtain the Mean-Variance-VaR efficient
surface by solving Problem \eqref{eq:Mean-Variance-VaR-single2}, similarly to \cite{roman2007mean}.
Basically, we minimize the portfolio variance by appropriately varying
the target level of the portfolio expected return $\eta$ and the target level of the portfolio VaR $z$.

\noindent
To obtain all the Pareto-optimal portfolios, we first determine an appropriate interval for $\eta$.
This is the interval $[\eta_{min},\,\eta_{max}]$,
where $\eta_{min}=\max\{\eta_{minV},\,\eta_{minVaR}\}$ and \\
$\eta_{max}=\mu_{P}(x_{max})$ with $x_{max} = \argmax_{x \in \Delta} \mu_{P}(x)$.
Here, $\eta_{minV}=\mu_{P}(x_{minV})$ with $x_{minV} = \argmin_{x \in \Delta} \sigma_{P}^{2}(x)$
and
$\eta_{minVaR}=\mu_{P}(x_{minVaR})$ with $x_{minVaR} = \argmin_{x \in \Delta} VaR_\varepsilon(x)$.

\noindent
Then, for a fixed level $\eta \in [\eta_{min},\,\eta_{max}]$, we define the appropriate interval of $z$, \\
$[z_{\min}(\eta), z_{\max}(\eta)]$,
whose values guarantee that the optimal portfolios of Problem \eqref{eq:Mean-Variance-VaR-single2} are non-dominated.
More precisely, $z_{\min}(\eta)=VaR_\varepsilon(x_{minVaR}(\eta))$, where $x_{minVaR}(\eta)$ is
the portfolio with minimum VaR whose expected return is bounded below by $\eta$,
namely it is the optimal solution of the following problem
\eqref{eq:minVaRstar}
\begin{equation}\label{eq:minVaRstar}
	\left\lbrace{
		\begin{array}{lll}
			\min & VaR_\varepsilon(x) &  \\
			\mbox{s.t.} &  &  \\
			& \mu_{P}(x) \ge \eta \\
			& x \in \Delta
		\end{array}
	}\right.
\end{equation}
On the other hand, $z_{\max}(\eta)=VaR_\varepsilon(x_{minV}(\eta))$, where $x_{minV}(\eta)$ is
the portfolio with minimum variance whose expected return is bounded below by $\eta$,
namely it is the optimal solution of the following problem \eqref{eq:minVstar}
\begin{equation}\label{eq:minVstar}
	\left\lbrace{
		\begin{array}{lll}
			\min & \sigma_{P}^{2}(x) &  \\
			\mbox{s.t.} &  &  \\
			& \mu_{P}(x) \ge \eta \\
			& x \in \Delta
		\end{array}
	}\right.
\end{equation}

\noindent
In Figure \ref{fig:VarianceVaRSurface_0,01} we report an example of the Pareto-optimal portfolios obtained from Model \eqref{eq:Mean-Variance-VaR-single2}
in the Variance-VaR plane for several fixed levels of the portfolio expected return $\eta$.
Note that by solving
Problem \eqref{eq:Mean-Variance-VaR-single2} for different levels of the portfolio expected return $\eta \in [\eta_{min},\,\eta_{max}]$
with $z=z_{\min}(\eta)$, we obtain the Mean-VaR efficient frontier (see the bold black dashed line).
On the other hand, when we solve Problem \eqref{eq:Mean-Variance-VaR-single2} for different values of $\eta \in [\eta_{min},\,\eta_{max}]$
with $z=z_{\max}(\eta)$, we achieve the Mean-Variance efficient frontier (see the red dashed line).

\begin{figure}
	\centering
	\includegraphics[width=0.8\linewidth]{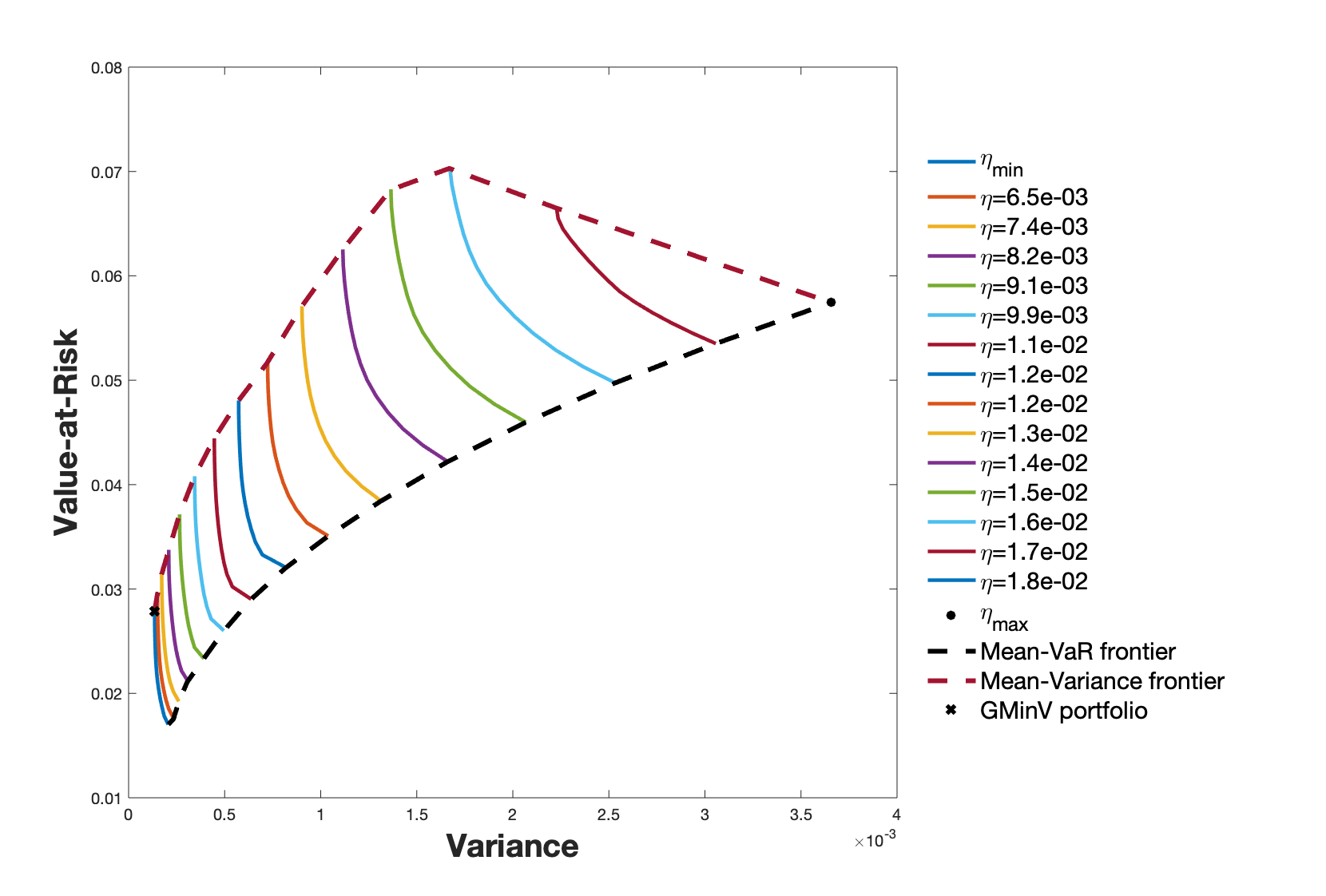}
	\caption{Example of the \emph{Mean-Variance-VaR} pareto-optimal portfolios (with $\varepsilon=1\%$)
		for several levels of the portfolio expected return $\eta$ in the Variance\,-\,VaR plane.}
	\label{fig:VarianceVaRSurface_0,01}
\end{figure}

\noindent
For a fixed level of $\eta \in [\eta_{min},\,\eta_{max}]$,
if we require stronger conditions on the portfolio VaR,
namely lower levels of $z(\eta)$,
we clearly obtain efficient portfolios with higher variance,
because the feasible region in \eqref{eq:Mean-Variance-VaR-single2} becomes smaller.
However, the in-sample increase in variance is relatively small for lower levels of required expected return $\eta$.
This behavior is confirmed in our out-of-sample analysis in Section \ref{subsec:outofsampleanalysis}.
Thus, it seems reasonable to complement the classical Mean-Variance model with
a restriction on the VaR level, particularly for low levels of $\eta$.

\noindent
We also observe that the number of selected assets in the optimal solutions
decreases when the required VaR becomes smaller (see also Table \ref{tab:NumTitoli}).
Furthermore, as the required target portfolio return $\eta$ increases,
both the portfolio variance and its VaR typically increase as shown in Figure \ref{fig:VarianceVaRSurface_0,01}.
We also note that for $\eta=\eta_{min}$ and $z=z_{\max}(\eta_{min})$,
we obtain
the Global Minimum Variance (GMinV) portfolio (see the bold x in Figure \ref{fig:VarianceVaRSurface_0,01}).
On the other hand,
when $\eta=\eta_{max}$, the efficient frontier degenerates into a single point:
the portfolio composed by the single asset with highest expected return.

\section{Empirical analysis}
\label{sec:Empiricalresults}

In this section we conduct an extensive analysis on several real-world datasets to examine both the practical applicability of the Mean-Variance-VaR model and its out-of-sample performance.
In Tables \ref{tab:WeeklyDatasets} and \ref{tab:DailyDatasets}, we list the datasets considered in this study,
which consist of weekly and daily linear returns, respectively.
The datasets of Table \ref{tab:WeeklyDatasets} are provided in \cite{BRUNI2016858}, while those of Table \ref{tab:DailyDatasets} are publicly available on the website \url{https://host.uniroma3.it/docenti/cesarone/DataSets.htm}.
Both have been used in other empirical analyses on portfolio selection \citep{bellini2021risk,bruni2017exact,carleo2017approximating,cesarone2019risk,cesarone2020optimization,cesarone2020stability,corsaro2021split,benati2021relative}.
\begin{table}[h!]\small
	\begin{center}
		\begin{tabular}{l l l l l l l l l l l l l l l l l l l}
			\hline
			\textbf{Index} & & & & & \textbf{$\#$assets} & & & & & & & \textbf{$\#$weeks} & &\textbf{Time interval}\\
			\hline
			\text{DowJones} & & & & & 28 & & & & & & & 1363 & & \text{Feb} 1990\,-\,\text{Apr} 2016\\
			\text{NASDAQ100} & & & & &82 & & & & & & &  596 & &  \text{Nov} 2004\,-\,\text{Apr} 2016\\
			\text{FTSE100} & & & & & 83 & & & & & & & 717 & & \text{Jul} 2002\,-\,\text{Apr} 2016\\
			\text{SP500} & & & & & 442 & & & & & & & 595 & & \text{Nov} 2004\,-\,\text{Apr} 2016\\
			\hline
		\end{tabular}
	\end{center}
	\caption{List of the weekly datasets analyzed}
	\label{tab:WeeklyDatasets}
\end{table}
\begin{table}[h!]\small
	\begin{center}
		\begin{tabular}{l l l l l l l l l l l l l l l l l l l}
			\hline
			\textbf{Index} & & & & &  \textbf{$\#$assets} & & & & & & & \textbf{$\#$days} && \textbf{Time interval}\\
			\hline
			\text{DowJones} & & & & & 28 & & & & & & & 6819 & & \text{Feb} 1990\,-\,\text{Apr} 2016\\
			\text{EuroStoxx 50} & & & & & 49 & & & & & & & 3885 & & \text{May} 2001\,-\,\text{Apr} 2016\\
			\text{Hang Seng} & & & & &  43 & & & & & & & 2707 & & \text{Nov} 2005\,-\,\text{Apr} 2016 \\
			\hline
		\end{tabular}
	\end{center}
	\caption{List of the daily datasets analyzed}
	\label{tab:DailyDatasets}
\end{table}

\noindent
We first perform some numerical tests to determine
the computational times required by Gurobi, one of the
best currently available MIQP solvers, to solve our model when varying the number $n$ of the assets in the investment universe, the number $T$ of the historical scenarios, and the confidence level $\varepsilon$.
For this purpose, we use the SP500 dataset listed in Table \ref{tab:WeeklyDatasets}.

\noindent
In Tables \ref{tab:ComputationaltimeSP500_0.01} and \ref{tab:ComputationaltimeSP500_0.05} we report the computational times (in seconds) for finding a Mean-Variance-VaR portfolio when $\varepsilon$ is fixed to 1\% and 5\%,
respectively.
The experiments are performed by varying $n$ from 20 to 442 and $T$ from 52 to 595, for each pair of $n<T$ that guarantees the nonsingularity of the covariance matrix.
As expected, the computational times tend to increase with $n$ and $T$,
but most notably when $\varepsilon=5\%$.
Indeed, in this case, the Gurobi MIQP solver typically spends more than eight hours for finding the optimal solution
for high values of $n$ and $T$.
On the other hand, for $\varepsilon=10\%$, the computational times tend to exceed one day thus becoming impractical,
as also pointed out by \cite{benati2007mixed}.\\
All the procedures have been implemented in MATLAB R2019b using the GUROBI 9.1 optimization solver, and have been executed on a laptop with an Intel(R) Core(TM) i7-8565U CPU @ 1.80GHz processor and 8,00 GB of RAM.
%
\begin{table}[htbp]
	\centering
	\caption{Computational times (in seconds) for solving the \emph{Mean-Variance-VaR} model with $\varepsilon$=1$\%$}
	\resizebox{1.02\textwidth}{!}{%
	\begin{tabular}{|c|c|cccccccccc|}
		\cmidrule{3-12}    \multicolumn{1}{c}{} &       & \multicolumn{10}{c|}{\textbf{n (\# of assets)}} \\
		\midrule
		\textbf{\# of years} & \textbf{T (\# of weeks)} & \textbf{n=20} & \textbf{n=40} & \textbf{n=80} & \textbf{n=120} & \textbf{n=150} & \textbf{n=200} & \textbf{n=250} & \textbf{n=300} & \textbf{n=400} & \textbf{n=442} \\
		\cmidrule{1-12}
		\textbf{1 Year} & \textbf{52} & 5.5   & 7.0   & -     & -     & -     & -     & -     & -     & -     & - \\
		\textbf{2 Years} & \textbf{104} & 5.6   & 6.1   & 6.6   & -     & -     & -     & -     & -     & -     & - \\
		\textbf{3 Years} & \textbf{156} & 5.9   & 7.8   & 8     & 8.8   & 8     & -     & -     & -     & -     & - \\
		\textbf{5 Years} & \textbf{260} & 6.2   & 8.1   & 8.6   & 11.1  & 11.4  & 13.2  & 15.5  & -     & -     & - \\
		\textbf{7 Years} & \textbf{364} & 7.4   & 9.2   & 11.2  & 15.3  & 15.4  & 18    & 25.4  & 27    & -     & - \\
		\textbf{10 Years} & \textbf{520} & 14    & 21.9  & 41.2  & 47.5  & 48.5  & 59.5  & 85.6  & 94    & 102   & 101 \\
		\textbf{11.5 Years} & \textbf{595} & 12.7  & 21.1  & 29.7  & 44.6  & 58.8  & 54.9  & 57.7  & 63    & 104   & 122 \\
		\bottomrule
	\end{tabular}%
}
	\label{tab:ComputationaltimeSP500_0.01}%
\end{table}%
%
\begin{table}[htbp]
	\centering
	\caption{Computational times (in seconds) for solving the \emph{Mean-Variance-VaR} model with $\varepsilon$=5$\%$}
	\resizebox{1.02\textwidth}{!}{%
	\begin{tabular}{|c|c|cccccccccc|}
		\cmidrule{3-12}    \multicolumn{1}{r}{} &       & \multicolumn{10}{c|}{\textbf{n (\# of assets)}} \\
		\cmidrule{1-12}
		\textbf{\# of years} & \textbf{T (\# of weeks)} & \textbf{n=20} & \textbf{n=40} & \textbf{n=80} & \textbf{n=120} & \textbf{n=150} & \textbf{n=200} & \textbf{n=250} & \textbf{n=300} & \textbf{n=400} & \textbf{n=442} \\
		\cmidrule{1-12}
		\textbf{1 Year} & \textbf{52} & 3     & 4     & -     & -     & -     & -     & -     & -     & -     & - \\
		\textbf{2 Years} & \textbf{104} & 4     & 7     & 6     & -     & -     & -     & -     & -     & -     & - \\
		\textbf{3 Years} & \textbf{156} & 6     & 12    & 31    & 98    & 82    & -     & -     & -     & -     & - \\
		\textbf{5 Years} & \textbf{260} & 20    & 52    & 506   & 2,649 & 3,034 & 8,180 & 9,171 & -     & -     & - \\
		\textbf{7 Years} & \textbf{364} & 55    & 850   & 3,854 & 7,886 & 14,505 & > \textbf{8 h} & > \textbf{8 h} & > \textbf{8 h} & -     & - \\
		\textbf{10 Years} & \textbf{520} & 634   & 4,074 & > \textbf{8 h} & > \textbf{8 h} & > \textbf{8 h} & > \textbf{8 h} & > \textbf{8 h} & > \textbf{8 h} & > \textbf{8 h} & > \textbf{8 h} \\
		\textbf{11.5 Years} & \textbf{595} & 1,242 & > \textbf{8 h} & > \textbf{8 h} & > \textbf{8 h} & > \textbf{8 h} & > \textbf{8 h} & > \textbf{8 h} & > \textbf{8 h} & > \textbf{8 h} & > \textbf{8 h} \\
		\bottomrule
	\end{tabular}%
}
	\label{tab:ComputationaltimeSP500_0.05}%
\end{table}%

\subsection{Out-of-sample performance analysis}
\label{subsec:outofsampleanalysis}

For the out-of-sample performance analysis, we adopt a rolling time window scheme (RTW) of evaluation,
namely we allow for the possibility of rebalancing the
portfolio composition during the holding period at fixed intervals.
Here we choose one financial month both as a rebalancing interval and as a holding period.
Furthermore,
on the basis of the computational times illustrated in Tables \ref{tab:ComputationaltimeSP500_0.01} and \ref{tab:ComputationaltimeSP500_0.05}, we choose
$\varepsilon=1\%$ and $5\%$,
in-sample windows of 2 years (namely 104 observations)
for weekly datasets (see Table \ref{tab:WeeklyDatasets}),
and in-sample windows of 10 months (namely 200 observations)
for daily datasets (see Table \ref{tab:DailyDatasets}).

\begin{figure}[h]
	\centering
	\includegraphics[width=0.8\linewidth]{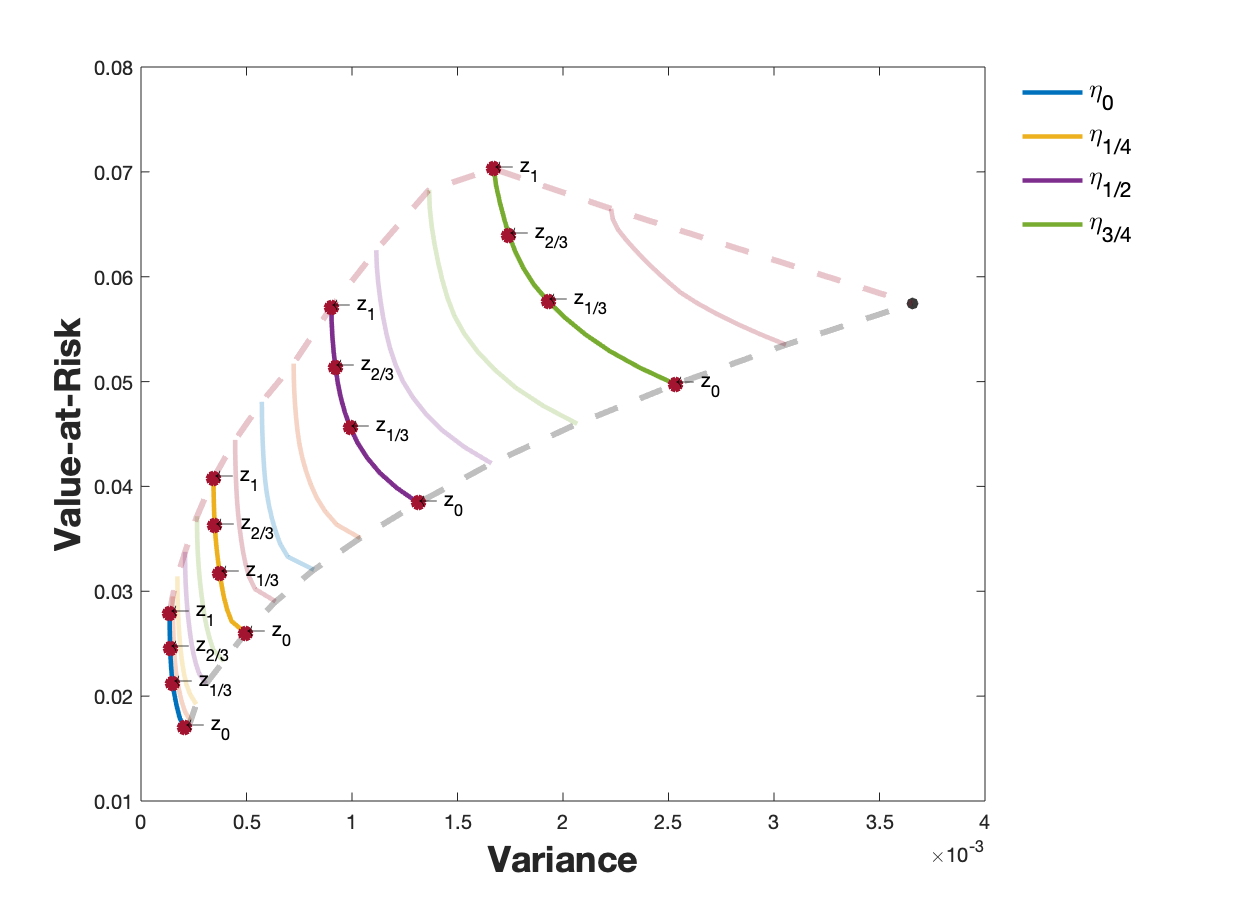}
	\caption{Example of the 16 \emph{Mean-Variance-VaR} efficient portfolios selected for the
		out-of-sample performance analysis}
	\label{fig: Curve_portfolios}
\end{figure}

\begin{figure}[htbp]
	\centering
	\subfigure{\label{fig:CumulativeReturnsfor_etamin_EX50_0.01}\includegraphics[scale=0.19]{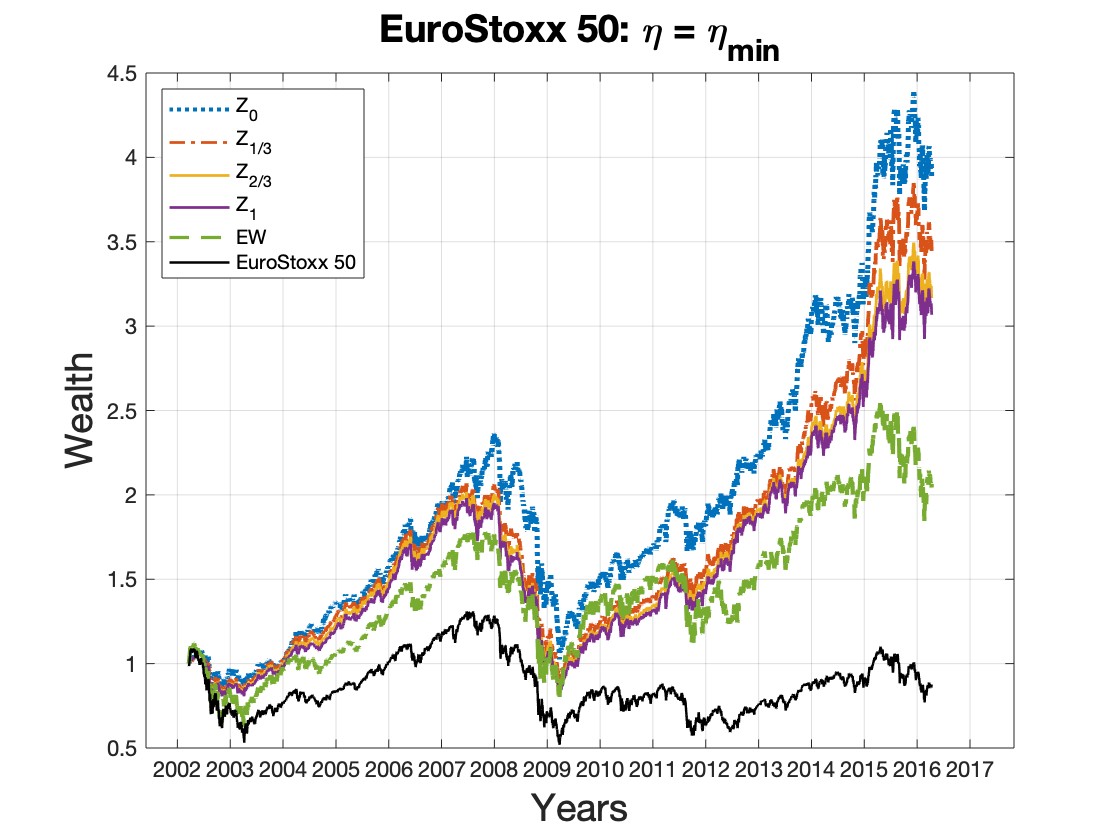}}
	\hfill
	\subfigure{\label{fig:CumulativeReturnsfor_eta1/4_EX50_0.01}
	\includegraphics[scale=0.19]{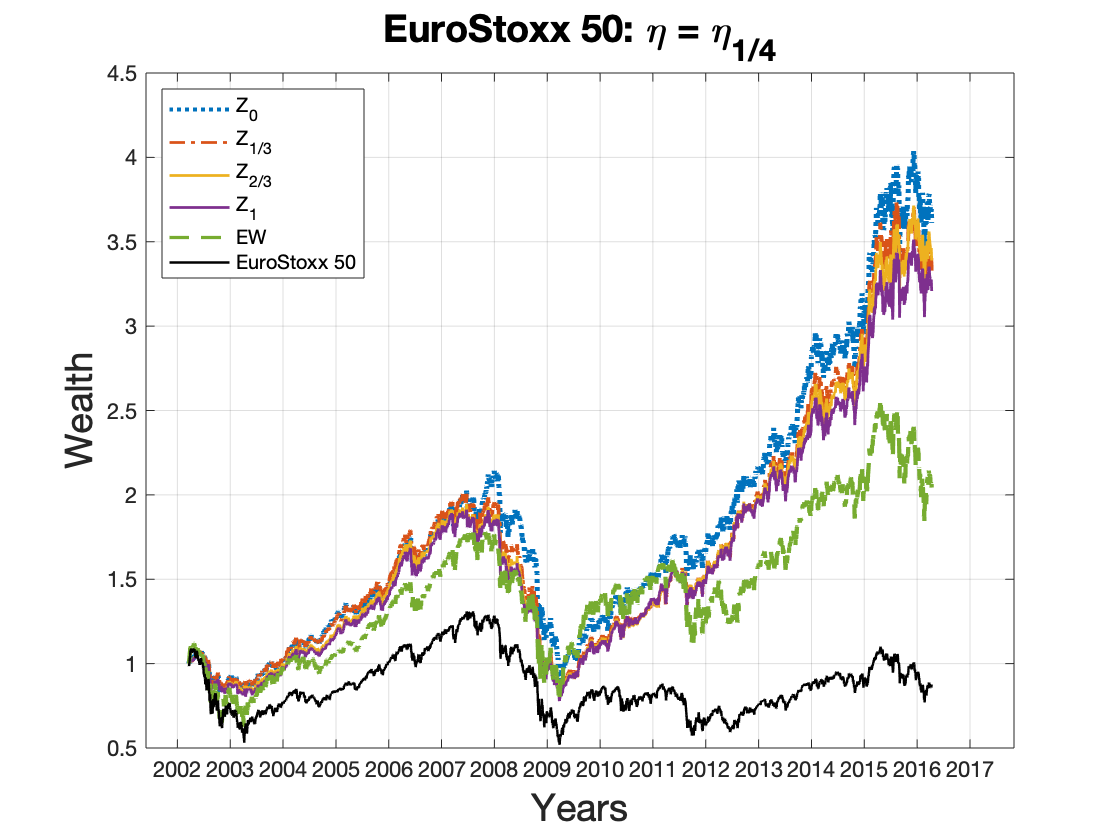}}
	\vfill
	\subfigure{\label{fig:CumulativeReturnsfor_eta1/2_EX50_0.01}
	\includegraphics[scale=0.19]{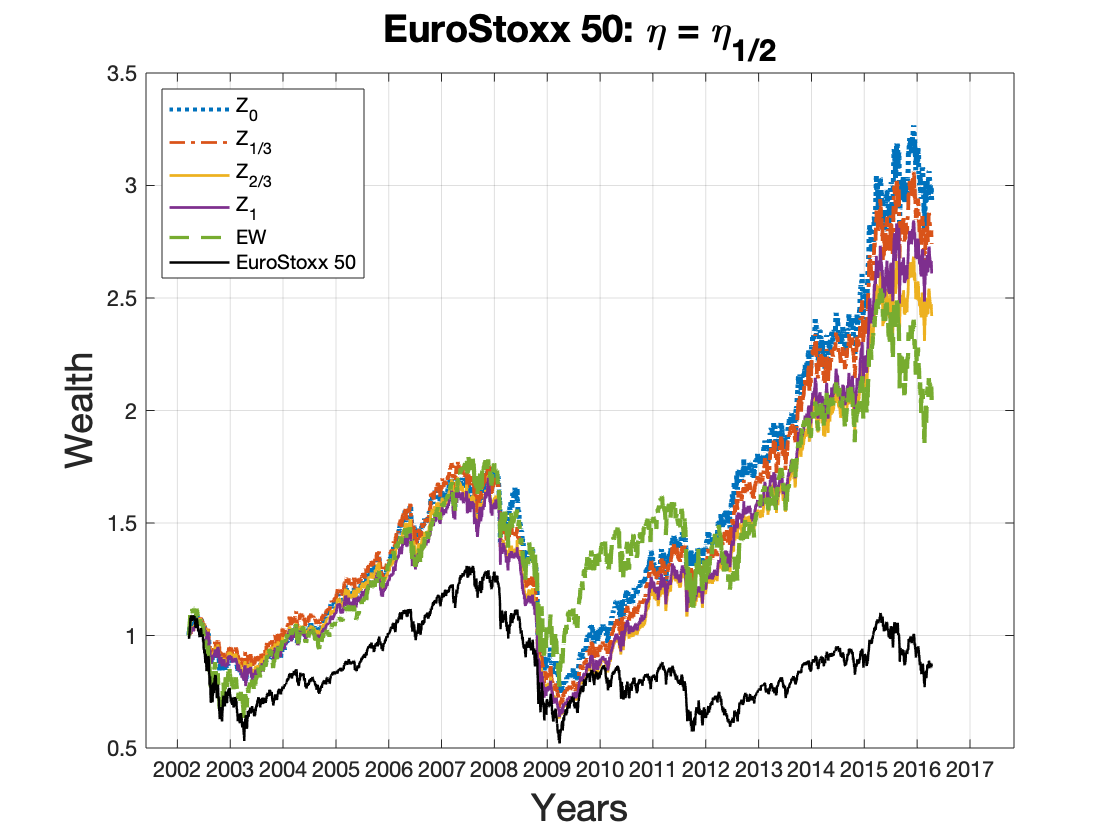}}
	\hfill
	\subfigure{\label{fig:CumulativeReturnsfor_eta_3/4_EX50_0.01}
	\includegraphics[scale=0.19]{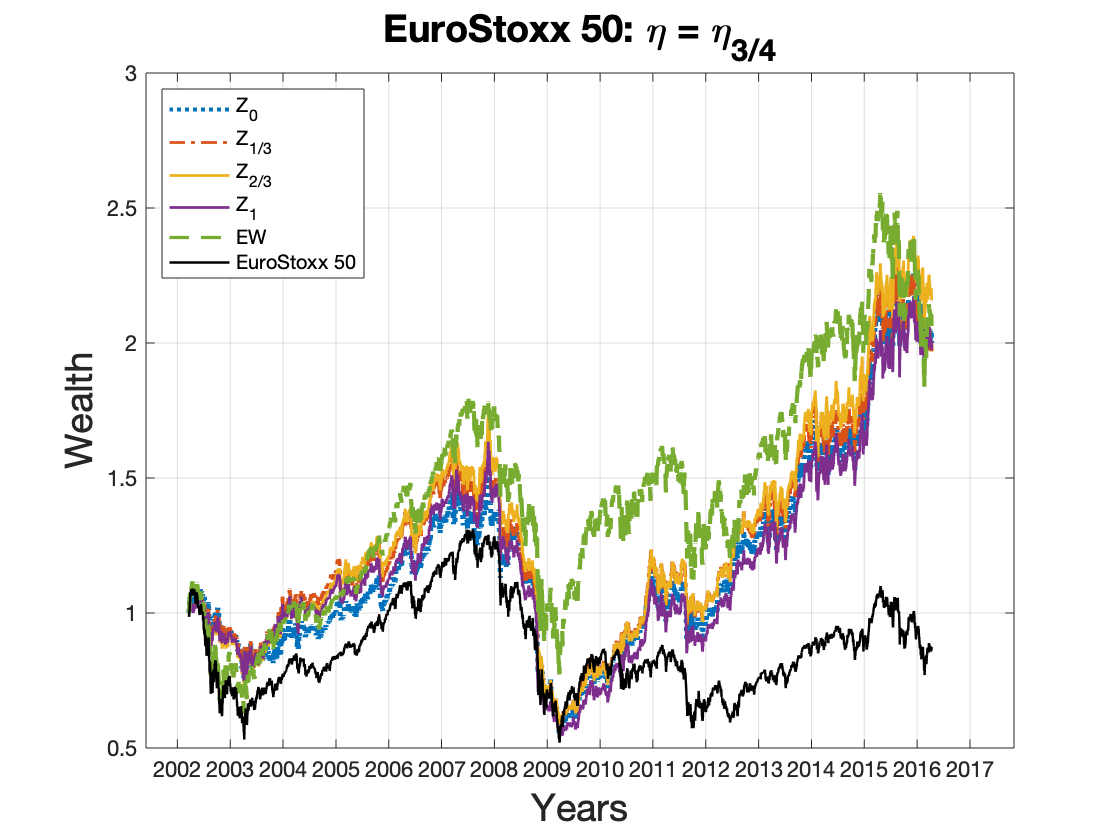}}
	\caption{Cumulative out-of-sample portfolio returns using different levels of $\eta$ and $\varepsilon=1\%$ for the EuroStoxx 50 daily dataset}
	\label{fig:CumulativeReturnsforfourleveloftargetreturn_eta_withepsilon=0.01_EX50}
\end{figure}

\begin{figure}[htbp!]
	\centering
	\subfigure{\label{fig:CumulativeReturnsfor_etamin_DJ_weekly_0.05}
	\includegraphics[scale=0.19]{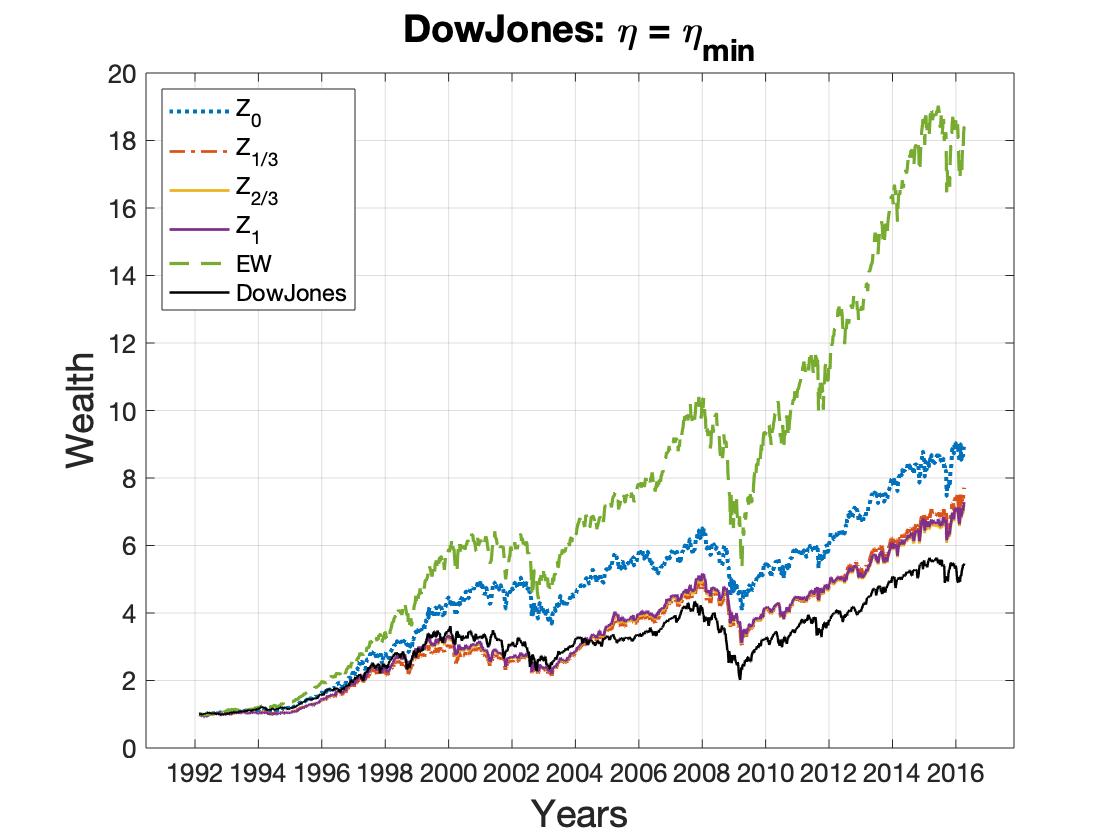}}
	\hfill
	\subfigure{\label{fig:CumulativeReturnsfor_eta1/4_DJ_weekly_0.05}
	\includegraphics[scale=0.19]{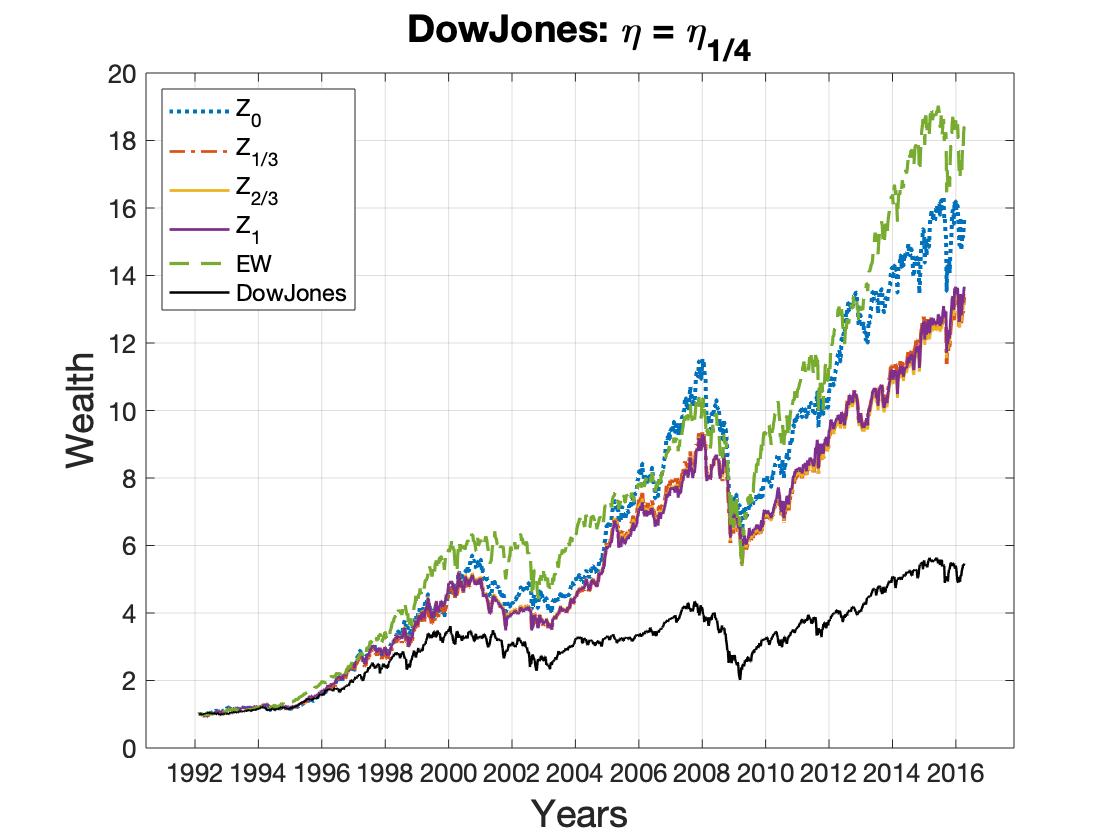}}
	\vfill
	\subfigure{\label{fig:CumulativeReturnsfor_eta1/2_DJ_weekly_0.05}
	\includegraphics[scale=0.19]{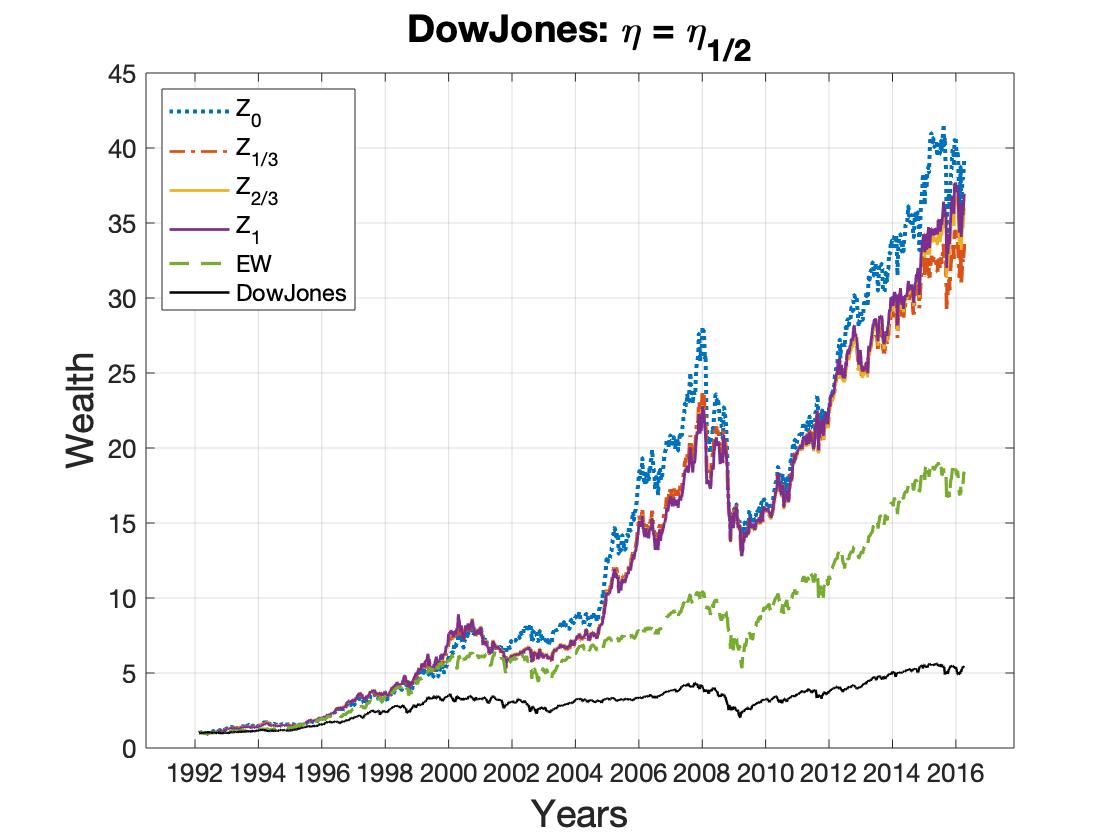}}
	\hfill
	\subfigure{\label{fig:CumulativeReturnsfor_eta3/4_DJ_weekly_0.05}
	\includegraphics[scale=0.19]{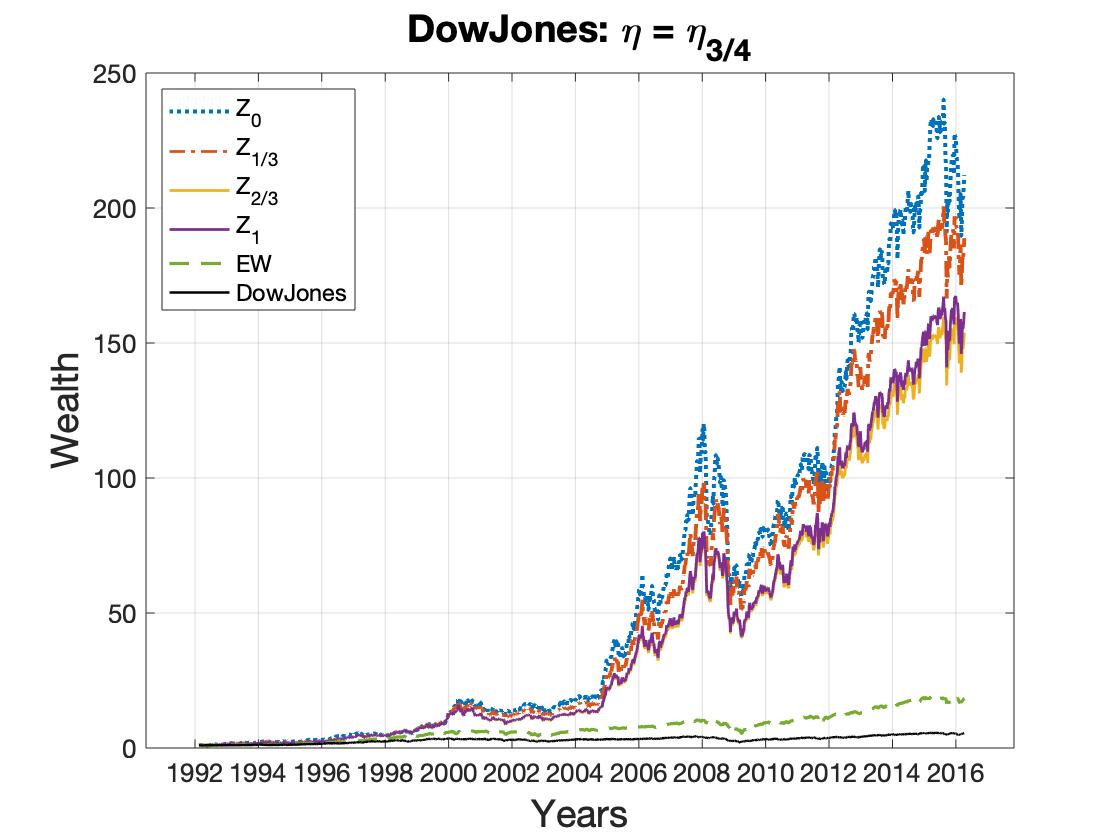}}
	\caption{Cumulative out-of-sample portfolio returns using different levels of $\eta$ and $\varepsilon=5\%$ for the DowJones weekly dataset} \label{fig:CumulativeReturnsforfourlevelof targetreturnetawithepsilon=0.05_DJ_weekly}
\end{figure}

\noindent
This section presents the out-of-sample analysis of 16
Pareto-optimal portfolios obtained from Problem \eqref{eq:Mean-Variance-VaR-single2} by appropriately varying the target levels of the portfolio expected return $\eta$ and of the portfolio VaR $z$ (see Section \ref{sec:FindingMean-Variance-VaR}).
More precisely, as shown in Figure \ref{fig: Curve_portfolios},
we consider 4 different levels of target return $\eta_{\alpha} = \eta_{\min} + \alpha\, (\eta_{\max} - \eta_{\min})$ with $\alpha\,=\,0, \,1/4, \,1/2, \,3/4$.
Furthermore, for a fixed level $\eta_{\alpha}$,
we choose 4 levels of the portfolio VaR $z_{\beta}$ in the interval $[z_{\min}(\eta_{\alpha}), z_{\max}(\eta_{\alpha})]$:
$z_{\beta}(\eta_{\alpha}) = z_{\min}(\eta_{\alpha}) + \beta \, (z_{\max}(\eta_{\alpha}) - z_{\min}(\eta_{\alpha}))$ with $\beta\,=\,0, \,1/3, \,2/3, \,1$.
Note that for $\beta=0$ we obtain the Mean-VaR optimal portfolios (see the bold black dashed line in Figure \ref{fig:VarianceVaRSurface_0,01}), while for $\beta=1$ we select the Mean-Variance optimal portfolios (see the red dashed line in Figure \ref{fig:VarianceVaRSurface_0,01}).
In Figure \ref{fig: Curve_portfolios} we report an example of the 16 Mean-Variance-VaR efficient portfolios in the Variance-VaR plane
for the DowJones daily dataset (see Table \ref{tab:DailyDatasets}).
Furthermore, to better understand the composition and diversification of these 16 Pareto-optimal portfolios,
in Table \ref{tab:NumTitoli} we show the number of selected assets.
We observe that, for a fixed target level of the portfolio expected return $\eta$, the number of selected assets
tends to
decrease when lower levels of VaR are required for the portfolio.
On the other hand, for a fixed level of the portfolio VaR,
the number of selected assets tends to decrease
when increasing the required level $\eta$ of the portfolio expected return.\\
\begin{table}[h!]
	\begin{center}
		\begin{tabular}{|c| c c c c|}
			\hline
			& \textbf{$z_0$} & \textbf{$z_{1/3}$} & \textbf{$z_{2/3}$} & \textbf{$z_1$}\\
			\hline
			\textbf{$\eta_{\min}$} & 4 & 9 & 9 & 9 \\
			\textbf{$\eta_{1/4}$} & 4 & 4 & 7 & 9 \\
			\textbf{$\eta_{1/2}$} & 3 & 4 & 4 & 5 \\
			\textbf{$\eta_{3/4}$} & 2 & 3 & 3 & 2 \\
			\hline
		\end{tabular}
	\end{center}
	\caption{Number of assets selected by each \emph{Mean-Variance-VaR} efficient portfolios}
	\label{tab:NumTitoli}
\end{table}
The out-of-sample performance of the 16 Mean-Variance-VaR efficient portfolios described above is also compared
with that of the Equally Weighted (EW) portfolio, using several performance measures commonly employed in the literature
\citep[see, e.g.,][]{cesarone2017minimum,bruni2017exact}.
More precisely,
for the out‐of‐sample portfolio
returns generated by each portfolio strategy, we evaluate the following performance measures: mean ($\mu$), Standard Deviation ($\sigma$), Sharpe ratio \citep{10.2307/2351741,sharpe1994sharpe}, Maximum Drawdown \citep[see, e.g.,][]{chekhlov2005drawdown}, Ulcer index \citep{martin1989investor}, Turnover \citep[see, e.g.,][]{hanturnover}, Sortino ratio \citep{sortino2001managing}, and Rachev ratio with confidence levels of $5\%$ and $10\%$ \citep{biglova2004different}.
%
\begin{table}[htbp]
	\centering
	\caption{Out-of-sample performance results for the EuroStoxx 50 daily dataset
		with $\varepsilon=1\%$}
	\resizebox{\textwidth}{!}{\begin{tabular}{|l|ccccccccccccccccc|}
			\cmidrule{3-18}    \multicolumn{1}{r}{} & \multicolumn{1}{r|}{} & \multicolumn{4}{c|}{\boldmath{$\eta_{min}$}} & \multicolumn{4}{c|}{\boldmath{$\eta_{1/4}$}} & \multicolumn{4}{c|}{\boldmath{$\eta_{1/2}$}} & \multicolumn{4}{c|}{\boldmath{$\eta_{3/4}$}} \\
			\cmidrule{2-18}    \multicolumn{1}{r|}{} & \multicolumn{1}{c|}{\textbf{EW}} & \multicolumn{1}{c|}{\boldmath{$z_0$}} & \multicolumn{1}{c|}{\boldmath{$z_{1/3}$}} & \multicolumn{1}{c|}{\boldmath{$z_{2/3}$}} & \multicolumn{1}{c|}{\boldmath{$z_1$}} & \multicolumn{1}{c|}{\boldmath{$z_0$}} & \multicolumn{1}{c|}{\boldmath{$z_{1/3}$}} & \multicolumn{1}{c|}{\boldmath{$z_{2/3}$}} & \multicolumn{1}{c|}{\boldmath{$z_1$}} & \multicolumn{1}{c|}{\boldmath{$z_0$}} & \multicolumn{1}{c|}{\boldmath{$z_{1/3}$}} & \multicolumn{1}{c|}{\boldmath{$z_{2/3}$}} & \multicolumn{1}{c|}{\boldmath{$z_1$}} & \multicolumn{1}{c|}{\boldmath{$z_0$}} & \multicolumn{1}{c|}{\boldmath{$z_{1/3}$}} & \multicolumn{1}{c|}{\boldmath{$z_{2/3}$}} & \boldmath{$z_1$} \\
			\midrule
			\boldmath{$\mu$}  & \multicolumn{1}{c|}{\cellcolor[rgb]{ .98,  .616,  .459}0.0301\%} & \multicolumn{1}{c|}{\cellcolor[rgb]{ .388,  .745,  .482}0.0419\%} & \multicolumn{1}{c|}{\cellcolor[rgb]{ .722,  .843,  .502}0.0382\%} & \multicolumn{1}{c|}{\cellcolor[rgb]{ .925,  .902,  .514}0.0359\%} & \multicolumn{1}{c|}{\cellcolor[rgb]{ 1,  .922,  .518}0.0350\%} & \multicolumn{1}{c|}{\cellcolor[rgb]{ .561,  .796,  .494}0.0400\%} & \multicolumn{1}{c|}{\cellcolor[rgb]{ .773,  .859,  .506}0.0376\%} & \multicolumn{1}{c|}{\cellcolor[rgb]{ .737,  .847,  .506}0.0380\%} & \multicolumn{1}{c|}{\cellcolor[rgb]{ .863,  .882,  .51}0.0366\%} & \multicolumn{1}{c|}{\cellcolor[rgb]{ .984,  .918,  .518}0.0352\%} & \multicolumn{1}{c|}{\cellcolor[rgb]{ .992,  .816,  .494}0.0334\%} & \multicolumn{1}{c|}{\cellcolor[rgb]{ .98,  .608,  .455}0.0300\%} & \multicolumn{1}{c|}{\cellcolor[rgb]{ .988,  .741,  .482}0.0321\%} & \multicolumn{1}{c|}{\cellcolor[rgb]{ .973,  .424,  .42}0.0270\%} & \multicolumn{1}{c|}{\cellcolor[rgb]{ .973,  .412,  .42}0.0268\%} & \multicolumn{1}{c|}{\cellcolor[rgb]{ .98,  .565,  .447}0.0293\%} & \cellcolor[rgb]{ .973,  .424,  .42}0.0270\% \\
			\midrule
			\boldmath{$\sigma$} & \multicolumn{1}{c|}{\cellcolor[rgb]{ .973,  .412,  .42}0.0146} & \multicolumn{1}{c|}{\cellcolor[rgb]{ .639,  .816,  .494}0.0101} & \multicolumn{1}{c|}{\cellcolor[rgb]{ .427,  .757,  .482}0.0096} & \multicolumn{1}{c|}{\cellcolor[rgb]{ .388,  .745,  .482}0.0095} & \multicolumn{1}{c|}{\cellcolor[rgb]{ .388,  .745,  .482}0.0095} & \multicolumn{1}{c|}{\cellcolor[rgb]{ .643,  .82,  .494}0.0101} & \multicolumn{1}{c|}{\cellcolor[rgb]{ .553,  .792,  .49}0.0099} & \multicolumn{1}{c|}{\cellcolor[rgb]{ .529,  .784,  .49}0.0099} & \multicolumn{1}{c|}{\cellcolor[rgb]{ .541,  .788,  .49}0.0099} & \multicolumn{1}{c|}{\cellcolor[rgb]{ 1,  .91,  .518}0.0110} & \multicolumn{1}{c|}{\cellcolor[rgb]{ 1,  .922,  .518}0.0109} & \multicolumn{1}{c|}{\cellcolor[rgb]{ 1,  .922,  .518}0.0109} & \multicolumn{1}{c|}{\cellcolor[rgb]{ 1,  .91,  .518}0.0110} & \multicolumn{1}{c|}{\cellcolor[rgb]{ .988,  .671,  .471}0.0128} & \multicolumn{1}{c|}{\cellcolor[rgb]{ .988,  .647,  .467}0.0129} & \multicolumn{1}{c|}{\cellcolor[rgb]{ .988,  .647,  .467}0.0129} & \cellcolor[rgb]{ .984,  .631,  .463}0.0130 \\
			\midrule
			\textbf{Sharpe} & \multicolumn{1}{c|}{\cellcolor[rgb]{ .973,  .412,  .42}0.0206} & \multicolumn{1}{c|}{\cellcolor[rgb]{ .388,  .745,  .482}0.0415} & \multicolumn{1}{c|}{\cellcolor[rgb]{ .51,  .78,  .49}0.0397} & \multicolumn{1}{c|}{\cellcolor[rgb]{ .639,  .82,  .498}0.0376} & \multicolumn{1}{c|}{\cellcolor[rgb]{ .698,  .835,  .502}0.0367} & \multicolumn{1}{c|}{\cellcolor[rgb]{ .514,  .784,  .49}0.0396} & \multicolumn{1}{c|}{\cellcolor[rgb]{ .62,  .812,  .498}0.0380} & \multicolumn{1}{c|}{\cellcolor[rgb]{ .576,  .8,  .494}0.0386} & \multicolumn{1}{c|}{\cellcolor[rgb]{ .678,  .831,  .502}0.0370} & \multicolumn{1}{c|}{\cellcolor[rgb]{ 1,  .922,  .518}0.0320} & \multicolumn{1}{c|}{\cellcolor[rgb]{ .996,  .859,  .502}0.0306} & \multicolumn{1}{c|}{\cellcolor[rgb]{ .988,  .718,  .478}0.0275} & \multicolumn{1}{c|}{\cellcolor[rgb]{ .992,  .796,  .49}0.0292} & \multicolumn{1}{c|}{\cellcolor[rgb]{ .973,  .435,  .424}0.0212} & \multicolumn{1}{c|}{\cellcolor[rgb]{ .973,  .416,  .42}0.0207} & \multicolumn{1}{c|}{\cellcolor[rgb]{ .976,  .502,  .435}0.0227} & \cellcolor[rgb]{ .973,  .416,  .42}0.0207 \\
			\midrule
			\textbf{Maximum Drawdown} & \multicolumn{1}{c|}{\cellcolor[rgb]{ .463,  .769,  .49}-0.5696} & \multicolumn{1}{c|}{\cellcolor[rgb]{ .388,  .745,  .482}-0.5662} & \multicolumn{1}{c|}{\cellcolor[rgb]{ .773,  .859,  .506}-0.5841} & \multicolumn{1}{c|}{\cellcolor[rgb]{ .996,  .918,  .514}-0.5952} & \multicolumn{1}{c|}{\cellcolor[rgb]{ 1,  .922,  .518}-0.5946} & \multicolumn{1}{c|}{\cellcolor[rgb]{ .69,  .835,  .502}-0.5802} & \multicolumn{1}{c|}{\cellcolor[rgb]{ .969,  .914,  .518}-0.5930} & \multicolumn{1}{c|}{\cellcolor[rgb]{ .973,  .914,  .518}-0.5933} & \multicolumn{1}{c|}{\cellcolor[rgb]{ .973,  .914,  .518}-0.5932} & \multicolumn{1}{c|}{\cellcolor[rgb]{ .694,  .835,  .502}-0.5803} & \multicolumn{1}{c|}{\cellcolor[rgb]{ .988,  .753,  .482}-0.6198} & \multicolumn{1}{c|}{\cellcolor[rgb]{ .984,  .678,  .471}-0.6316} & \multicolumn{1}{c|}{\cellcolor[rgb]{ .988,  .722,  .478}-0.6250} & \multicolumn{1}{c|}{\cellcolor[rgb]{ .98,  .62,  .459}-0.6402} & \multicolumn{1}{c|}{\cellcolor[rgb]{ .98,  .573,  .447}-0.6477} & \multicolumn{1}{c|}{\cellcolor[rgb]{ .973,  .451,  .427}-0.6659} & \cellcolor[rgb]{ .973,  .412,  .42}-0.6722 \\
			\midrule
			\textbf{Ulcer} & \multicolumn{1}{c|}{\cellcolor[rgb]{ .635,  .816,  .494}0.1916} & \multicolumn{1}{c|}{\cellcolor[rgb]{ .388,  .745,  .482}0.1799} & \multicolumn{1}{c|}{\cellcolor[rgb]{ .824,  .871,  .506}0.2004} & \multicolumn{1}{c|}{\cellcolor[rgb]{ .961,  .91,  .514}0.2069} & \multicolumn{1}{c|}{\cellcolor[rgb]{ 1,  .922,  .518}0.2087} & \multicolumn{1}{c|}{\cellcolor[rgb]{ .682,  .827,  .498}0.1939} & \multicolumn{1}{c|}{\cellcolor[rgb]{ 1,  .918,  .518}0.2096} & \multicolumn{1}{c|}{\cellcolor[rgb]{ .933,  .902,  .514}0.2057} & \multicolumn{1}{c|}{\cellcolor[rgb]{ .91,  .894,  .51}0.2045} & \multicolumn{1}{c|}{\cellcolor[rgb]{ .82,  .867,  .506}0.2002} & \multicolumn{1}{c|}{\cellcolor[rgb]{ .996,  .847,  .506}0.2197} & \multicolumn{1}{c|}{\cellcolor[rgb]{ .992,  .745,  .486}0.2343} & \multicolumn{1}{c|}{\cellcolor[rgb]{ .992,  .776,  .49}0.2297} & \multicolumn{1}{c|}{\cellcolor[rgb]{ .988,  .667,  .471}0.2453} & \multicolumn{1}{c|}{\cellcolor[rgb]{ .984,  .612,  .459}0.2532} & \multicolumn{1}{c|}{\cellcolor[rgb]{ .976,  .478,  .435}0.2721} & \cellcolor[rgb]{ .973,  .412,  .42}0.2811 \\
			\midrule
			\textbf{Turnover} & \multicolumn{1}{c|}{0} & \multicolumn{1}{c|}{\cellcolor[rgb]{ .851,  .878,  .506}0.5698} & \multicolumn{1}{c|}{\cellcolor[rgb]{ .533,  .784,  .49}0.4208} & \multicolumn{1}{c|}{\cellcolor[rgb]{ .439,  .757,  .482}0.3759} & \multicolumn{1}{c|}{\cellcolor[rgb]{ .388,  .745,  .482}0.3508} & \multicolumn{1}{c|}{\cellcolor[rgb]{ .973,  .914,  .514}0.6282} & \multicolumn{1}{c|}{\cellcolor[rgb]{ .792,  .859,  .502}0.5423} & \multicolumn{1}{c|}{\cellcolor[rgb]{ .694,  .831,  .498}0.4968} & \multicolumn{1}{c|}{\cellcolor[rgb]{ .655,  .82,  .494}0.4771} & \multicolumn{1}{c|}{\cellcolor[rgb]{ .992,  .718,  .478}0.7213} & \multicolumn{1}{c|}{\cellcolor[rgb]{ .996,  .808,  .498}0.6856} & \multicolumn{1}{c|}{\cellcolor[rgb]{ 1,  .894,  .514}0.6512} & \multicolumn{1}{c|}{\cellcolor[rgb]{ 1,  .886,  .514}0.6549} & \multicolumn{1}{c|}{\cellcolor[rgb]{ .973,  .412,  .42}0.8402} & \multicolumn{1}{c|}{\cellcolor[rgb]{ .976,  .435,  .424}0.8324} & \multicolumn{1}{c|}{\cellcolor[rgb]{ .98,  .49,  .435}0.8097} & \cellcolor[rgb]{ .976,  .459,  .431}0.8220 \\
			\midrule
			\textbf{Sortino} & \multicolumn{1}{c|}{\cellcolor[rgb]{ .973,  .435,  .424}0.0297} & \multicolumn{1}{c|}{\cellcolor[rgb]{ .388,  .745,  .482}0.0590} & \multicolumn{1}{c|}{\cellcolor[rgb]{ .525,  .788,  .494}0.0559} & \multicolumn{1}{c|}{\cellcolor[rgb]{ .663,  .824,  .498}0.0528} & \multicolumn{1}{c|}{\cellcolor[rgb]{ .722,  .843,  .502}0.0514} & \multicolumn{1}{c|}{\cellcolor[rgb]{ .522,  .784,  .49}0.0560} & \multicolumn{1}{c|}{\cellcolor[rgb]{ .639,  .82,  .498}0.0532} & \multicolumn{1}{c|}{\cellcolor[rgb]{ .592,  .804,  .494}0.0543} & \multicolumn{1}{c|}{\cellcolor[rgb]{ .69,  .835,  .502}0.0521} & \multicolumn{1}{c|}{\cellcolor[rgb]{ 1,  .922,  .518}0.0449} & \multicolumn{1}{c|}{\cellcolor[rgb]{ .996,  .855,  .502}0.0428} & \multicolumn{1}{c|}{\cellcolor[rgb]{ .988,  .706,  .475}0.0382} & \multicolumn{1}{c|}{\cellcolor[rgb]{ .992,  .788,  .49}0.0408} & \multicolumn{1}{c|}{\cellcolor[rgb]{ .973,  .431,  .424}0.0295} & \multicolumn{1}{c|}{\cellcolor[rgb]{ .973,  .412,  .42}0.0288} & \multicolumn{1}{c|}{\cellcolor[rgb]{ .976,  .498,  .435}0.0317} & \cellcolor[rgb]{ .973,  .412,  .42}0.0289 \\
			\midrule
			\textbf{Rachev 5\%} & \multicolumn{1}{c|}{\cellcolor[rgb]{ .388,  .745,  .482}0.9974} & \multicolumn{1}{c|}{\cellcolor[rgb]{ .851,  .878,  .51}0.9436} & \multicolumn{1}{c|}{\cellcolor[rgb]{ 1,  .922,  .518}0.9262} & \multicolumn{1}{c|}{\cellcolor[rgb]{ .992,  .776,  .486}0.9213} & \multicolumn{1}{c|}{\cellcolor[rgb]{ .984,  .647,  .463}0.9169} & \multicolumn{1}{c|}{\cellcolor[rgb]{ .953,  .91,  .518}0.9318} & \multicolumn{1}{c|}{\cellcolor[rgb]{ .973,  .412,  .42}0.9087} & \multicolumn{1}{c|}{\cellcolor[rgb]{ .984,  .647,  .463}0.9169} & \multicolumn{1}{c|}{\cellcolor[rgb]{ .996,  .851,  .502}0.9239} & \multicolumn{1}{c|}{\cellcolor[rgb]{ .953,  .91,  .518}0.9320} & \multicolumn{1}{c|}{\cellcolor[rgb]{ .996,  .89,  .51}0.9252} & \multicolumn{1}{c|}{\cellcolor[rgb]{ .988,  .706,  .475}0.9188} & \multicolumn{1}{c|}{\cellcolor[rgb]{ .992,  .824,  .498}0.9230} & \multicolumn{1}{c|}{\cellcolor[rgb]{ .91,  .898,  .514}0.9370} & \multicolumn{1}{c|}{\cellcolor[rgb]{ .914,  .898,  .514}0.9363} & \multicolumn{1}{c|}{\cellcolor[rgb]{ .816,  .871,  .51}0.9477} & \cellcolor[rgb]{ .898,  .894,  .514}0.9384 \\
			\midrule
			\textbf{Rachev 10\%} & \multicolumn{1}{c|}{\cellcolor[rgb]{ .388,  .745,  .482}0.9939} & \multicolumn{1}{c|}{\cellcolor[rgb]{ .851,  .878,  .51}0.9636} & \multicolumn{1}{c|}{\cellcolor[rgb]{ .984,  .655,  .467}0.9448} & \multicolumn{1}{c|}{\cellcolor[rgb]{ .98,  .557,  .447}0.9414} & \multicolumn{1}{c|}{\cellcolor[rgb]{ .976,  .486,  .431}0.9390} & \multicolumn{1}{c|}{\cellcolor[rgb]{ .965,  .914,  .518}0.9560} & \multicolumn{1}{c|}{\cellcolor[rgb]{ .973,  .412,  .42}0.9365} & \multicolumn{1}{c|}{\cellcolor[rgb]{ .984,  .655,  .467}0.9447} & \multicolumn{1}{c|}{\cellcolor[rgb]{ .988,  .769,  .486}0.9485} & \multicolumn{1}{c|}{\cellcolor[rgb]{ .882,  .89,  .514}0.9614} & \multicolumn{1}{c|}{\cellcolor[rgb]{ .992,  .843,  .502}0.9511} & \multicolumn{1}{c|}{\cellcolor[rgb]{ .984,  .671,  .467}0.9453} & \multicolumn{1}{c|}{\cellcolor[rgb]{ .992,  .922,  .518}0.9543} & \multicolumn{1}{c|}{\cellcolor[rgb]{ 1,  .922,  .518}0.9536} & \multicolumn{1}{c|}{\cellcolor[rgb]{ .867,  .886,  .514}0.9625} & \multicolumn{1}{c|}{\cellcolor[rgb]{ .788,  .863,  .506}0.9676} & \cellcolor[rgb]{ .867,  .886,  .514}0.9624 \\
			\midrule
			\textbf{Time (in seconds)} & \multicolumn{17}{c|}{1863} \\
			\bottomrule
	\end{tabular}}%
	\label{tab:Daily_EuroStoxx50_0.01}%
\end{table}%
\\
In Table \ref{tab:Daily_EuroStoxx50_0.01} we provide the computational results obtained by the 16 analyzed portfolio strategies
and by the benchmark EW portfolio
on the EuroStoxx 50 daily dataset (see Table \ref{tab:DailyDatasets}), with $\varepsilon=1\%$.
The rank of the performance results is shown in different colors.
More precisely, for each row the colors range from deep-green to deep-red, where deep-green represents the best performance, while deep-red represents the worst one.
We observe that the EW portfolio shows the highest volatilities, the lowest Sharpe ratio and one of the worst Sortino ratios.
Furthermore, we notice that by considering stronger conditions on the portfolio VaR,
the Mean-Variance-VaR portfolios tend to improve their out-of sample performance.
This is more evident when choosing low levels of the portfolio expected return $\eta$,
for which the Pareto-optimal portfolios with low levels of $z$
achieve the best performance in terms of mean, Sharpe ratio, and Sortino ratio.
This behavior is also confirmed by the trend of the cumulative out-of-sample portfolio returns, reported in Figure \ref{fig:CumulativeReturnsforfourleveloftargetreturn_eta_withepsilon=0.01_EX50}.
Although the EW portfolio seems to be preferable to the Mean-Variance-VaR portfolios with $\eta=\eta_{3/4}$,
for lower levels of  the required portfolio expected return, the EW portfolio tends to be dominated in terms of the cumulative returns.
Furthermore, there seems to be a behavioral pattern of portfolios with lower levels of $z$, namely $z_{0}$ and $z_{1/3}$,
which have better performance w.r.t. the Mean-Variance portfolios (namely, the Mean-Variance-VaR portfolios with $z_1$).
This is more remarkable for the lowest level, $\eta_{min}$, of the portfolio expected return.
%
\begin{table}[htbp]
	\centering
	\caption{Out-of-sample performance results for the DowJones weekly dataset
		with $\varepsilon=5\%$}
	\resizebox{\textwidth}{!}{\begin{tabular}{|l|ccccccccccccccccc|}
			\cmidrule{3-18}    \multicolumn{1}{r}{} & \multicolumn{1}{r|}{} & \multicolumn{4}{c|}{\boldmath{$\eta_{min}$}} & \multicolumn{4}{c|}{\boldmath{$\eta_{1/4}$}} & \multicolumn{4}{c|}{\boldmath{$\eta_{1/2}$}} & \multicolumn{4}{c|}{\boldmath{$\eta_{3/4}$}} \\
			\cmidrule{2-18}    \multicolumn{1}{r|}{} & \multicolumn{1}{c|}{\textbf{EW}} & \multicolumn{1}{c|}{\boldmath{$z_0$}} & \multicolumn{1}{c|}{\boldmath{$z_{1/3}$}} & \multicolumn{1}{c|}{\boldmath{$z_{2/3}$}} & \multicolumn{1}{c|}{\boldmath{$z_1$}} & \multicolumn{1}{c|}{\boldmath{$z_0$}} & \multicolumn{1}{c|}{\boldmath{$z_{1/3}$}} & \multicolumn{1}{c|}{\boldmath{$z_{2/3}$}} & \multicolumn{1}{c|}{\boldmath{$z_1$}} & \multicolumn{1}{c|}{\boldmath{$z_0$}} & \multicolumn{1}{c|}{\boldmath{$z_{1/3}$}} & \multicolumn{1}{c|}{\boldmath{$z_{2/3}$}} & \multicolumn{1}{c|}{\boldmath{$z_1$}} & \multicolumn{1}{c|}{\boldmath{$z_0$}} & \multicolumn{1}{c|}{\boldmath{$z_{1/3}$}} & \multicolumn{1}{c|}{\boldmath{$z_{2/3}$}} & \boldmath{$z_1$} \\
			\midrule
			\boldmath{$\mu$} & \multicolumn{1}{c|}{\cellcolor[rgb]{ 1,  .922,  .518}0.0026} & \multicolumn{1}{c|}{\cellcolor[rgb]{ .976,  .537,  .443}0.0020} & \multicolumn{1}{c|}{\cellcolor[rgb]{ .973,  .412,  .42}0.0018} & \multicolumn{1}{c|}{\cellcolor[rgb]{ .973,  .412,  .42}0.0018} & \multicolumn{1}{c|}{\cellcolor[rgb]{ .973,  .412,  .42}0.0018} & \multicolumn{1}{c|}{\cellcolor[rgb]{ .996,  .855,  .502}0.0025} & \multicolumn{1}{c|}{\cellcolor[rgb]{ .988,  .729,  .478}0.0023} & \multicolumn{1}{c|}{\cellcolor[rgb]{ .988,  .729,  .478}0.0023} & \multicolumn{1}{c|}{\cellcolor[rgb]{ .988,  .729,  .478}0.0023} & \multicolumn{1}{c|}{\cellcolor[rgb]{ .788,  .863,  .506}0.0034} & \multicolumn{1}{c|}{\cellcolor[rgb]{ .843,  .878,  .51}0.0032} & \multicolumn{1}{c|}{\cellcolor[rgb]{ .843,  .878,  .51}0.0032} & \multicolumn{1}{c|}{\cellcolor[rgb]{ .843,  .878,  .51}0.0032} & \multicolumn{1}{c|}{\cellcolor[rgb]{ .388,  .745,  .482}0.0049} & \multicolumn{1}{c|}{\cellcolor[rgb]{ .416,  .753,  .486}0.0048} & \multicolumn{1}{c|}{\cellcolor[rgb]{ .471,  .769,  .49}0.0046} & \cellcolor[rgb]{ .443,  .761,  .486}0.0047 \\
			\midrule
			\boldmath{$\sigma$} & \multicolumn{1}{c|}{\cellcolor[rgb]{ 1,  .922,  .518}0.0242} & \multicolumn{1}{c|}{\cellcolor[rgb]{ .533,  .784,  .49}0.0210} & \multicolumn{1}{c|}{\cellcolor[rgb]{ .4,  .749,  .482}0.0201} & \multicolumn{1}{c|}{\cellcolor[rgb]{ .4,  .749,  .482}0.0201} & \multicolumn{1}{c|}{\cellcolor[rgb]{ .388,  .745,  .482}0.0200} & \multicolumn{1}{c|}{\cellcolor[rgb]{ .898,  .89,  .51}0.0235} & \multicolumn{1}{c|}{\cellcolor[rgb]{ .737,  .843,  .502}0.0224} & \multicolumn{1}{c|}{\cellcolor[rgb]{ .737,  .843,  .502}0.0224} & \multicolumn{1}{c|}{\cellcolor[rgb]{ .737,  .843,  .502}0.0224} & \multicolumn{1}{c|}{\cellcolor[rgb]{ .992,  .71,  .478}0.0294} & \multicolumn{1}{c|}{\cellcolor[rgb]{ .996,  .784,  .494}0.0276} & \multicolumn{1}{c|}{\cellcolor[rgb]{ .996,  .792,  .494}0.0274} & \multicolumn{1}{c|}{\cellcolor[rgb]{ .996,  .792,  .494}0.0274} & \multicolumn{1}{c|}{\cellcolor[rgb]{ .973,  .412,  .42}0.0367} & \multicolumn{1}{c|}{\cellcolor[rgb]{ .976,  .455,  .427}0.0357} & \multicolumn{1}{c|}{\cellcolor[rgb]{ .976,  .459,  .431}0.0356} & \cellcolor[rgb]{ .976,  .459,  .431}0.0356 \\
			\midrule
			\textbf{Sharpe} & \multicolumn{1}{c|}{\cellcolor[rgb]{ 1,  .922,  .518}0.1077} & \multicolumn{1}{c|}{\cellcolor[rgb]{ .976,  .545,  .443}0.0934} & \multicolumn{1}{c|}{\cellcolor[rgb]{ .973,  .471,  .427}0.0906} & \multicolumn{1}{c|}{\cellcolor[rgb]{ .973,  .412,  .42}0.0883} & \multicolumn{1}{c|}{\cellcolor[rgb]{ .973,  .424,  .42}0.0888} & \multicolumn{1}{c|}{\cellcolor[rgb]{ .992,  .843,  .502}0.1048} & \multicolumn{1}{c|}{\cellcolor[rgb]{ .992,  .8,  .494}0.1032} & \multicolumn{1}{c|}{\cellcolor[rgb]{ .992,  .808,  .494}0.1035} & \multicolumn{1}{c|}{\cellcolor[rgb]{ .992,  .816,  .494}0.1038} & \multicolumn{1}{c|}{\cellcolor[rgb]{ .859,  .882,  .51}0.1140} & \multicolumn{1}{c|}{\cellcolor[rgb]{ .839,  .875,  .51}0.1149} & \multicolumn{1}{c|}{\cellcolor[rgb]{ .773,  .859,  .506}0.1178} & \multicolumn{1}{c|}{\cellcolor[rgb]{ .757,  .855,  .506}0.1184} & \multicolumn{1}{c|}{\cellcolor[rgb]{ .396,  .749,  .486}0.1343} & \multicolumn{1}{c|}{\cellcolor[rgb]{ .388,  .745,  .482}0.1346} & \multicolumn{1}{c|}{\cellcolor[rgb]{ .49,  .776,  .49}0.1302} & \cellcolor[rgb]{ .463,  .769,  .49}0.1314 \\
			\midrule
			\textbf{Maximum Drawdown} & \multicolumn{1}{c|}{\cellcolor[rgb]{ .984,  .635,  .463}-0.4928} & \multicolumn{1}{c|}{\cellcolor[rgb]{ .388,  .745,  .482}-0.3739} & \multicolumn{1}{c|}{\cellcolor[rgb]{ .443,  .761,  .486}-0.3798} & \multicolumn{1}{c|}{\cellcolor[rgb]{ .58,  .804,  .494}-0.3948} & \multicolumn{1}{c|}{\cellcolor[rgb]{ .608,  .808,  .498}-0.3976} & \multicolumn{1}{c|}{\cellcolor[rgb]{ .996,  .851,  .502}-0.4534} & \multicolumn{1}{c|}{\cellcolor[rgb]{ .796,  .863,  .506}-0.4183} & \multicolumn{1}{c|}{\cellcolor[rgb]{ .718,  .843,  .502}-0.4097} & \multicolumn{1}{c|}{\cellcolor[rgb]{ .62,  .812,  .498}-0.3991} & \multicolumn{1}{c|}{\cellcolor[rgb]{ .973,  .435,  .424}-0.5298} & \multicolumn{1}{c|}{\cellcolor[rgb]{ .992,  .835,  .498}-0.4565} & \multicolumn{1}{c|}{\cellcolor[rgb]{ 1,  .922,  .518}-0.4410} & \multicolumn{1}{c|}{\cellcolor[rgb]{ .976,  .914,  .518}-0.4380} & \multicolumn{1}{c|}{\cellcolor[rgb]{ .973,  .412,  .42}-0.5344} & \multicolumn{1}{c|}{\cellcolor[rgb]{ .988,  .718,  .478}-0.4778} & \multicolumn{1}{c|}{\cellcolor[rgb]{ .984,  .698,  .475}-0.4816} & \cellcolor[rgb]{ .984,  .682,  .471}-0.4842 \\
			\midrule
			\textbf{Ulcer} & \multicolumn{1}{c|}{\cellcolor[rgb]{ .388,  .745,  .482}0.0926} & \multicolumn{1}{c|}{\cellcolor[rgb]{ .388,  .745,  .482}0.0925} & \multicolumn{1}{c|}{\cellcolor[rgb]{ .471,  .769,  .486}0.0982} & \multicolumn{1}{c|}{\cellcolor[rgb]{ .698,  .831,  .498}0.1139} & \multicolumn{1}{c|}{\cellcolor[rgb]{ .729,  .843,  .502}0.1162} & \multicolumn{1}{c|}{\cellcolor[rgb]{ 1,  .922,  .518}0.1346} & \multicolumn{1}{c|}{\cellcolor[rgb]{ .843,  .875,  .506}0.1239} & \multicolumn{1}{c|}{\cellcolor[rgb]{ .839,  .875,  .506}0.1236} & \multicolumn{1}{c|}{\cellcolor[rgb]{ .808,  .863,  .506}0.1214} & \multicolumn{1}{c|}{\cellcolor[rgb]{ .984,  .592,  .455}0.1553} & \multicolumn{1}{c|}{\cellcolor[rgb]{ .988,  .706,  .478}0.1482} & \multicolumn{1}{c|}{\cellcolor[rgb]{ .992,  .71,  .478}0.1478} & \multicolumn{1}{c|}{\cellcolor[rgb]{ .988,  .682,  .475}0.1496} & \multicolumn{1}{c|}{\cellcolor[rgb]{ .98,  .522,  .443}0.1596} & \multicolumn{1}{c|}{\cellcolor[rgb]{ .988,  .659,  .467}0.1511} & \multicolumn{1}{c|}{\cellcolor[rgb]{ .973,  .412,  .42}0.1663} & \cellcolor[rgb]{ .976,  .455,  .431}0.1637 \\
			\midrule
			\textbf{Turnover} & \multicolumn{1}{c|}{0} & \multicolumn{1}{c|}{\cellcolor[rgb]{ .988,  .698,  .478}0.5488} & \multicolumn{1}{c|}{\cellcolor[rgb]{ .827,  .871,  .506}0.3960} & \multicolumn{1}{c|}{\cellcolor[rgb]{ .482,  .773,  .486}0.3253} & \multicolumn{1}{c|}{\cellcolor[rgb]{ .388,  .745,  .482}0.3054} & \multicolumn{1}{c|}{\cellcolor[rgb]{ .973,  .412,  .42}0.6995} & \multicolumn{1}{c|}{\cellcolor[rgb]{ 1,  .898,  .514}0.4455} & \multicolumn{1}{c|}{\cellcolor[rgb]{ .761,  .851,  .502}0.3821} & \multicolumn{1}{c|}{\cellcolor[rgb]{ .725,  .839,  .498}0.3748} & \multicolumn{1}{c|}{\cellcolor[rgb]{ .976,  .482,  .435}0.6631} & \multicolumn{1}{c|}{\cellcolor[rgb]{ .996,  .808,  .498}0.4916} & \multicolumn{1}{c|}{\cellcolor[rgb]{ 1,  .918,  .518}0.4352} & \multicolumn{1}{c|}{\cellcolor[rgb]{ .976,  .914,  .514}0.4270} & \multicolumn{1}{c|}{\cellcolor[rgb]{ .992,  .725,  .482}0.5355} & \multicolumn{1}{c|}{\cellcolor[rgb]{ 1,  .898,  .514}0.4451} & \multicolumn{1}{c|}{\cellcolor[rgb]{ .965,  .91,  .514}0.4244} & \cellcolor[rgb]{ .941,  .902,  .514}0.4197 \\
			\midrule
			\textbf{Sortino} & \multicolumn{1}{c|}{\cellcolor[rgb]{ 1,  .922,  .518}0.1634} & \multicolumn{1}{c|}{\cellcolor[rgb]{ .98,  .569,  .447}0.1394} & \multicolumn{1}{c|}{\cellcolor[rgb]{ .973,  .478,  .431}0.1332} & \multicolumn{1}{c|}{\cellcolor[rgb]{ .973,  .412,  .42}0.1286} & \multicolumn{1}{c|}{\cellcolor[rgb]{ .973,  .42,  .42}0.1294} & \multicolumn{1}{c|}{\cellcolor[rgb]{ .992,  .808,  .494}0.1559} & \multicolumn{1}{c|}{\cellcolor[rgb]{ .992,  .792,  .49}0.1548} & \multicolumn{1}{c|}{\cellcolor[rgb]{ .992,  .8,  .494}0.1552} & \multicolumn{1}{c|}{\cellcolor[rgb]{ .992,  .808,  .494}0.1557} & \multicolumn{1}{c|}{\cellcolor[rgb]{ .878,  .886,  .514}0.1731} & \multicolumn{1}{c|}{\cellcolor[rgb]{ .847,  .878,  .51}0.1755} & \multicolumn{1}{c|}{\cellcolor[rgb]{ .8,  .867,  .51}0.1792} & \multicolumn{1}{c|}{\cellcolor[rgb]{ .784,  .859,  .506}0.1804} & \multicolumn{1}{c|}{\cellcolor[rgb]{ .4,  .749,  .486}0.2102} & \multicolumn{1}{c|}{\cellcolor[rgb]{ .388,  .745,  .482}0.2111} & \multicolumn{1}{c|}{\cellcolor[rgb]{ .502,  .78,  .49}0.2025} & \cellcolor[rgb]{ .475,  .773,  .49}0.2044 \\
			\midrule
			\textbf{Rachev 5\%} & \multicolumn{1}{c|}{\cellcolor[rgb]{ .753,  .851,  .506}1.0997} & \multicolumn{1}{c|}{\cellcolor[rgb]{ .996,  .902,  .514}1.0726} & \multicolumn{1}{c|}{\cellcolor[rgb]{ .976,  .541,  .443}0.9951} & \multicolumn{1}{c|}{\cellcolor[rgb]{ .973,  .435,  .424}0.9729} & \multicolumn{1}{c|}{\cellcolor[rgb]{ .973,  .412,  .42}0.9671} & \multicolumn{1}{c|}{\cellcolor[rgb]{ .984,  .635,  .463}1.0155} & \multicolumn{1}{c|}{\cellcolor[rgb]{ .992,  .78,  .49}1.0465} & \multicolumn{1}{c|}{\cellcolor[rgb]{ .988,  .737,  .482}1.0371} & \multicolumn{1}{c|}{\cellcolor[rgb]{ .988,  .741,  .482}1.0385} & \multicolumn{1}{c|}{\cellcolor[rgb]{ 1,  .922,  .518}1.0764} & \multicolumn{1}{c|}{\cellcolor[rgb]{ .812,  .871,  .51}1.0942} & \multicolumn{1}{c|}{\cellcolor[rgb]{ 1,  .922,  .518}1.0766} & \multicolumn{1}{c|}{\cellcolor[rgb]{ .976,  .918,  .518}1.0789} & \multicolumn{1}{c|}{\cellcolor[rgb]{ .518,  .784,  .49}1.1218} & \multicolumn{1}{c|}{\cellcolor[rgb]{ .388,  .745,  .482}1.1338} & \multicolumn{1}{c|}{\cellcolor[rgb]{ .608,  .812,  .498}1.1133} & \cellcolor[rgb]{ .62,  .812,  .498}1.1122 \\
			\midrule
			\textbf{Rachev 10\%} & \multicolumn{1}{c|}{\cellcolor[rgb]{ .996,  .894,  .51}1.1040} & \multicolumn{1}{c|}{\cellcolor[rgb]{ 1,  .922,  .518}1.1073} & \multicolumn{1}{c|}{\cellcolor[rgb]{ .976,  .529,  .439}1.0532} & \multicolumn{1}{c|}{\cellcolor[rgb]{ .973,  .435,  .424}1.0402} & \multicolumn{1}{c|}{\cellcolor[rgb]{ .973,  .412,  .42}1.0364} & \multicolumn{1}{c|}{\cellcolor[rgb]{ .984,  .694,  .471}1.0757} & \multicolumn{1}{c|}{\cellcolor[rgb]{ .992,  .804,  .494}1.0914} & \multicolumn{1}{c|}{\cellcolor[rgb]{ .988,  .749,  .482}1.0835} & \multicolumn{1}{c|}{\cellcolor[rgb]{ .988,  .745,  .482}1.0828} & \multicolumn{1}{c|}{\cellcolor[rgb]{ .776,  .859,  .506}1.1395} & \multicolumn{1}{c|}{\cellcolor[rgb]{ .733,  .847,  .506}1.1454} & \multicolumn{1}{c|}{\cellcolor[rgb]{ .839,  .875,  .51}1.1306} & \multicolumn{1}{c|}{\cellcolor[rgb]{ .851,  .878,  .51}1.1286} & \multicolumn{1}{c|}{\cellcolor[rgb]{ .427,  .757,  .486}1.1891} & \multicolumn{1}{c|}{\cellcolor[rgb]{ .388,  .745,  .482}1.1942} & \multicolumn{1}{c|}{\cellcolor[rgb]{ .557,  .796,  .494}1.1707} & \cellcolor[rgb]{ .545,  .792,  .494}1.1721 \\
			\midrule
			\textbf{Time (in seconds)} & \multicolumn{17}{c|}{11727} \\
			\bottomrule
	\end{tabular}}%
	\label{tab:Weekly_DowJones_0.05}%
\end{table}%
\\
In Table \ref{tab:Weekly_DowJones_0.05}, for each portfolio strategy we show the out-of-sample performance results obtained on the DowJones weekly dataset (see Table \ref{tab:WeeklyDatasets}), with $\varepsilon\,=\,5\%$.
In this case, for the highest level of the portfolio expected return $\eta_{3/4}$, the portfolios with lower levels of VaR, namely $z_0$ and $z_{1/3}$,  show the highest Mean, Sharpe ratio, Sortino ratio  and Rachev ratios.
We highlight that, when comparing the efficient Mean-Variance-VaR portfolios with the Mean-Variance ones,
we also observe a general improvement of the
out-of-sample performance, particularly for low levels of $z$.
This behavior is also confirmed by the trend of the cumulative out-of-sample portfolio returns, reported in Figure \ref{fig:CumulativeReturnsforfourlevelof targetreturnetawithepsilon=0.05_DJ_weekly}.
In this case, the EW portfolio seems to be preferred to the Mean-Variance-VaR portfolios with $\eta=\eta_{min}$, while for higher levels of  the required portfolio return, the Mean-Variance-VaR portfolios tend to dominate the EW portfolio in terms of cumulative returns.
As in the previous case, portfolios with lower levels of $z$, namely $z_{0}$ and $z_{1/3}$, seem to exhibit a behavioral pattern,
showing better performance w.r.t. the Mean-Variance portfolios (namely, the Mean-Variance-VaR portfolios with $z_1$).
This is more noticeable for the highest level of the portfolio expected return
$\eta_{3/4}$.
\\
For a better assessment of the performance of Mean-Variance-VaR efficient portfolios,
we now report a summary of our computational results on all the datasets listed
in Tables \ref{tab:WeeklyDatasets} and \ref{tab:DailyDatasets}.
The detailed out-of-sample results can be found in the Appendix.
In Tables \ref{tab:summarytable_EW_0.01} and \ref{tab:summarytable_EW_0.05} we summarize the number of datasets where the Mean-Variance-VaR efficient portfolios achieve an equal or better performance than that of the EW portfolio, when $\varepsilon=1\%$ and $5\%$ respectively.
On the other hand, in Tables \ref{tab:summarytable_MV_0.01} and \ref{tab:summarytable_MV_0.05} we report the number of datasets where the Mean-Variance-VaR efficient portfolios achieve an equal or better performance than that of the Mean-Variance portfolios, when $\varepsilon=1\%$ and $5\%$ respectively.\\
In each table we use the green marker when the best performances are obtained in
at least 50\% of the total cases.\\
We point out that, under stronger conditions on the portfolio VaR, the performance of the optimal portfolios of our model seems to be tipically better than that of the EW portfolio (see Tables \ref{tab:summarytable_EW_0.01} and \ref{tab:summarytable_EW_0.05}) in terms of mean, Sharpe ratio, Sortino ratio, and Maximum Drawdown, for both levels $\varepsilon=1\%,5\%$.
Similarly, the performance of the efficient portfolios of our model seems to be generally better than that of the classical Mean-Variance portfolios (see Table \ref{tab:summarytable_MV_0.01}) when $\varepsilon=1\%$,
and, for low expected return levels, also when $\varepsilon=5\%$.
%
\begin{table}[htbp]
	\centering
	\caption{Number of datasets out of six where the Mean-Variance-VaR efficient portfolios achieve equal or better performance than that of the EW portfolio, when $\varepsilon=1\%$}
	\resizebox{0.99\textwidth}{!}{%
	\begin{tabular}{|l|c|c|c|c|c|c|c|c|c|c|c|c|}
		\cmidrule{2-13}    \multicolumn{1}{r|}{} & \multicolumn{1}{r}{} & \multicolumn{1}{c}{\boldmath{$\eta_{min}$}} &       & \multicolumn{1}{r}{} & \multicolumn{1}{c}{\boldmath{$\eta_{1/4}$}} &       & \multicolumn{1}{r}{} & \multicolumn{1}{c}{\boldmath{$\eta_{1/2}$}} &       & \multicolumn{1}{r}{} & \multicolumn{1}{c}{\boldmath{$\eta_{3/4}$}} &  \\
		\cmidrule{2-13}    \multicolumn{1}{r|}{} & \boldmath{$z_0$} & \boldmath{$z_{1/3}$} & \boldmath{$z_{2/3}$} & \boldmath{$z_0$} & \boldmath{$z_{1/3}$} & \boldmath{$z_{2/3}$} & \boldmath{$z_0$} & \boldmath{$z_{1/3}$} & \boldmath{$z_{2/3}$} & \boldmath{$z_0$} & \boldmath{$z_{1/3}$} & \boldmath{$z_{2/3}$} \\
		\midrule
		\boldmath{$\mu$} & 2     & 2     & 2     & \cellcolor[rgb]{ .439,  .678,  .278}4 & \cellcolor[rgb]{ .439,  .678,  .278}4 & \cellcolor[rgb]{ .439,  .678,  .278}4 & \cellcolor[rgb]{ .439,  .678,  .278}6 & \cellcolor[rgb]{ .439,  .678,  .278}6 & \cellcolor[rgb]{ .439,  .678,  .278}5 & \cellcolor[rgb]{ .439,  .678,  .278}5 & \cellcolor[rgb]{ .439,  .678,  .278}5 & \cellcolor[rgb]{ .439,  .678,  .278}5 \\
		\midrule
		\boldmath{$\sigma$} & \cellcolor[rgb]{ .439,  .678,  .278}6 & \cellcolor[rgb]{ .439,  .678,  .278}6 & \cellcolor[rgb]{ .439,  .678,  .278}6 & \cellcolor[rgb]{ .439,  .678,  .278}6 & \cellcolor[rgb]{ .439,  .678,  .278}6 & \cellcolor[rgb]{ .439,  .678,  .278}6 & 2     & 2     & 2     & 1     & 1     & 1 \\
		\midrule
		\textbf{Sharpe} & \cellcolor[rgb]{ .439,  .678,  .278}3 & \cellcolor[rgb]{ .439,  .678,  .278}3 & \cellcolor[rgb]{ .439,  .678,  .278}4 & \cellcolor[rgb]{ .439,  .678,  .278}4 & \cellcolor[rgb]{ .439,  .678,  .278}4 & \cellcolor[rgb]{ .439,  .678,  .278}4 & \cellcolor[rgb]{ .439,  .678,  .278}5 & \cellcolor[rgb]{ .439,  .678,  .278}6 & \cellcolor[rgb]{ .439,  .678,  .278}6 & \cellcolor[rgb]{ .439,  .678,  .278}5 & \cellcolor[rgb]{ .439,  .678,  .278}5 & \cellcolor[rgb]{ .439,  .678,  .278}5 \\
		\midrule
		\textbf{Maximum Drawdown} & \cellcolor[rgb]{ .439,  .678,  .278}5 & \cellcolor[rgb]{ .439,  .678,  .278}5 & \cellcolor[rgb]{ .439,  .678,  .278}5 & \cellcolor[rgb]{ .439,  .678,  .278}3 & \cellcolor[rgb]{ .439,  .678,  .278}5 & \cellcolor[rgb]{ .439,  .678,  .278}5 & 2     & \cellcolor[rgb]{ .439,  .678,  .278}4 & \cellcolor[rgb]{ .439,  .678,  .278}5 & 1     & 1     & 1 \\
		\midrule
		\textbf{Ulcer} & 2     & \cellcolor[rgb]{ .439,  .678,  .278}3 & 2     & 2     & 2     & 2     & 2     & 2     & 2     & 1     & 1     & 1 \\
		\midrule
		\textbf{Turnover} & -     & -     & -     & -     & -     & -     & -     & -     & -     & -     & -     & - \\
		\midrule
		\textbf{Sortino} & \cellcolor[rgb]{ .439,  .678,  .278}3 & \cellcolor[rgb]{ .439,  .678,  .278}3 & \cellcolor[rgb]{ .439,  .678,  .278}3 & \cellcolor[rgb]{ .439,  .678,  .278}4 & \cellcolor[rgb]{ .439,  .678,  .278}4 & \cellcolor[rgb]{ .439,  .678,  .278}4 & \cellcolor[rgb]{ .439,  .678,  .278}5 & \cellcolor[rgb]{ .439,  .678,  .278}6 & \cellcolor[rgb]{ .439,  .678,  .278}6 & \cellcolor[rgb]{ .439,  .678,  .278}4 & \cellcolor[rgb]{ .439,  .678,  .278}4 & \cellcolor[rgb]{ .439,  .678,  .278}5 \\
		\midrule
		\textbf{Rachev 5\%} & 1     & 1     & 0     & 1     & 2     & 1     & 2     & 2     & 2     & \cellcolor[rgb]{ .439,  .678,  .278}3 & \cellcolor[rgb]{ .439,  .678,  .278}3 & \cellcolor[rgb]{ .439,  .678,  .278}3 \\
		\midrule
		\textbf{Rachev 10\%} & 2     & 2     & 1     & \cellcolor[rgb]{ .439,  .678,  .278}3 & 2     & 2     & \cellcolor[rgb]{ .439,  .678,  .278}4 & \cellcolor[rgb]{ .439,  .678,  .278}4 & \cellcolor[rgb]{ .439,  .678,  .278}4 & \cellcolor[rgb]{ .439,  .678,  .278}5 & \cellcolor[rgb]{ .439,  .678,  .278}5 & \cellcolor[rgb]{ .439,  .678,  .278}5 \\
		\bottomrule
	\end{tabular}%
}
	\label{tab:summarytable_EW_0.01}%
\end{table}%
%
\begin{table}[htbp]
	\centering
	\caption{Number of datasets out of six where the Mean-Variance-VaR efficient portfolios achieve equal or better performance than that of the EW portfolio, when $\varepsilon=5\%$}
	\resizebox{0.99\textwidth}{!}{%
		\begin{tabular}{|l|c|c|c|c|c|c|c|c|c|c|c|c|}
		\cmidrule{2-13}    \multicolumn{1}{r|}{} & \multicolumn{1}{r}{} & \multicolumn{1}{c}{\boldmath{$\eta_{min}$}} &       & \multicolumn{1}{r}{} & \multicolumn{1}{c}{\boldmath{$\eta_{1/4}$}} &       & \multicolumn{1}{r}{} & \multicolumn{1}{c}{\boldmath{$\eta_{1/2}$}} &       & \multicolumn{1}{r}{} & \multicolumn{1}{c}{\boldmath{$\eta_{3/4}$}} &  \\
		\cmidrule{2-13}    \multicolumn{1}{r|}{} & \boldmath{$z_0$} & \boldmath{$z_{1/3}$} & \boldmath{$z_{2/3}$} & \boldmath{$z_0$} & \boldmath{$z_{1/3}$} & \boldmath{$z_{2/3}$} & \boldmath{$z_0$} & \boldmath{$z_{1/3}$} & \boldmath{$z_{2/3}$} & \boldmath{$z_0$} & \boldmath{$z_{1/3}$} & \boldmath{$z_{2/3}$} \\
		\midrule
		\boldmath{$\mu$} & 1     & 1     & 1     & \cellcolor[rgb]{ .439,  .678,  .278}4 & \cellcolor[rgb]{ .439,  .678,  .278}4 & \cellcolor[rgb]{ .439,  .678,  .278}4 & \cellcolor[rgb]{ .439,  .678,  .278}6 & \cellcolor[rgb]{ .439,  .678,  .278}5 & \cellcolor[rgb]{ .439,  .678,  .278}5 & \cellcolor[rgb]{ .439,  .678,  .278}5 & \cellcolor[rgb]{ .439,  .678,  .278}5 & \cellcolor[rgb]{ .439,  .678,  .278}5 \\
		\midrule
		\boldmath{$\sigma$} & \cellcolor[rgb]{ .439,  .678,  .278}6 & \cellcolor[rgb]{ .439,  .678,  .278}6 & \cellcolor[rgb]{ .439,  .678,  .278}6 & \cellcolor[rgb]{ .439,  .678,  .278}6 & \cellcolor[rgb]{ .439,  .678,  .278}6 & \cellcolor[rgb]{ .439,  .678,  .278}6 & 2     & 2     & 2     & 1     & 1     & 1 \\
		\midrule
		\textbf{Sharpe} & \cellcolor[rgb]{ .439,  .678,  .278}4 & \cellcolor[rgb]{ .439,  .678,  .278}4 & \cellcolor[rgb]{ .439,  .678,  .278}4 & \cellcolor[rgb]{ .439,  .678,  .278}4 & \cellcolor[rgb]{ .439,  .678,  .278}5 & \cellcolor[rgb]{ .439,  .678,  .278}5 & \cellcolor[rgb]{ .439,  .678,  .278}6 & \cellcolor[rgb]{ .439,  .678,  .278}6 & \cellcolor[rgb]{ .439,  .678,  .278}6 & \cellcolor[rgb]{ .439,  .678,  .278}5 & \cellcolor[rgb]{ .439,  .678,  .278}4 & \cellcolor[rgb]{ .439,  .678,  .278}4 \\
		\midrule
		\textbf{Maximum Drawdown} & \cellcolor[rgb]{ .439,  .678,  .278}5 & \cellcolor[rgb]{ .439,  .678,  .278}5 & \cellcolor[rgb]{ .439,  .678,  .278}5 & \cellcolor[rgb]{ .439,  .678,  .278}5 & \cellcolor[rgb]{ .439,  .678,  .278}5 & \cellcolor[rgb]{ .439,  .678,  .278}5 & \cellcolor[rgb]{ .439,  .678,  .278}3 & \cellcolor[rgb]{ .439,  .678,  .278}5 & \cellcolor[rgb]{ .439,  .678,  .278}5 & 1     & 2     & 2 \\
		\midrule
		\textbf{Ulcer} & \cellcolor[rgb]{ .439,  .678,  .278}4 & \cellcolor[rgb]{ .439,  .678,  .278}3 & \cellcolor[rgb]{ .439,  .678,  .278}3 & 2     & \cellcolor[rgb]{ .439,  .678,  .278}3 & 2     & 2     & 2     & 2     & 1     & 1     & 1 \\
		\midrule
		\textbf{Turnover} & -     & -     & -     & -     & -     & -     & -     & -     & -     & -     & -     & - \\
		\midrule
		\textbf{Sortino} & \cellcolor[rgb]{ .439,  .678,  .278}4 & \cellcolor[rgb]{ .439,  .678,  .278}4 & \cellcolor[rgb]{ .439,  .678,  .278}4 & \cellcolor[rgb]{ .439,  .678,  .278}4 & \cellcolor[rgb]{ .439,  .678,  .278}5 & \cellcolor[rgb]{ .439,  .678,  .278}5 & \cellcolor[rgb]{ .439,  .678,  .278}6 & \cellcolor[rgb]{ .439,  .678,  .278}6 & \cellcolor[rgb]{ .439,  .678,  .278}6 & \cellcolor[rgb]{ .439,  .678,  .278}5 & \cellcolor[rgb]{ .439,  .678,  .278}4 & \cellcolor[rgb]{ .439,  .678,  .278}4 \\
		\midrule
		\textbf{Rachev 5\%} & 0     & 0     & 0     & 0     & 0     & 1     & \cellcolor[rgb]{ .439,  .678,  .278}3 & 2     & 2     & \cellcolor[rgb]{ .439,  .678,  .278}4 & \cellcolor[rgb]{ .439,  .678,  .278}3 & \cellcolor[rgb]{ .439,  .678,  .278}3 \\
		\midrule
		\textbf{Rachev 10\%} & 2     & 1     & 0     & 1     & 2     & 2     & \cellcolor[rgb]{ .439,  .678,  .278}5 & \cellcolor[rgb]{ .439,  .678,  .278}5 & \cellcolor[rgb]{ .439,  .678,  .278}5 & \cellcolor[rgb]{ .439,  .678,  .278}5 & \cellcolor[rgb]{ .439,  .678,  .278}5 & \cellcolor[rgb]{ .439,  .678,  .278}5 \\
		\bottomrule
	\end{tabular}%
}
	\label{tab:summarytable_EW_0.05}%
\end{table}%
%
\begin{table}[htbp]
	\centering
	\caption{Number of datasets out of six where the Mean-Variance-VaR efficient portfolios achieve equal or better performance than that of the Mean-Variance portfolios, when $\varepsilon=1\%$}
	\resizebox{0.99\textwidth}{!}{%
	\begin{tabular}{|l|c|c|c|c|c|c|c|c|c|c|c|c|}
		\cmidrule{2-13}    \multicolumn{1}{r|}{} & \multicolumn{1}{r}{} & \multicolumn{1}{c}{\boldmath{$\eta_{min}$}} &       & \multicolumn{1}{r}{} & \multicolumn{1}{c}{\boldmath{$\eta_{1/4}$}} &       & \multicolumn{1}{r}{} & \multicolumn{1}{c}{\boldmath{$\eta_{1/2}$}} &       & \multicolumn{1}{r}{} & \multicolumn{1}{c}{\boldmath{$\eta_{3/4}$}} &  \\
		\cmidrule{2-13}    \multicolumn{1}{r|}{} & \boldmath{$z_0$} & \boldmath{$z_{1/3}$} & \boldmath{$z_{2/3}$} & \boldmath{$z_0$} & \boldmath{$z_{1/3}$} & \boldmath{$z_{2/3}$} & \boldmath{$z_0$} & \boldmath{$z_{1/3}$} & \boldmath{$z_{2/3}$} & \boldmath{$z_0$} & \boldmath{$z_{1/3}$} & \boldmath{$z_{2/3}$} \\
		\midrule
		\boldmath{$\mu$} & \cellcolor[rgb]{ .439,  .678,  .278}5 & \cellcolor[rgb]{ .439,  .678,  .278}6 & \cellcolor[rgb]{ .439,  .678,  .278}6 & \cellcolor[rgb]{ .439,  .678,  .278}4 & \cellcolor[rgb]{ .439,  .678,  .278}5 & \cellcolor[rgb]{ .439,  .678,  .278}6 & \cellcolor[rgb]{ .439,  .678,  .278}4 & \cellcolor[rgb]{ .439,  .678,  .278}4 & \cellcolor[rgb]{ .439,  .678,  .278}4 & \cellcolor[rgb]{ .439,  .678,  .278}3 & \cellcolor[rgb]{ .439,  .678,  .278}3 & \cellcolor[rgb]{ .439,  .678,  .278}5 \\
		\midrule
		\boldmath{$\sigma$} & 0     & 1     & \cellcolor[rgb]{ .439,  .678,  .278}3 & 0     & 0     & 2     & 0     & 2     & \cellcolor[rgb]{ .439,  .678,  .278}5 & 1     & 2     & \cellcolor[rgb]{ .439,  .678,  .278}3 \\
		\midrule
		\textbf{Sharpe} & 2     & \cellcolor[rgb]{ .439,  .678,  .278}4 & \cellcolor[rgb]{ .439,  .678,  .278}6 & \cellcolor[rgb]{ .439,  .678,  .278}4 & \cellcolor[rgb]{ .439,  .678,  .278}5 & \cellcolor[rgb]{ .439,  .678,  .278}6 & \cellcolor[rgb]{ .439,  .678,  .278}3 & \cellcolor[rgb]{ .439,  .678,  .278}4 & \cellcolor[rgb]{ .439,  .678,  .278}4 & 1     & \cellcolor[rgb]{ .439,  .678,  .278}3 & 2 \\
		\midrule
		\textbf{Maximum Drawdown} & \cellcolor[rgb]{ .439,  .678,  .278}3 & \cellcolor[rgb]{ .439,  .678,  .278}3 & \cellcolor[rgb]{ .439,  .678,  .278}3 & 2     & \cellcolor[rgb]{ .439,  .678,  .278}5 & \cellcolor[rgb]{ .439,  .678,  .278}4 & \cellcolor[rgb]{ .439,  .678,  .278}3 & \cellcolor[rgb]{ .439,  .678,  .278}3 & \cellcolor[rgb]{ .439,  .678,  .278}3 & 2     & \cellcolor[rgb]{ .439,  .678,  .278}3 & \cellcolor[rgb]{ .439,  .678,  .278}4 \\
		\midrule
		\textbf{Ulcer} & \cellcolor[rgb]{ .439,  .678,  .278}3 & \cellcolor[rgb]{ .439,  .678,  .278}4 & \cellcolor[rgb]{ .439,  .678,  .278}4 & 2     & \cellcolor[rgb]{ .439,  .678,  .278}4 & \cellcolor[rgb]{ .439,  .678,  .278}4 & \cellcolor[rgb]{ .439,  .678,  .278}3 & 2     & 2     & 2     & \cellcolor[rgb]{ .439,  .678,  .278}5 & \cellcolor[rgb]{ .439,  .678,  .278}4 \\
		\midrule
		\textbf{Turnover} & 0     & 0     & 0     & 0     & 0     & 0     & 0     & 0     & 1     & 0     & 0     & 1 \\
		\midrule
		\textbf{Sortino} & 2     & \cellcolor[rgb]{ .439,  .678,  .278}4 & \cellcolor[rgb]{ .439,  .678,  .278}6 & \cellcolor[rgb]{ .439,  .678,  .278}4 & \cellcolor[rgb]{ .439,  .678,  .278}5 & \cellcolor[rgb]{ .439,  .678,  .278}6 & \cellcolor[rgb]{ .439,  .678,  .278}3 & \cellcolor[rgb]{ .439,  .678,  .278}4 & \cellcolor[rgb]{ .439,  .678,  .278}4 & 1     & 2     & 1 \\
		\midrule
		\textbf{Rachev 5\%} & \cellcolor[rgb]{ .439,  .678,  .278}4 & \cellcolor[rgb]{ .439,  .678,  .278}5 & \cellcolor[rgb]{ .439,  .678,  .278}4 & \cellcolor[rgb]{ .439,  .678,  .278}3 & \cellcolor[rgb]{ .439,  .678,  .278}5 & \cellcolor[rgb]{ .439,  .678,  .278}3 & \cellcolor[rgb]{ .439,  .678,  .278}3 & \cellcolor[rgb]{ .439,  .678,  .278}4 & 2     & \cellcolor[rgb]{ .439,  .678,  .278}3 & 2     & 2 \\
		\midrule
		\textbf{Rachev 10\%} & \cellcolor[rgb]{ .439,  .678,  .278}5 & \cellcolor[rgb]{ .439,  .678,  .278}5 & \cellcolor[rgb]{ .439,  .678,  .278}4 & \cellcolor[rgb]{ .439,  .678,  .278}3 & \cellcolor[rgb]{ .439,  .678,  .278}4 & \cellcolor[rgb]{ .439,  .678,  .278}4 & \cellcolor[rgb]{ .439,  .678,  .278}4 & \cellcolor[rgb]{ .439,  .678,  .278}4 & 2     & \cellcolor[rgb]{ .439,  .678,  .278}3 & 1     & \cellcolor[rgb]{ .439,  .678,  .278}3 \\
		\bottomrule
	\end{tabular}%
}
	\label{tab:summarytable_MV_0.01}%
\end{table}%
%
\begin{table}[htbp]
	\centering
	\caption{Number of datasets out of six where the Mean-Variance-VaR efficient portfolios achieve equal or better performance than that of the Mean-Variance portfolios, when $\varepsilon=5\%$}
	\resizebox{0.99\textwidth}{!}{%
	\begin{tabular}{|l|c|c|c|c|c|c|c|c|c|c|c|c|}
		\cmidrule{2-13}    \multicolumn{1}{r|}{} & \multicolumn{1}{r}{} & \multicolumn{1}{c}{\boldmath{$\eta_{min}$}} &       & \multicolumn{1}{r}{} & \multicolumn{1}{c}{\boldmath{$\eta_{1/4}$}} &       & \multicolumn{1}{r}{} & \multicolumn{1}{c}{\boldmath{$\eta_{1/2}$}} &       & \multicolumn{1}{r}{} & \multicolumn{1}{c}{\boldmath{$\eta_{3/4}$}} &  \\
		\cmidrule{2-13}    \multicolumn{1}{r|}{} & \boldmath{$z_0$} & \boldmath{$z_{1/3}$} & \boldmath{$z_{2/3}$} & \boldmath{$z_0$} & \boldmath{$z_{1/3}$} & \boldmath{$z_{2/3}$} & \boldmath{$z_0$} & \boldmath{$z_{1/3}$} & \boldmath{$z_{2/3}$} & \boldmath{$z_0$} & \boldmath{$z_{1/3}$} & \boldmath{$z_{2/3}$} \\
		\midrule
		\boldmath{$\mu$} & \cellcolor[rgb]{ .439,  .678,  .278}5 & \cellcolor[rgb]{ .439,  .678,  .278}5 & 2     & \cellcolor[rgb]{ .439,  .678,  .278}4 & \cellcolor[rgb]{ .439,  .678,  .278}3 & \cellcolor[rgb]{ .439,  .678,  .278}5 & \cellcolor[rgb]{ .439,  .678,  .278}4 & \cellcolor[rgb]{ .439,  .678,  .278}3 & \cellcolor[rgb]{ .439,  .678,  .278}4 & \cellcolor[rgb]{ .439,  .678,  .278}4 & \cellcolor[rgb]{ .439,  .678,  .278}3 & 2 \\
		\midrule
		\boldmath{$\sigma$} & 0     & \cellcolor[rgb]{ .439,  .678,  .278}3 & \cellcolor[rgb]{ .439,  .678,  .278}5 & 0     & 2     & \cellcolor[rgb]{ .439,  .678,  .278}4 & 0     & 1     & \cellcolor[rgb]{ .439,  .678,  .278}3 & 0     & 0     & 2 \\
		\midrule
		\textbf{Sharpe} & \cellcolor[rgb]{ .439,  .678,  .278}4 & \cellcolor[rgb]{ .439,  .678,  .278}4 & 1     & 1     & 2     & \cellcolor[rgb]{ .439,  .678,  .278}3 & 2     & 1     & 2     & \cellcolor[rgb]{ .439,  .678,  .278}4 & 2     & 1 \\
		\midrule
		\textbf{Maximum Drawdown} & \cellcolor[rgb]{ .439,  .678,  .278}4 & \cellcolor[rgb]{ .439,  .678,  .278}4 & \cellcolor[rgb]{ .439,  .678,  .278}4 & 2     & 2     & 2     & 2     & 2     & 1     & 1     & \cellcolor[rgb]{ .439,  .678,  .278}3 & 2 \\
		\midrule
		\textbf{Ulcer} & \cellcolor[rgb]{ .439,  .678,  .278}4 & \cellcolor[rgb]{ .439,  .678,  .278}5 & \cellcolor[rgb]{ .439,  .678,  .278}4 & 2     & 2     & \cellcolor[rgb]{ .439,  .678,  .278}3 & 2     & \cellcolor[rgb]{ .439,  .678,  .278}3 & \cellcolor[rgb]{ .439,  .678,  .278}3 & \cellcolor[rgb]{ .439,  .678,  .278}3 & \cellcolor[rgb]{ .439,  .678,  .278}3 & 1 \\
		\midrule
		\textbf{Turnover} & 0     & 0     & 0     & 0     & 0     & 0     & 0     & 0     & 0     & 0     & 0     & 0 \\
		\midrule
		\textbf{Sortino} & \cellcolor[rgb]{ .439,  .678,  .278}4 & \cellcolor[rgb]{ .439,  .678,  .278}4 & 1     & 2     & 2     & \cellcolor[rgb]{ .439,  .678,  .278}3 & 2     & 1     & 2     & \cellcolor[rgb]{ .439,  .678,  .278}4 & 2     & 1 \\
		\midrule
		\textbf{Rachev 5\%} & \cellcolor[rgb]{ .439,  .678,  .278}4 & \cellcolor[rgb]{ .439,  .678,  .278}3 & \cellcolor[rgb]{ .439,  .678,  .278}4 & 2     & \cellcolor[rgb]{ .439,  .678,  .278}3 & \cellcolor[rgb]{ .439,  .678,  .278}3 & \cellcolor[rgb]{ .439,  .678,  .278}4 & \cellcolor[rgb]{ .439,  .678,  .278}4 & 2     & \cellcolor[rgb]{ .439,  .678,  .278}4 & \cellcolor[rgb]{ .439,  .678,  .278}5 & 2 \\
		\midrule
		\textbf{Rachev 10\%} & \cellcolor[rgb]{ .439,  .678,  .278}4 & \cellcolor[rgb]{ .439,  .678,  .278}4 & \cellcolor[rgb]{ .439,  .678,  .278}3 & 2     & \cellcolor[rgb]{ .439,  .678,  .278}5 & \cellcolor[rgb]{ .439,  .678,  .278}4 & \cellcolor[rgb]{ .439,  .678,  .278}4 & 2     & 2     & \cellcolor[rgb]{ .439,  .678,  .278}3 & 2     & 0 \\
		\bottomrule
	\end{tabular}%
}
	\label{tab:summarytable_MV_0.05}%
\end{table}%

\newpage
\section{Conclusions and future research}
\label{sec:ConclusiveRemarks}
In  this paper we have proposed a tri-objective portfolio selection model which adds conditions on the portfolio VaR to the classical Mean-Variance approach.
The Mean-Variance-VaR model is formulated as an MIQP problem, and is solved using Gurobi.
We have described appropriate combinations of the parameters $\varepsilon$, $n$, and $T$,
for which we can obtain an optimal solution in a reasonable time.
Our extensive empirical analysis based on several real-world datasets shows promising results in terms of several performance measures.
Indeed, it seems that the tri-objective optimal portfolios can generally achieve equal or better results than those of the Mean-Variance and of the Mean-VaR portfolios.
Further and future research might be directed to
extend this approach to other Mean-Risk models,
to investigate the stability of the Pareto-optimal solutions obtained by our model \citep[see, e.g.,][]{cesarone2020stability}, and to extend the empirical analysis to other datasets.

{\footnotesize
	\bibliographystyle{spbasic}
	\bibliography{BIB_CMT_M-V-VaR}
}

\newpage
\appendix

\section{Additional out-of-sample performance results}

\begin{figure}[h!]
\centering
\subfigure{\label{fig:CumulativeReturnsfor_etamin_DJ_weekly_0.01}
\includegraphics[scale=0.19]{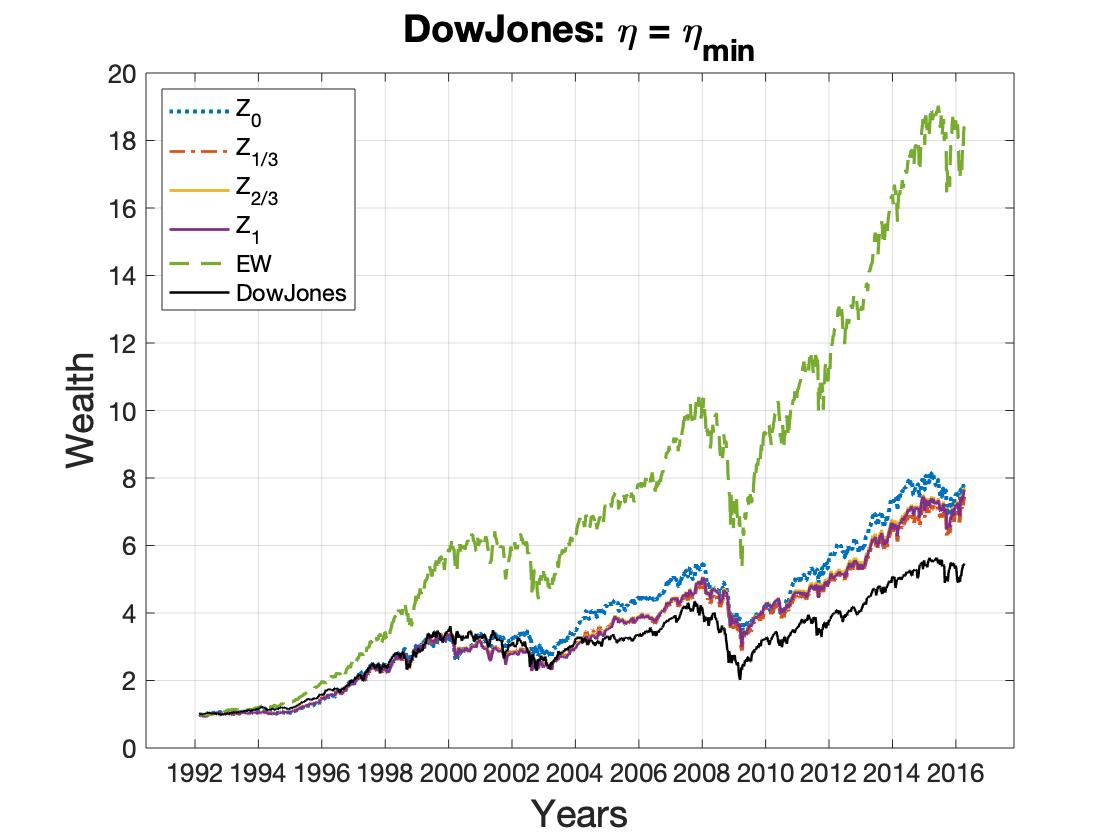}}
\hfill
\subfigure{\label{fig:CumulativeReturnsfor_eta1/4_DJ_weekly_0.01}
\includegraphics[scale=0.19]{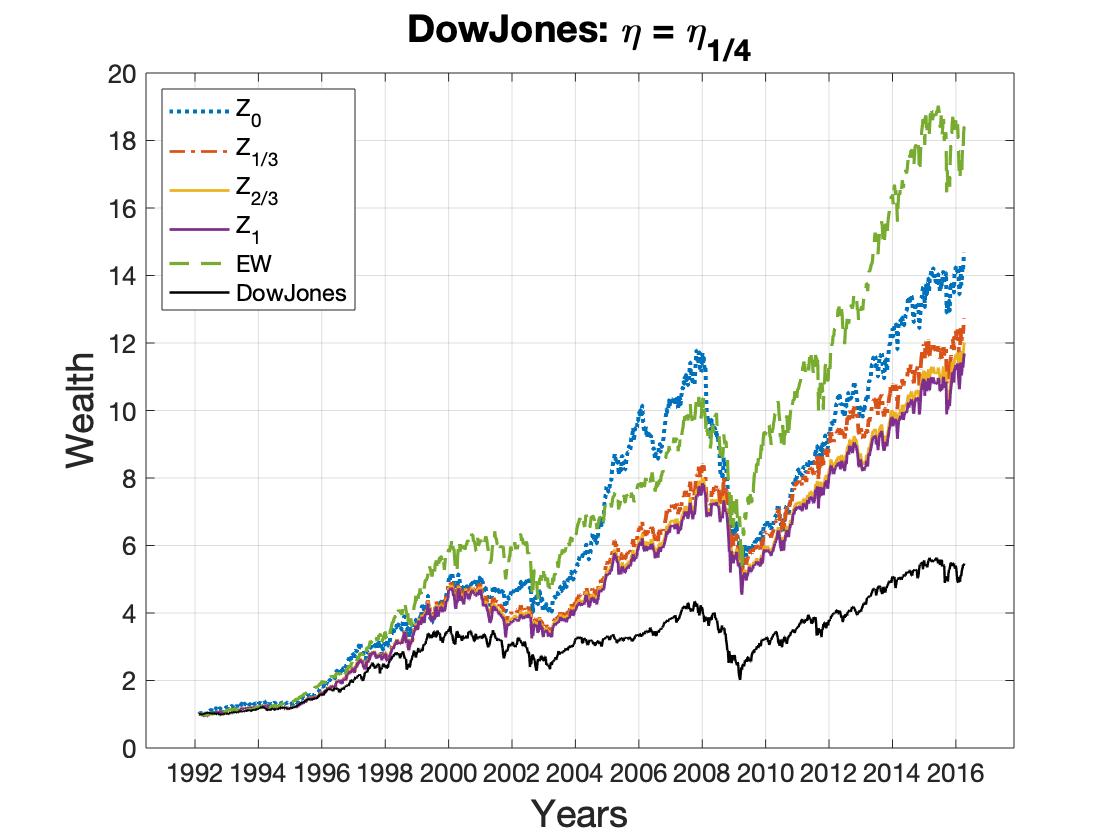}}
\vfill
\subfigure{\label{fig:CumulativeReturnsfor_eta1/2_DJ_weekly_0.01}
\includegraphics[scale=0.19]{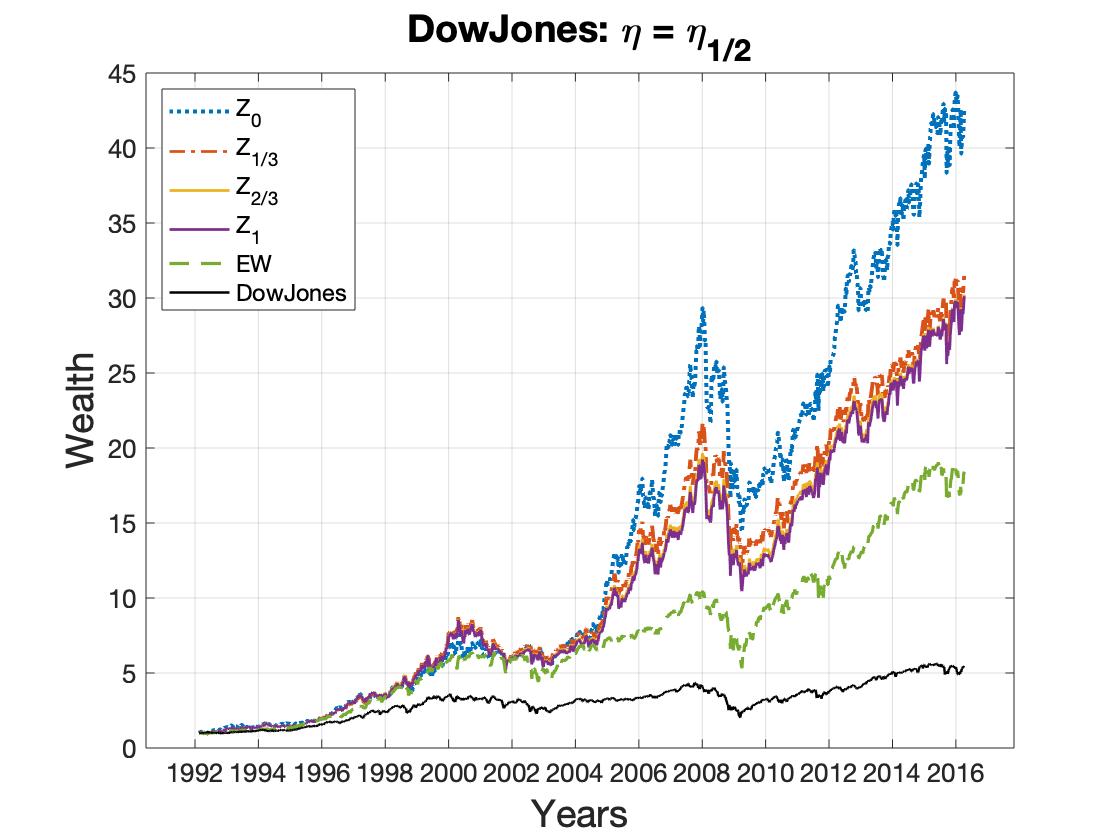}}
\hfill
\subfigure{\label{fig:CumulativeReturnsfor_eta3/4_DJ_weekly_0.01}
\includegraphics[scale=0.19]{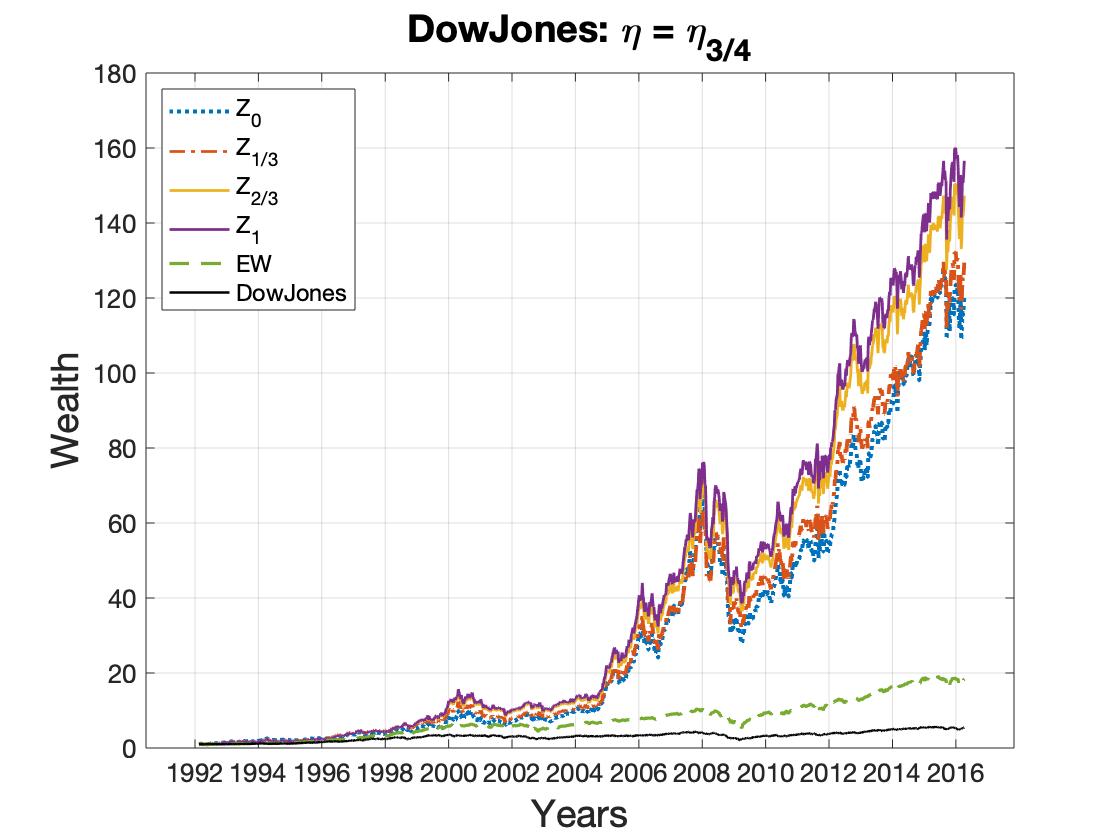}}
\caption{Cumulative out-of-sample portfolio returns using different levels of $\eta$ and $\varepsilon=1\%$ for the DowJones weekly dataset}
\label{fig:CumulativeReturnsforfourlevelof targetreturnetawithepsilon=0.01_DJ_weekly}
\end{figure}

\begin{figure}[h!]
\centering
\subfigure{\label{fig:CumulativeReturnsfor_etamin_NASDAQ100_weekly_0.01}
\includegraphics[scale=0.19]{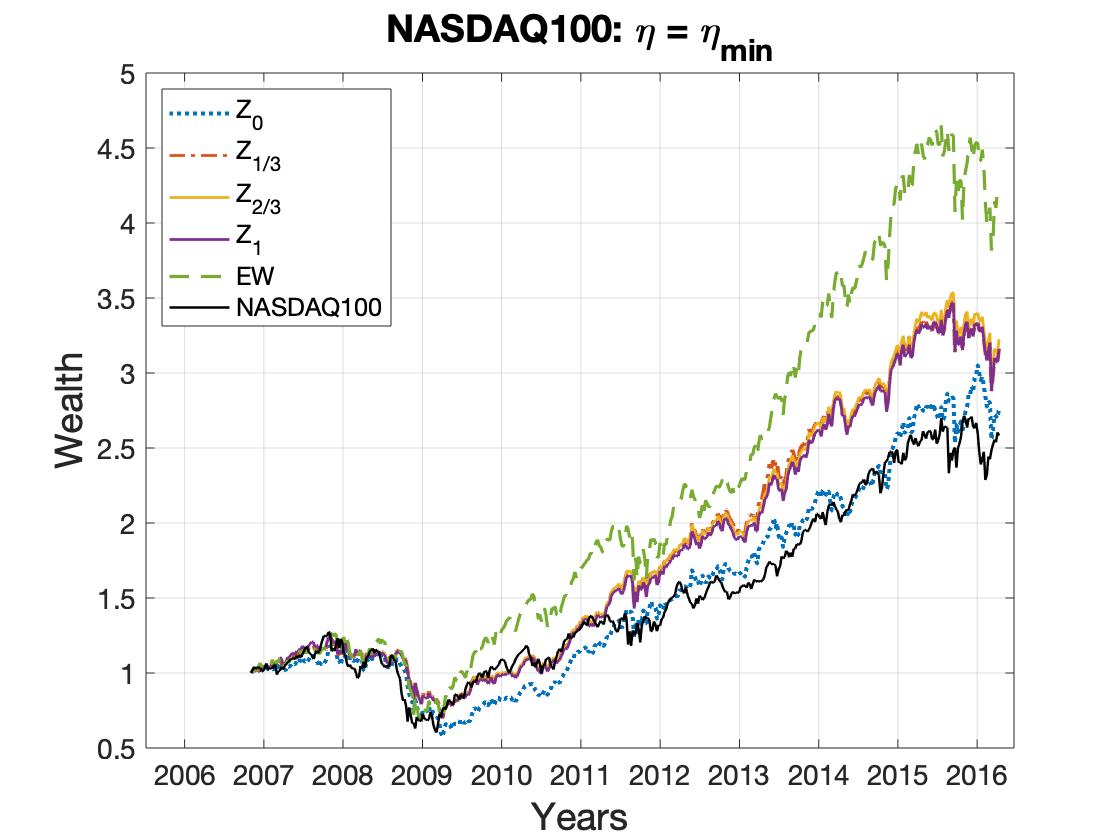}}
\hfill
\subfigure{\label{fig:CumulativeReturnsfor_eta1/4_NASDAQ100_weekly_0.01}
\includegraphics[scale=0.19]{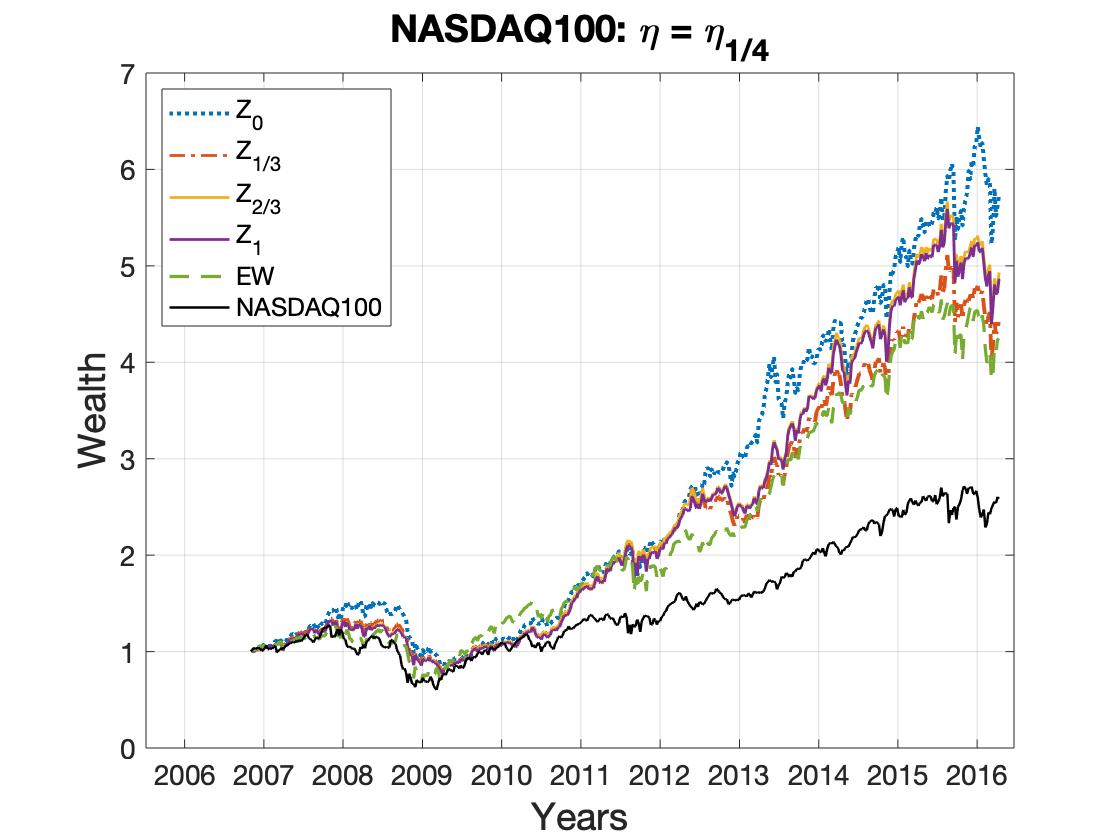}}
\vfill
\subfigure{\label{fig:CumulativeReturnsfor_eta1/2_NASDAQ100_weekly_0.01}
\includegraphics[scale=0.19]{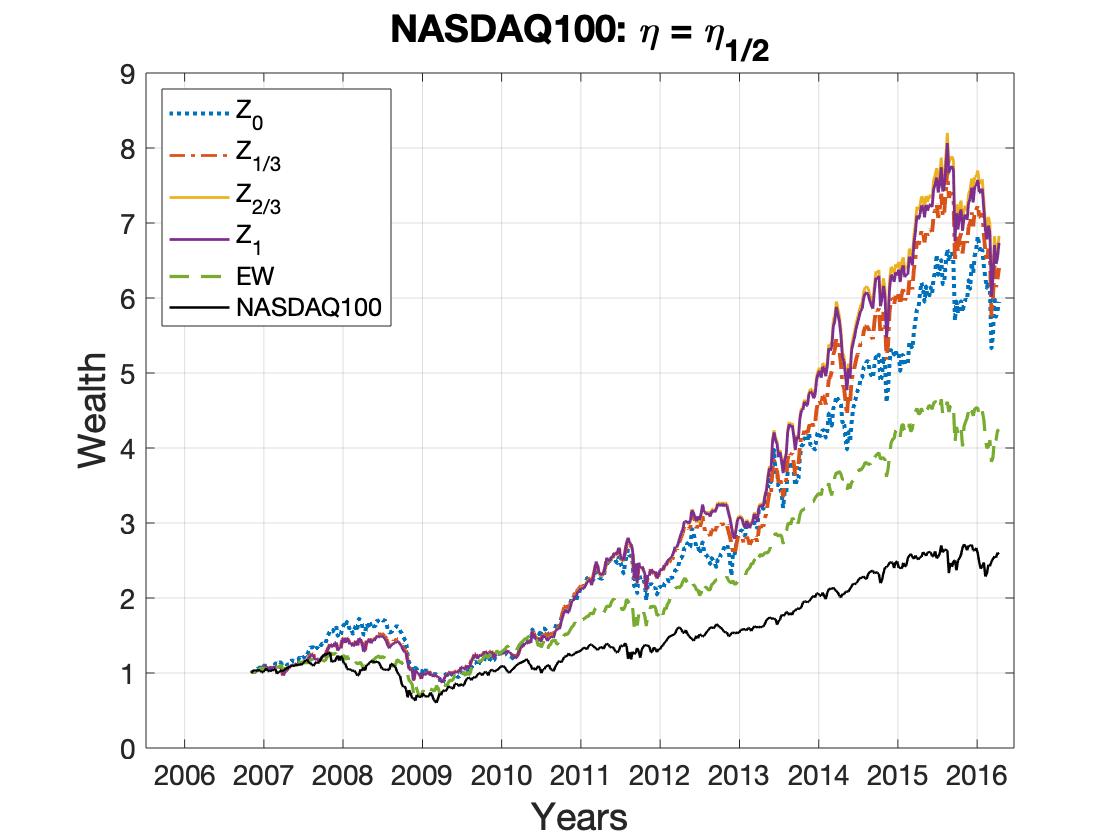}}
\hfill
\subfigure{\label{fig:CumulativeReturnsfor_eta3/4_NASDAQ100_weekly_0.01}
\includegraphics[scale=0.19]{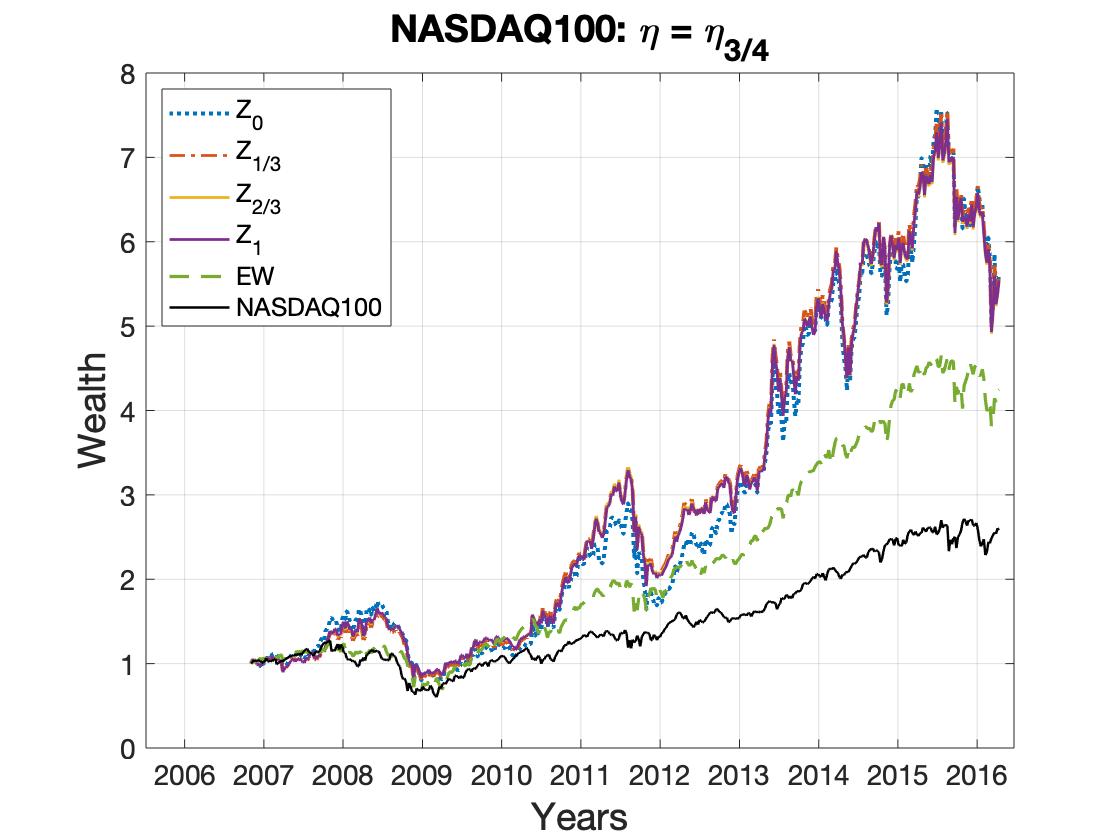}}
\caption{Cumulative out-of-sample portfolio returns using different levels of $\eta$ and $\varepsilon=1\%$ for the NASDAQ100 weekly dataset}
\label{fig:CumulativeReturnsforfourlevelof targetreturnetawithepsilon=0.01_NASDAQ100_weekly}
\end{figure}

\begin{figure}[h!]
\centering
\subfigure{\label{fig:CumulativeReturnsfor_etamin_NASDAQ100_weekly_0.05}
\includegraphics[scale=0.19]{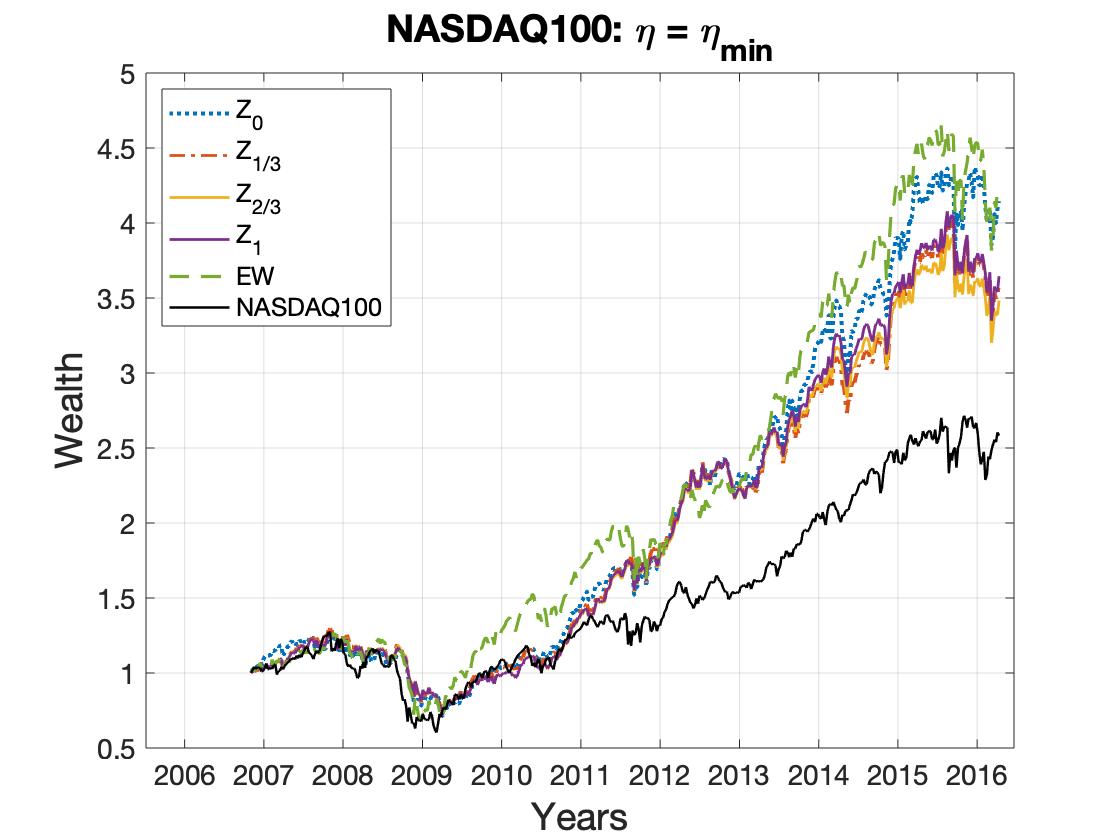}}
\hfill
\subfigure{\label{fig:CumulativeReturnsfor_eta1/4_NASDAQ100_weekly_0.05}
\includegraphics[scale=0.19]{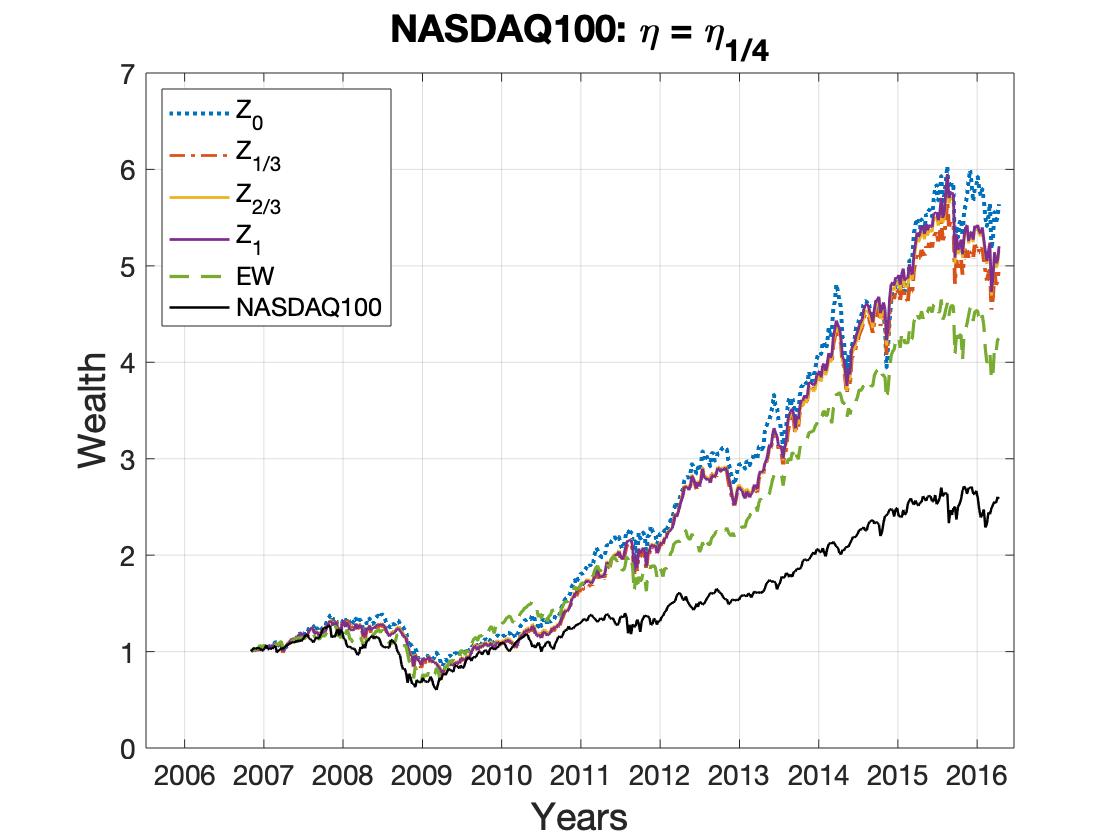}}
\vfill
\subfigure{\label{fig:CumulativeReturnsfor_eta1/2_NASDAQ100_weekly_0.05}
\includegraphics[scale=0.19]{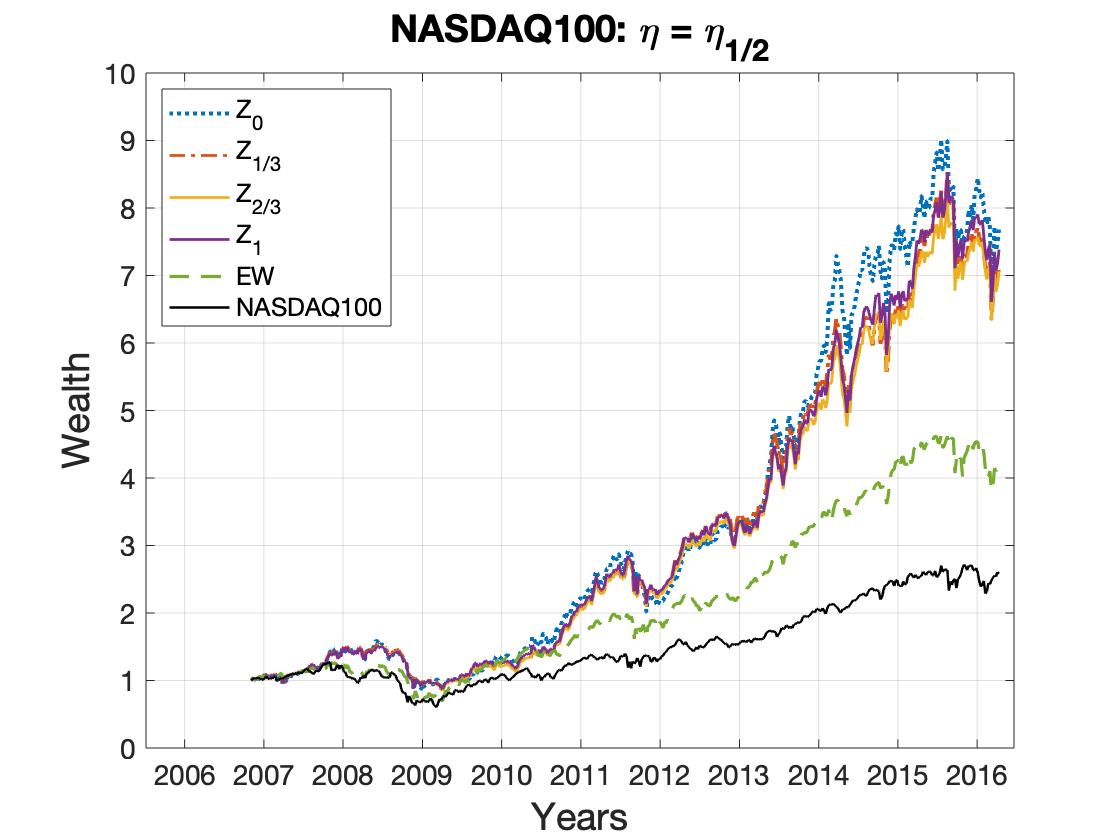}}
\hfill
\subfigure{\label{fig:CumulativeReturnsfor_eta3/4_NASDAQ100_weekly_0.05}
\includegraphics[scale=0.19]{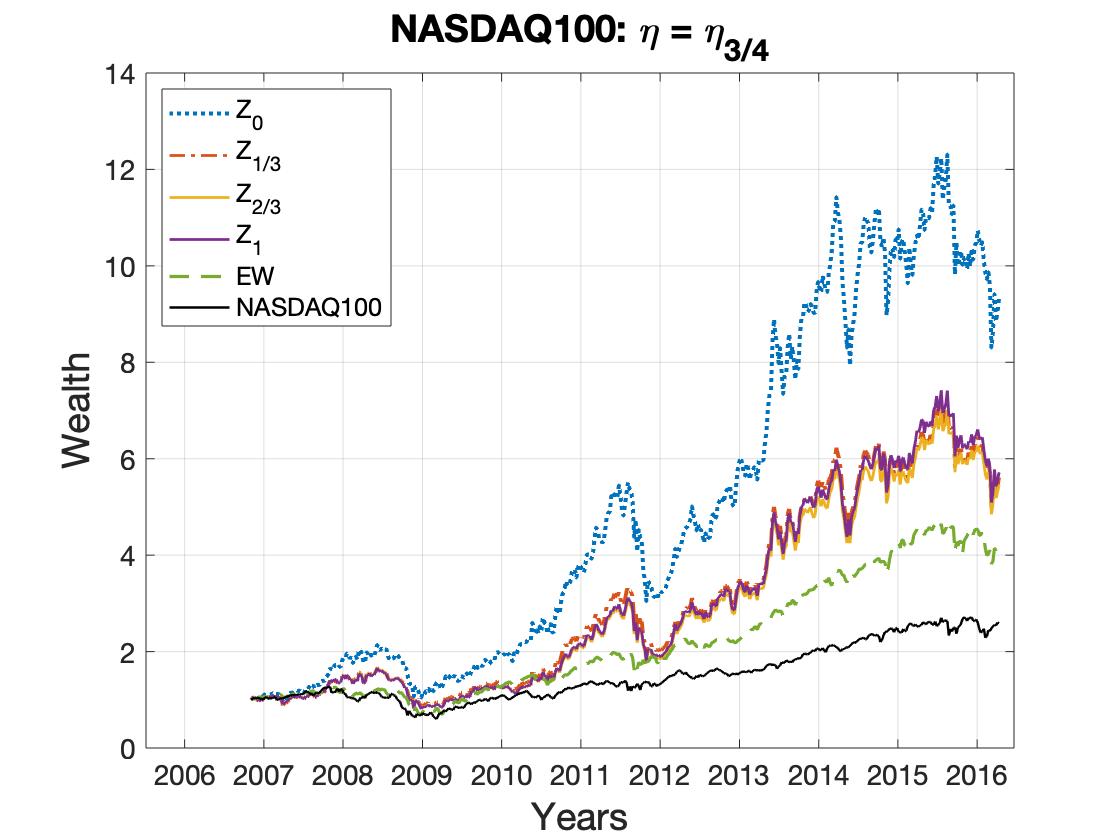}}
\caption{Cumulative out-of-sample portfolio returns using different levels of $\eta$ and $\varepsilon=5\%$ for the NASDAQ100 weekly dataset}
\label{fig:CumulativeReturnsforfourlevelof targetreturnetawithepsilon=0.05_NASDAQ100_weekly}
\end{figure}

\begin{figure}[h!]
\centering
\subfigure{\label{fig:CumulativeReturnsfor_etamin_FTSE100_weekly_0.01}
\includegraphics[scale=0.19]{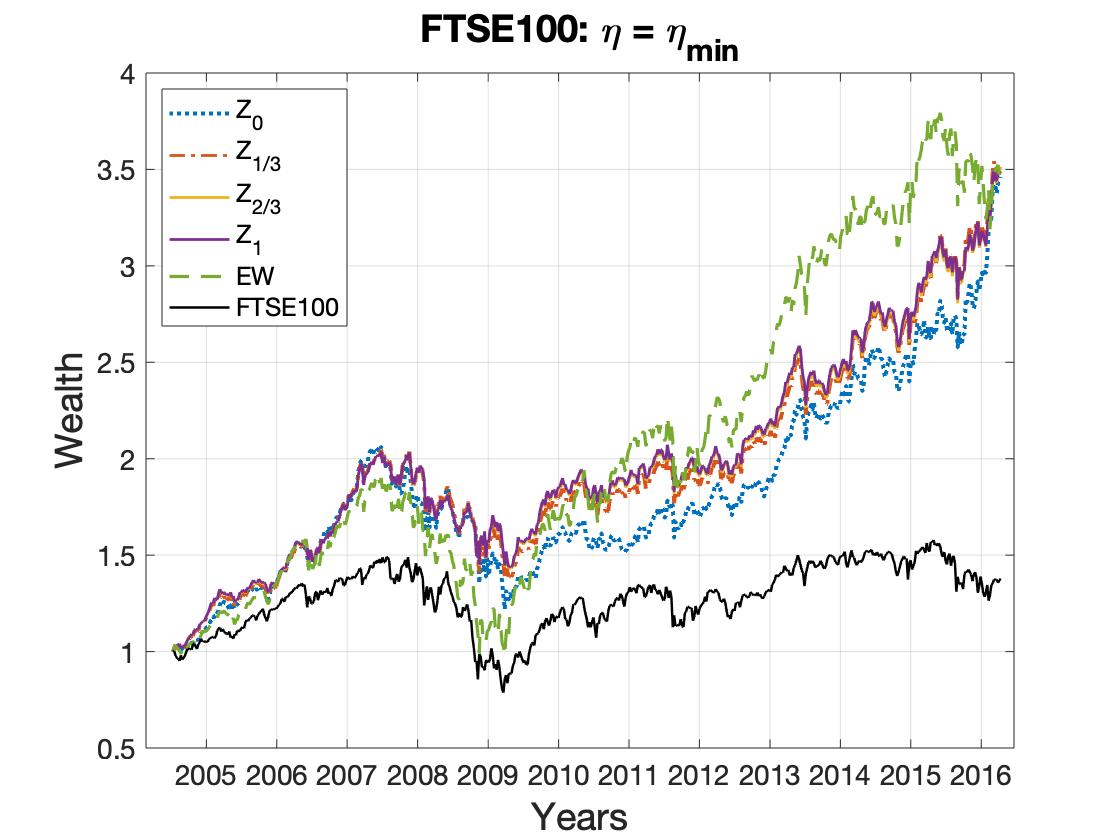}}
\hfill
\subfigure{\label{fig:CumulativeReturnsfor_eta1/4_FTSE100_weekly_0.01}
\includegraphics[scale=0.19]{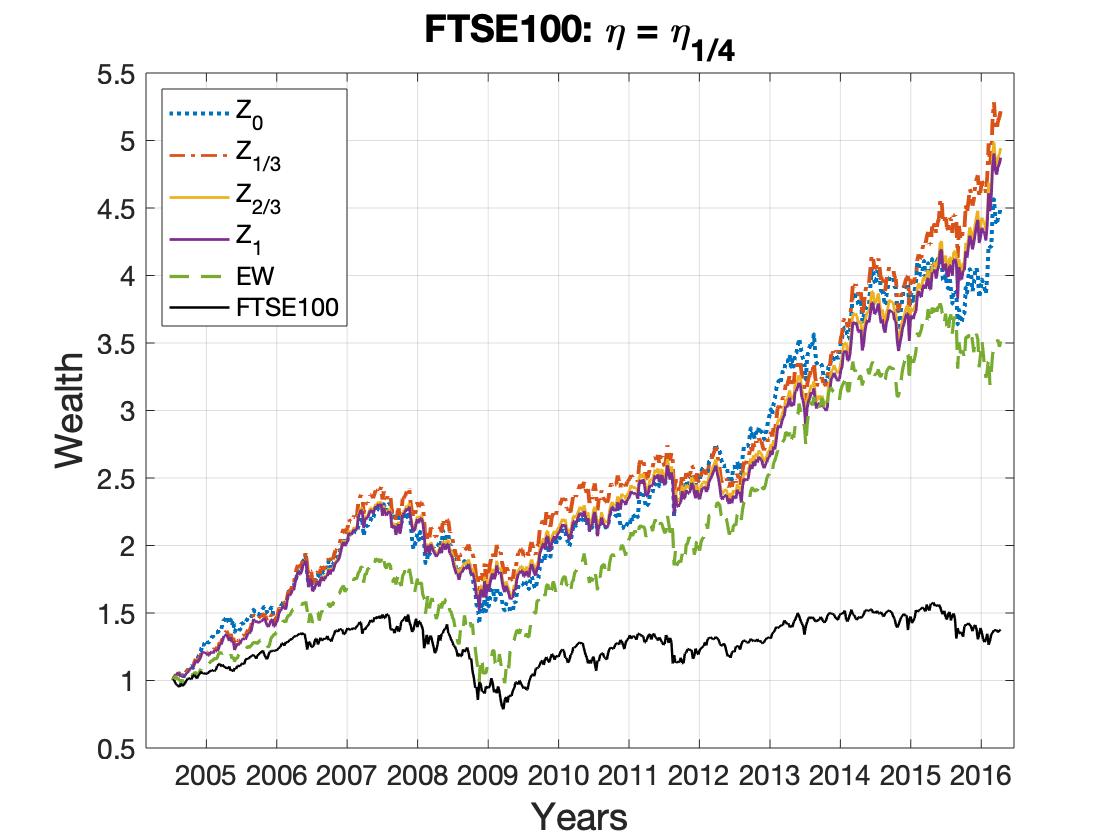}}
\vfill
\subfigure{\label{fig:CumulativeReturnsfor_eta1/2_FTSE100_weekly_0.01}
\includegraphics[scale=0.19]{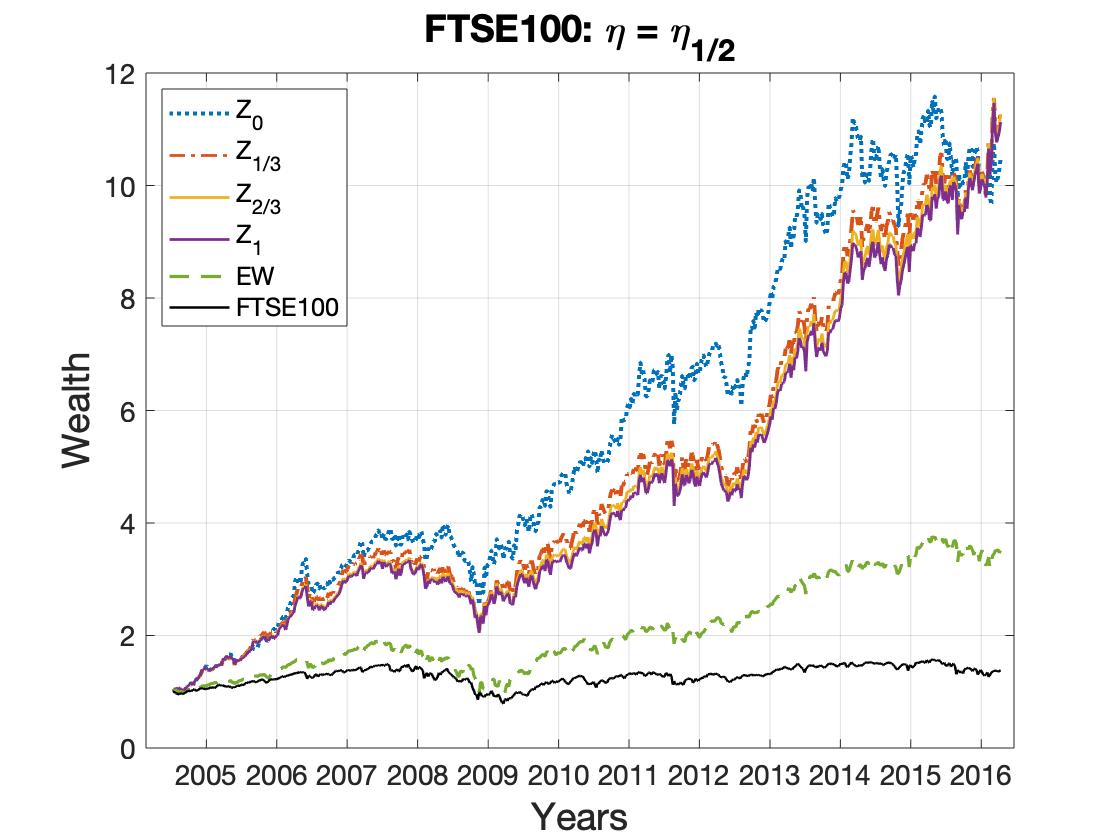}}
\hfill
\subfigure{\label{fig:CumulativeReturnsfor_eta3/4_FTSE100_weekly_0.01}
\includegraphics[scale=0.19]{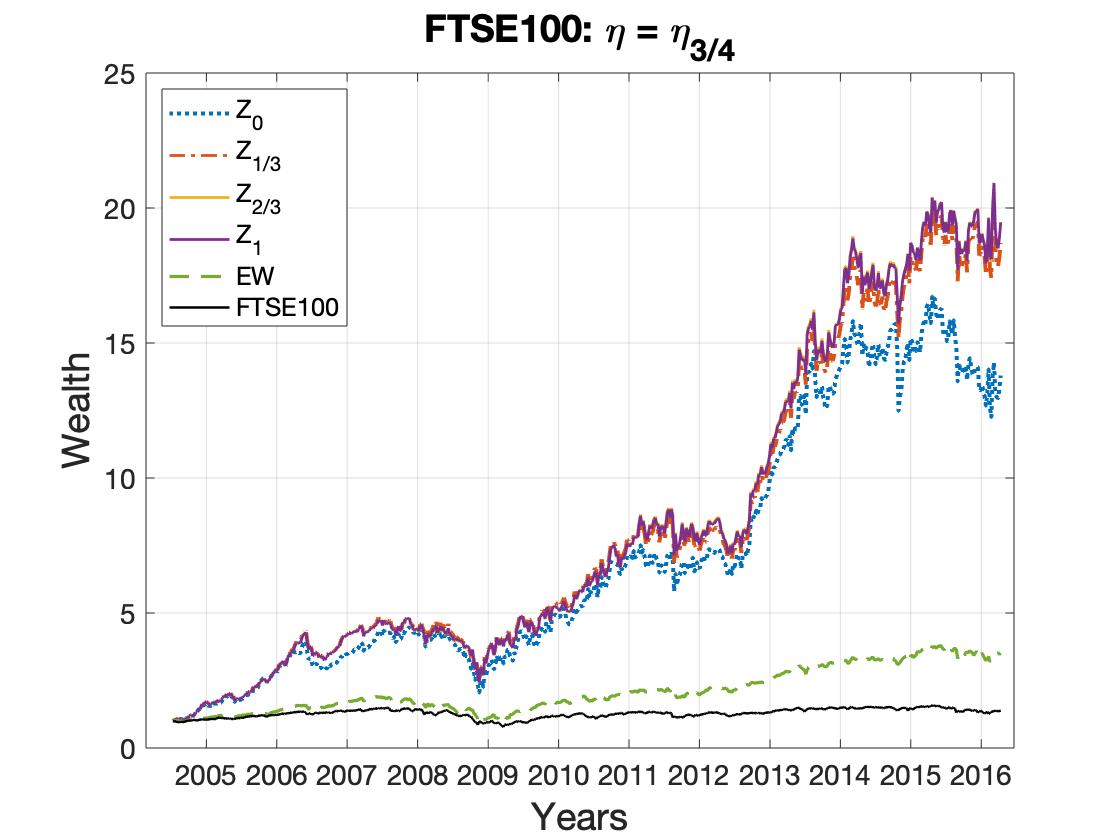}}
\caption{Cumulative out-of-sample portfolio returns using different levels of $\eta$ and $\varepsilon=1\%$ for the FTSE100 weekly dataset}
\label{fig:CumulativeReturnsforfourlevelof targetreturnetawithepsilon=0.01_FTSE100_weekly}
\end{figure}

\begin{figure}[h!]
\centering
\subfigure{\label{fig:CumulativeReturnsfor_etamin_FTSE100_weekly_0.05}
\includegraphics[scale=0.19]{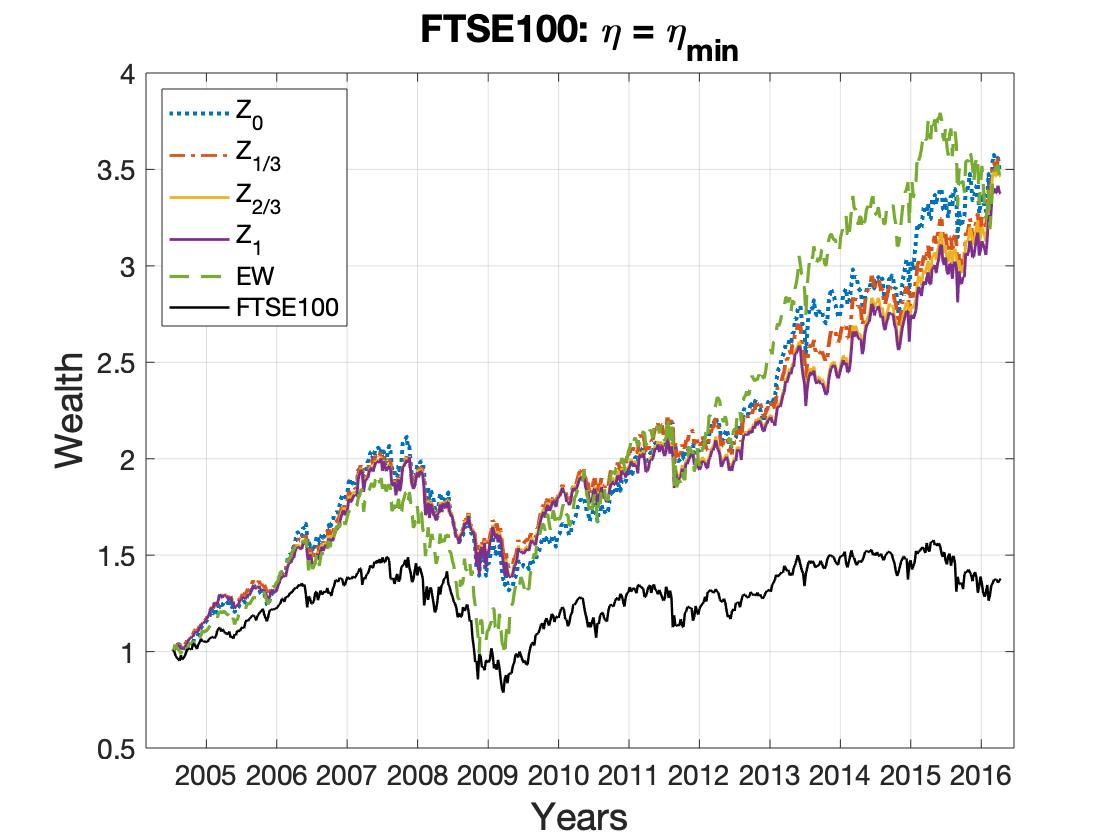}}
\hfill
\subfigure{\label{fig:CumulativeReturnsfor_eta1/4_FTSE100_weekly_0.05}
\includegraphics[scale=0.19]{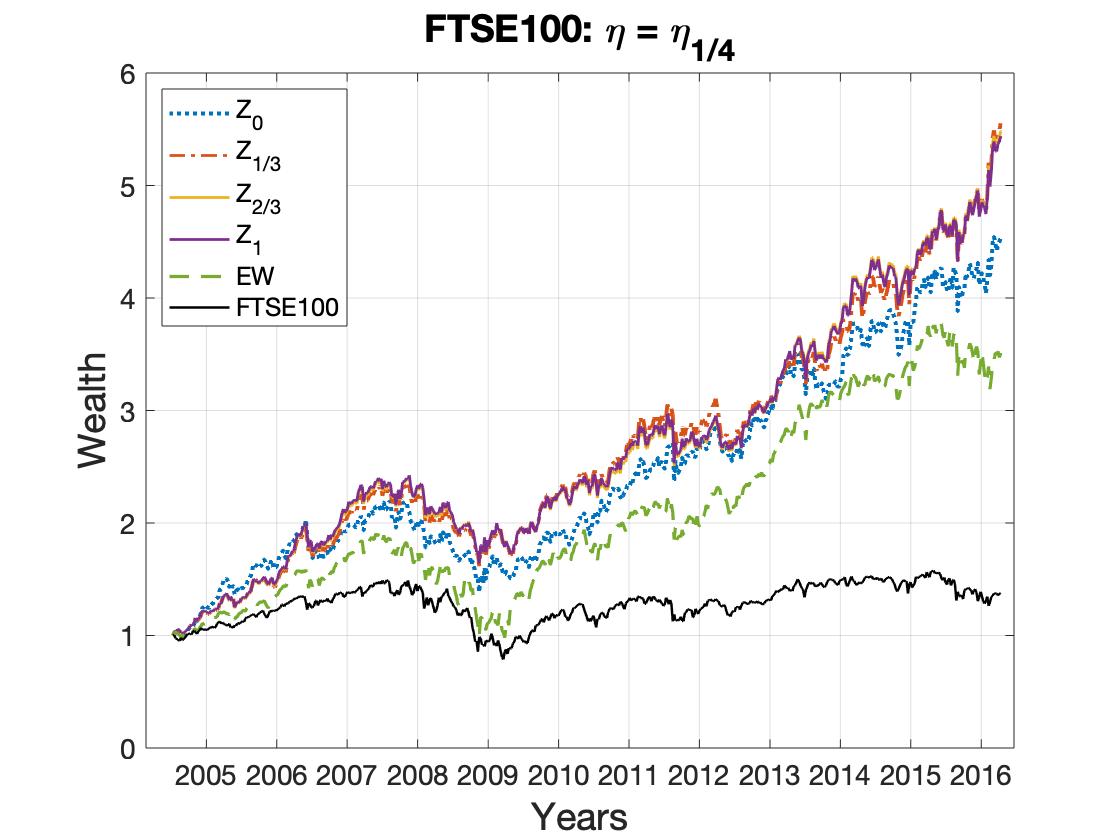}}
\vfill
\subfigure{\label{fig:CumulativeReturnsfor_eta1/2_FTSE100_weekly_0.05}
\includegraphics[scale=0.19]{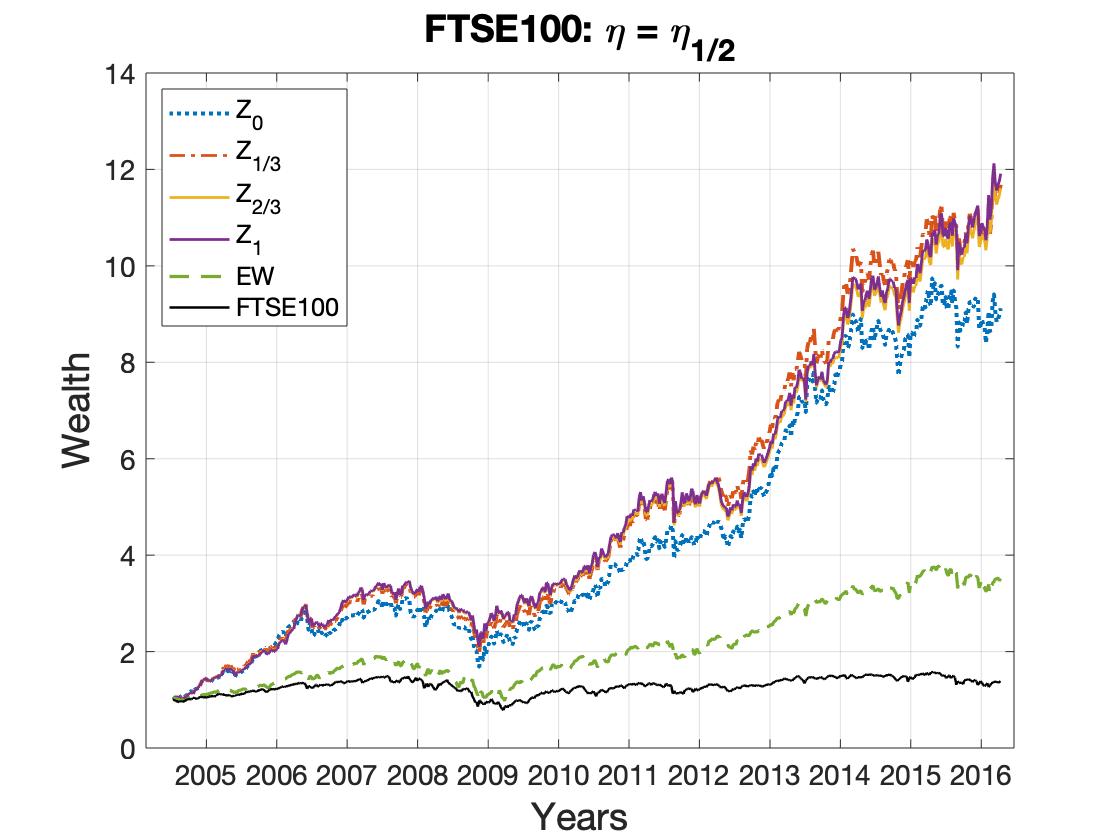}}
\hfill
\subfigure{\label{fig:CumulativeReturnsfor_eta3/4_FTSE100_weekly_0.05}
\includegraphics[scale=0.19]{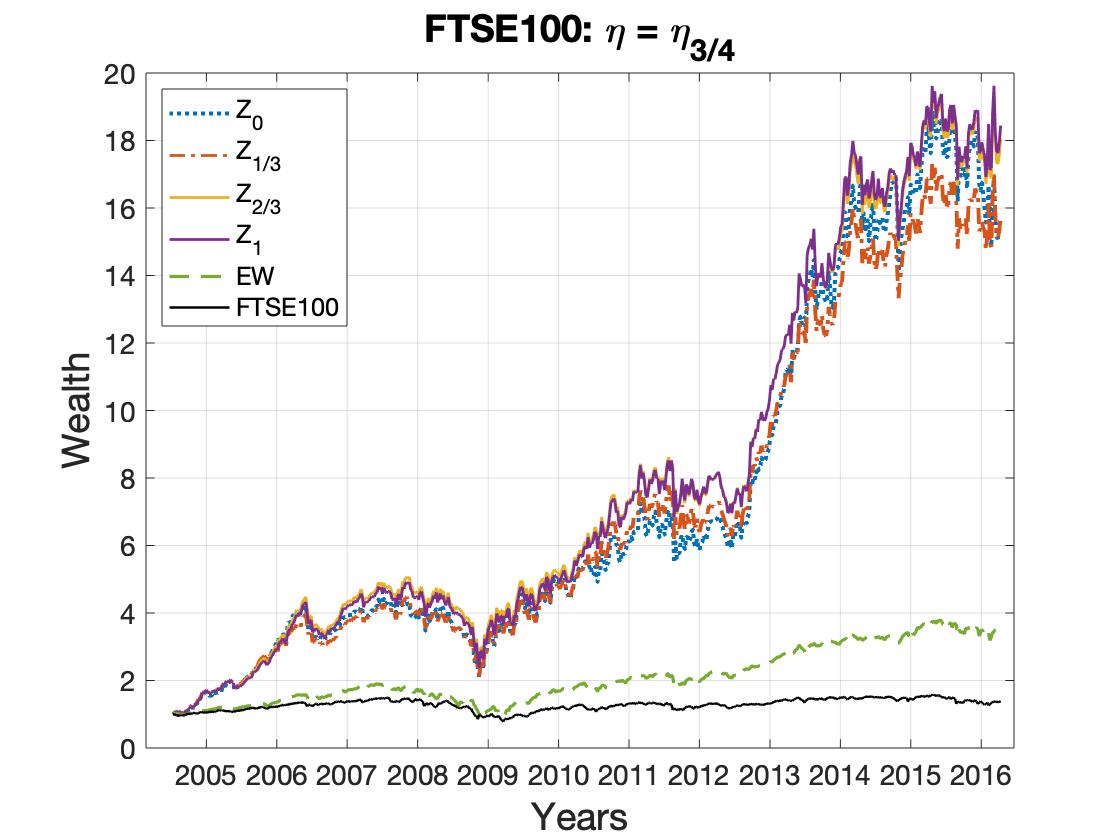}}
\caption{Cumulative out-of-sample portfolio returns using different levels of $\eta$ and $\varepsilon=5\%$ for the FTSE100 weekly dataset}
\label{fig:CumulativeReturnsforfourlevelof targetreturnetawithepsilon=0.05_FTSE100_weekly}
\end{figure}

\begin{figure}[h!]
\centering
\subfigure{\label{fig:CumulativeReturnsfor_etamin_EX50_daily_0.05}
\includegraphics[scale=0.19]{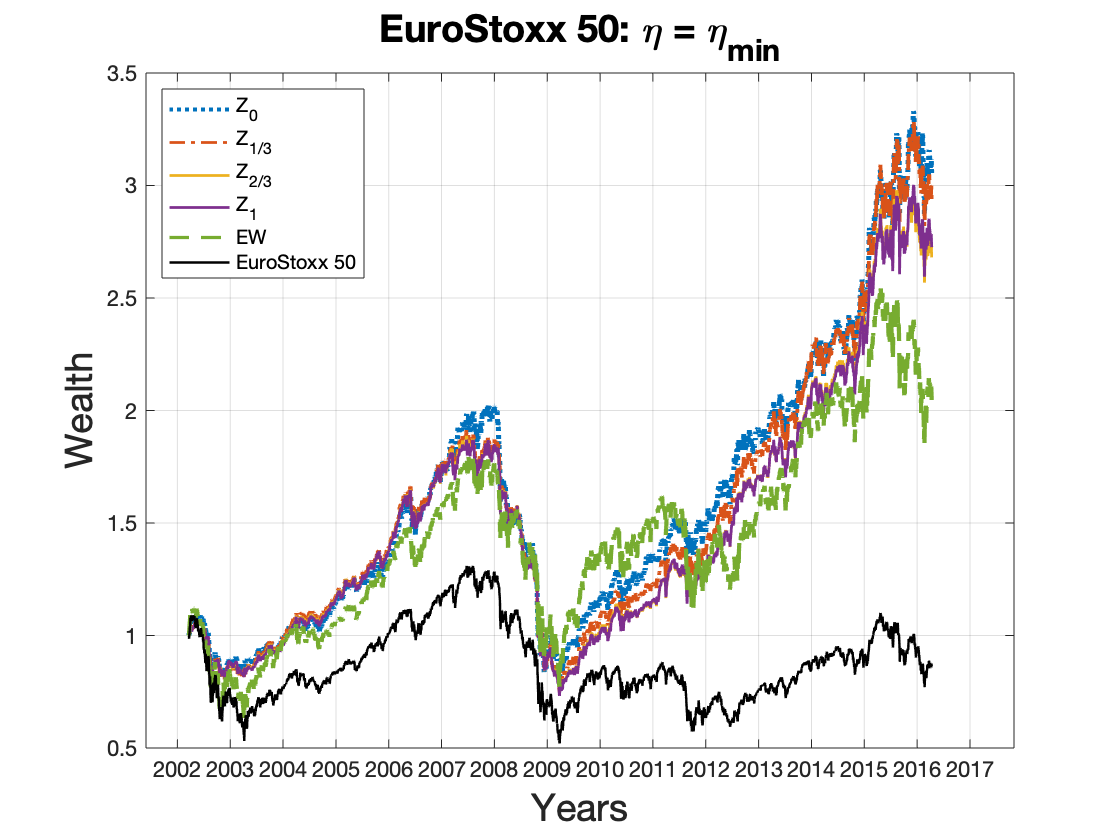}}
\hfill
\subfigure{\label{fig:CumulativeReturnsfor_eta1/4_EX50_daily_0.05}
\includegraphics[scale=0.19]{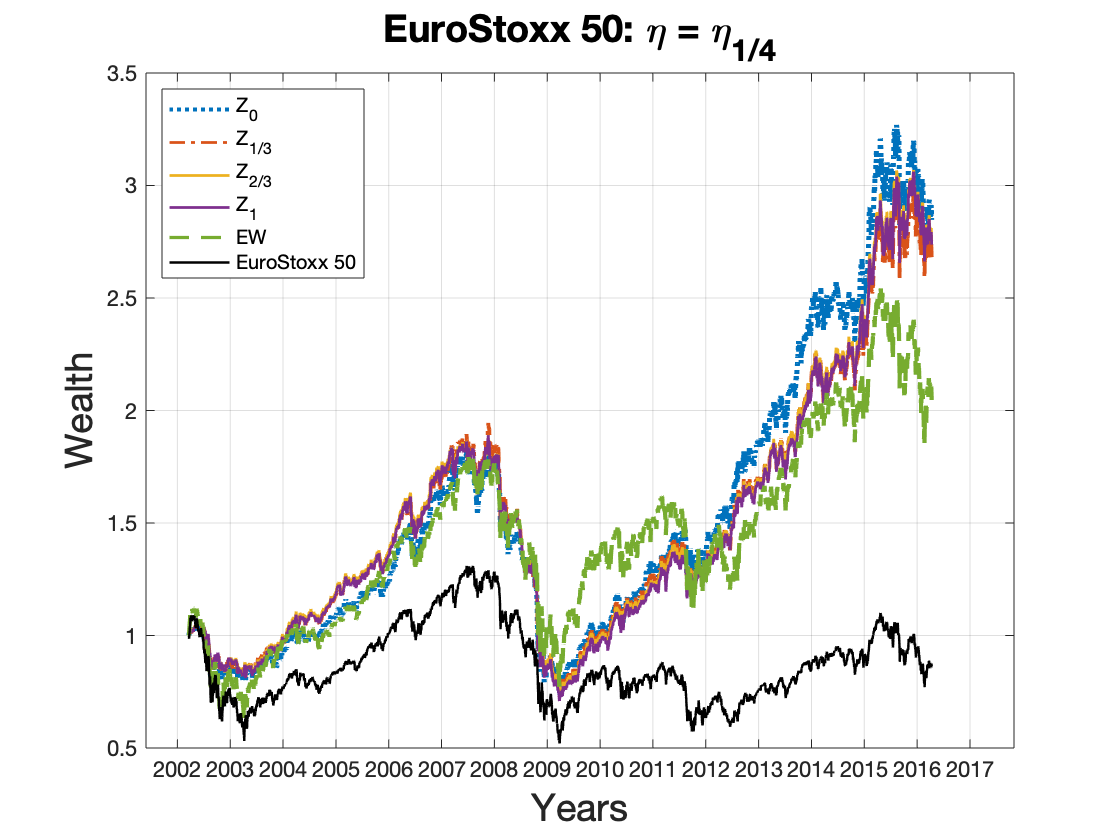}}
\vfill
\subfigure{\label{fig:CumulativeReturnsfor_eta1/2_EX50_daily_0.05}
\includegraphics[scale=0.19]{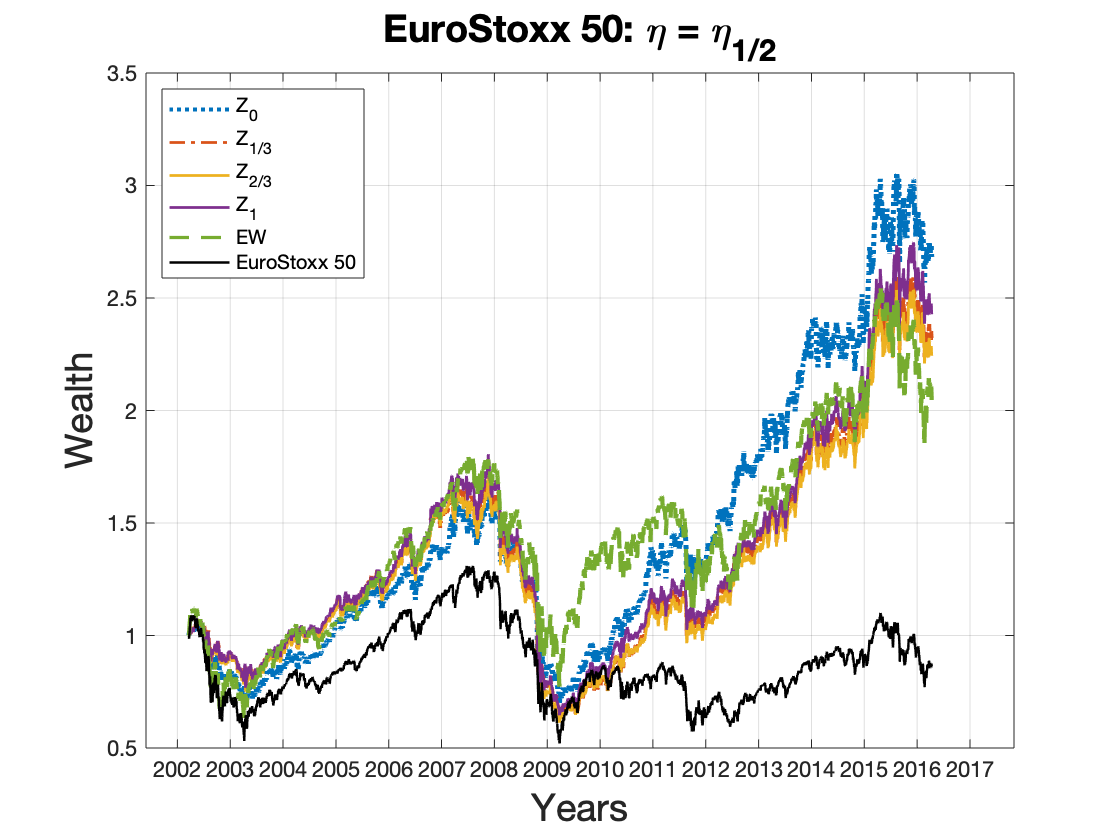}}
\hfill
\subfigure{\label{fig:CumulativeReturnsfor_eta3/4_EX50_daily_0.05}
\includegraphics[scale=0.19]{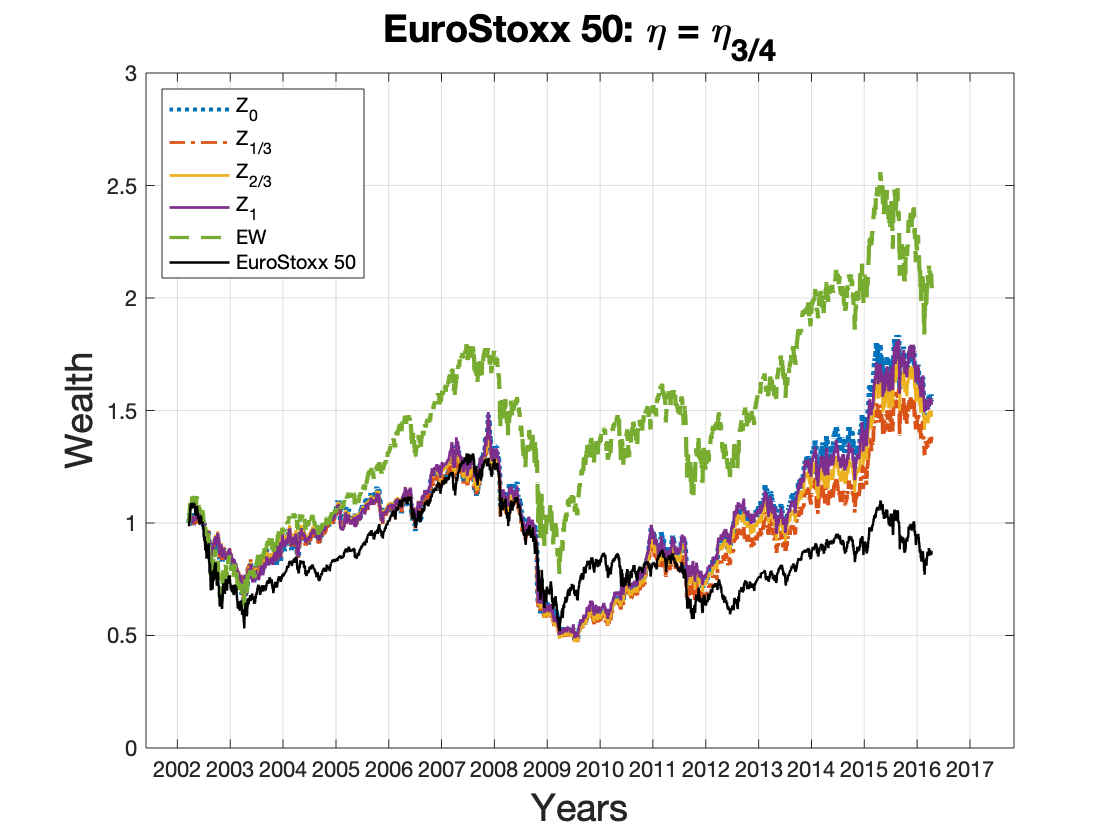}}
\caption{Cumulative out-of-sample portfolio returns using different levels of $\eta$ and $\varepsilon=5\%$ for the EuroStoxx 50 daily dataset}
\label{fig:CumulativeReturnsforfourlevelof targetreturnetawithepsilon=0.05_EuroStoxx50_daily}
\end{figure}

\begin{figure}[h]
\centering
\subfigure{\label{fig:CumulativeReturnsfor_etamin_DJ_daily_0.01}
\includegraphics[scale=0.19]{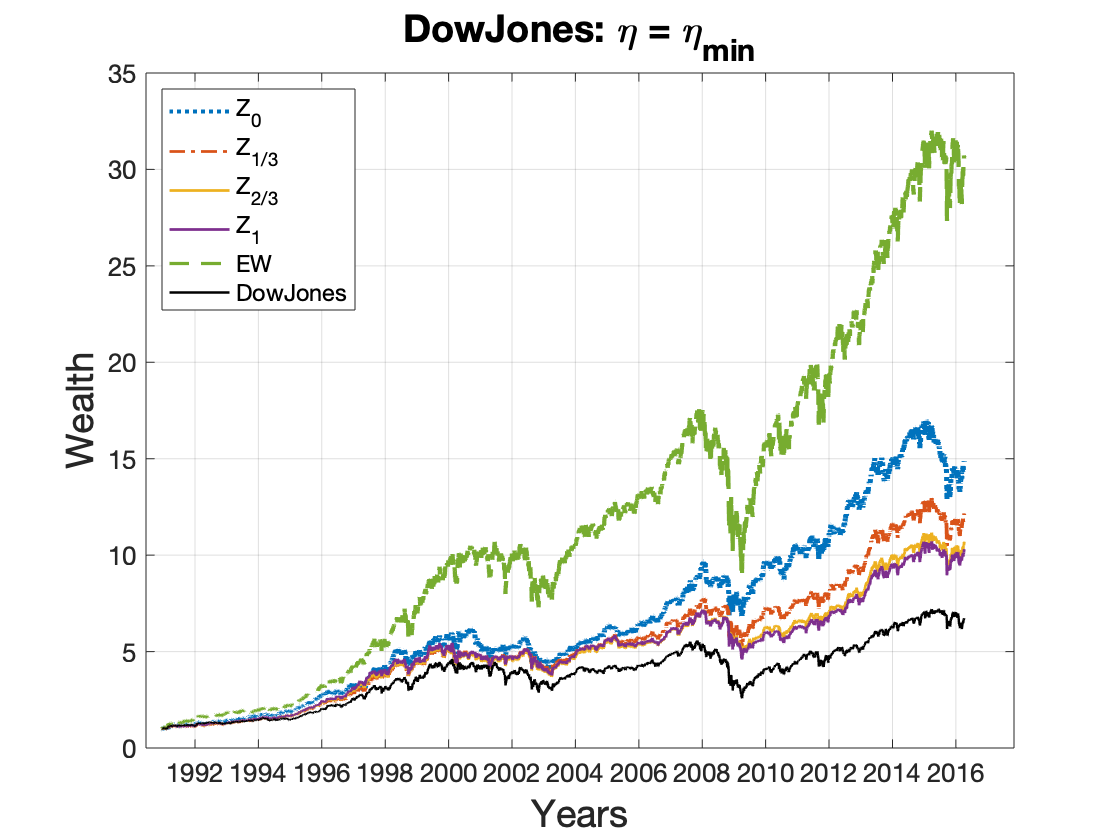}}
\hfill
\subfigure{\label{fig:CumulativeReturnsfor_eta1/4_DJ_daily_0.01}
\includegraphics[scale=0.19]{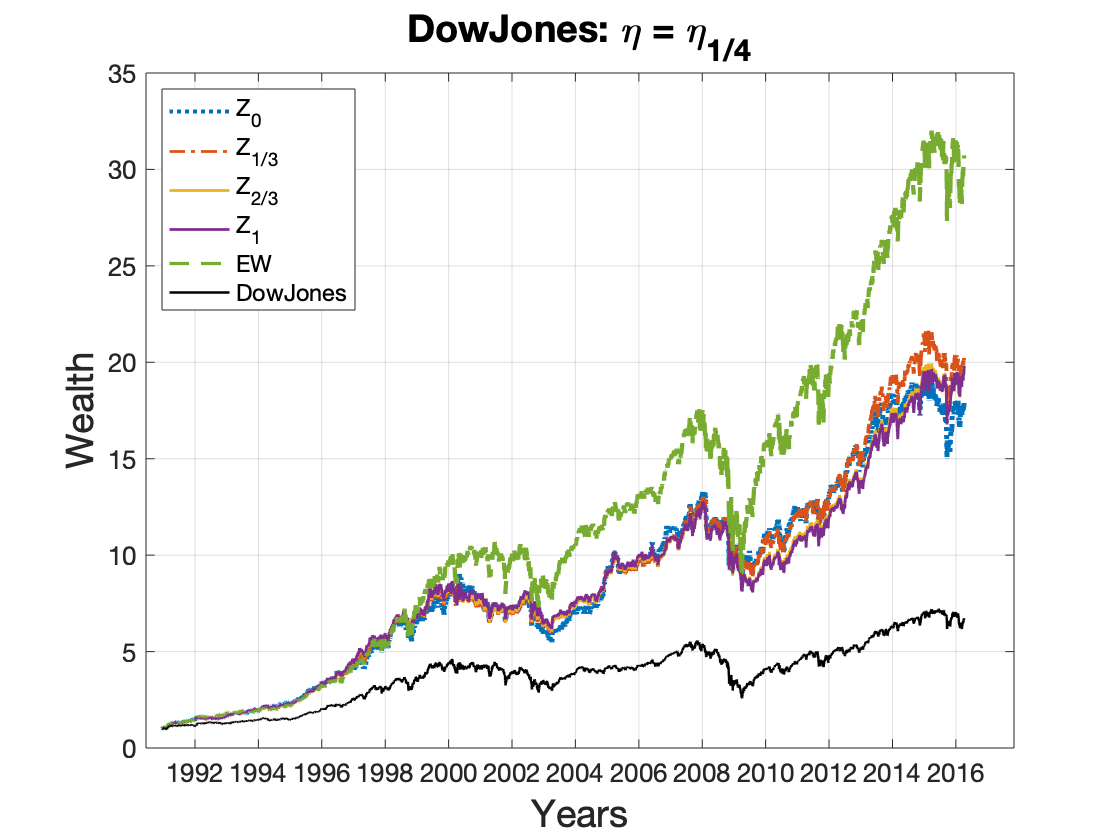}}
\vfill
\subfigure{\label{fig:CumulativeReturnsfor_eta1/2_DJ_daily_0.01}
\includegraphics[scale=0.19]{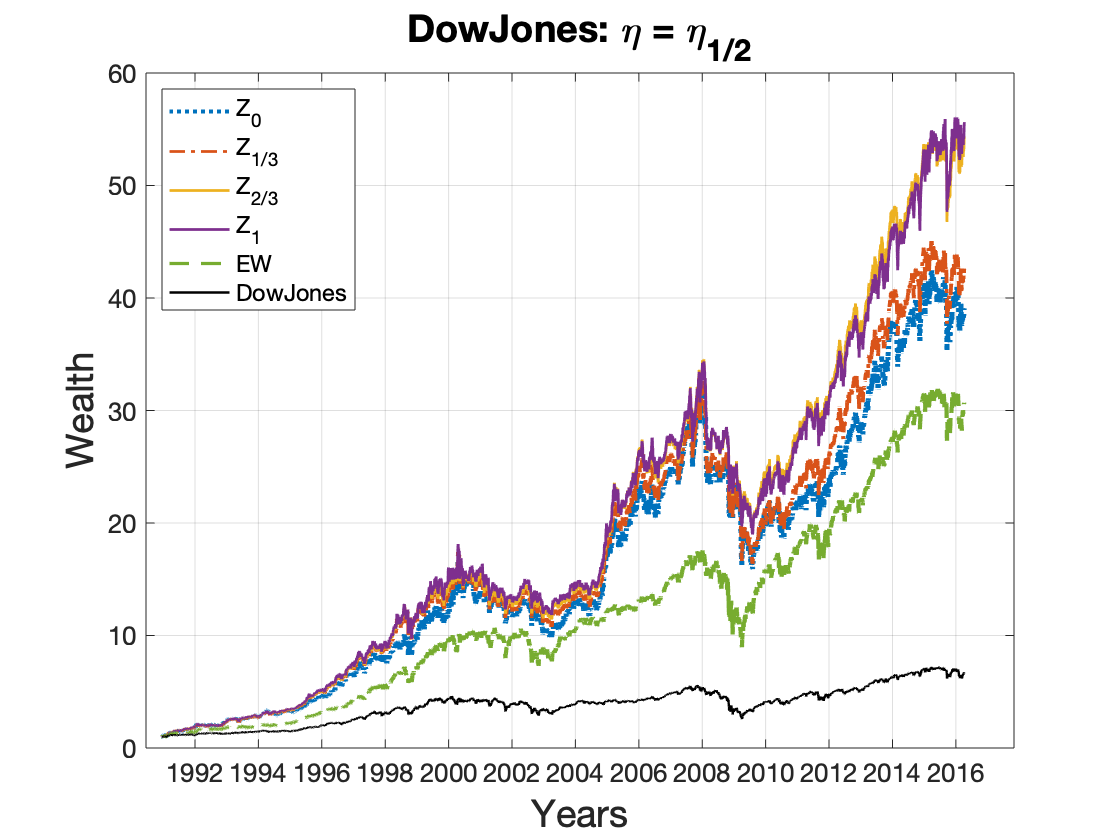}}
\hfill
\subfigure{\label{fig:CumulativeReturnsfor_eta3/4_DJ_daily_0.01}
\includegraphics[scale=0.19]{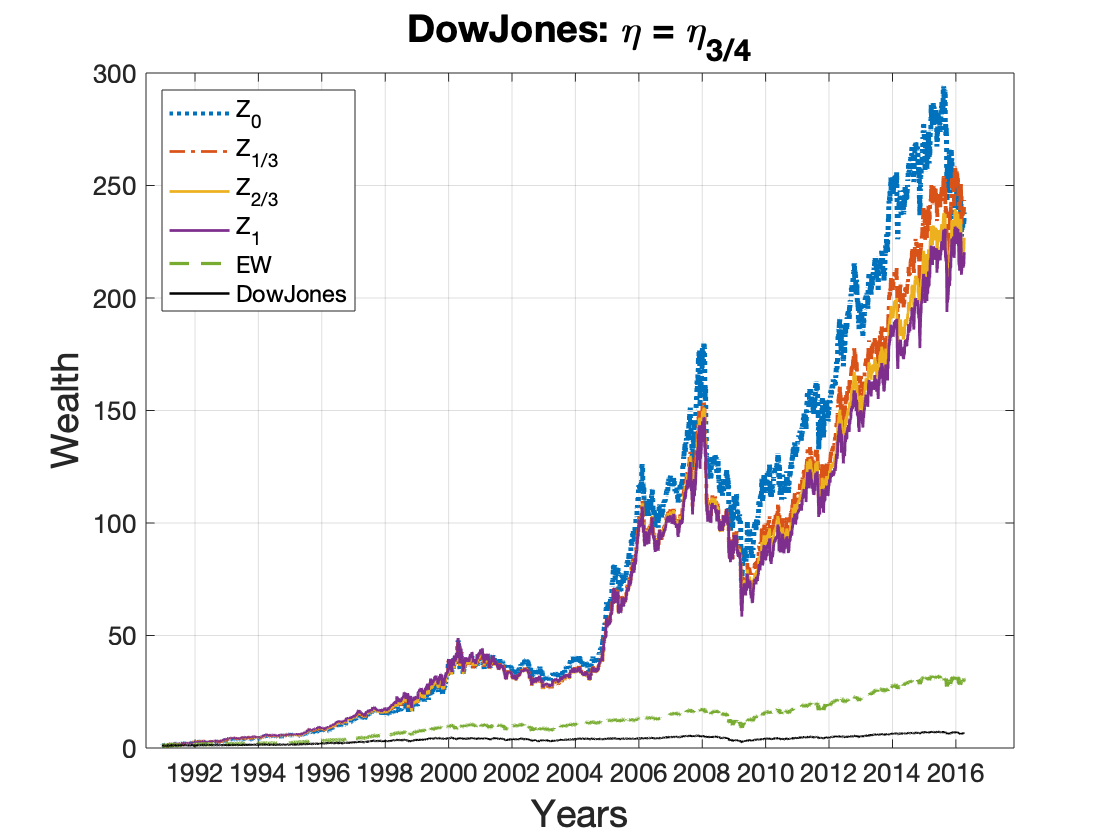}}
\caption{Cumulative out-of-sample portfolio returns using different levels of $\eta$ and $\varepsilon=1\%$ for the DowJones daily dataset}
\label{fig:CumulativeReturnsforfourlevelof targetreturnetawithepsilon=0.01_DJ_daily}
\end{figure}

\begin{figure}[h]
\centering
\subfigure{\label{fig:CumulativeReturnsfor_etamin_DJ_daily_0.05}
\includegraphics[scale=0.19]{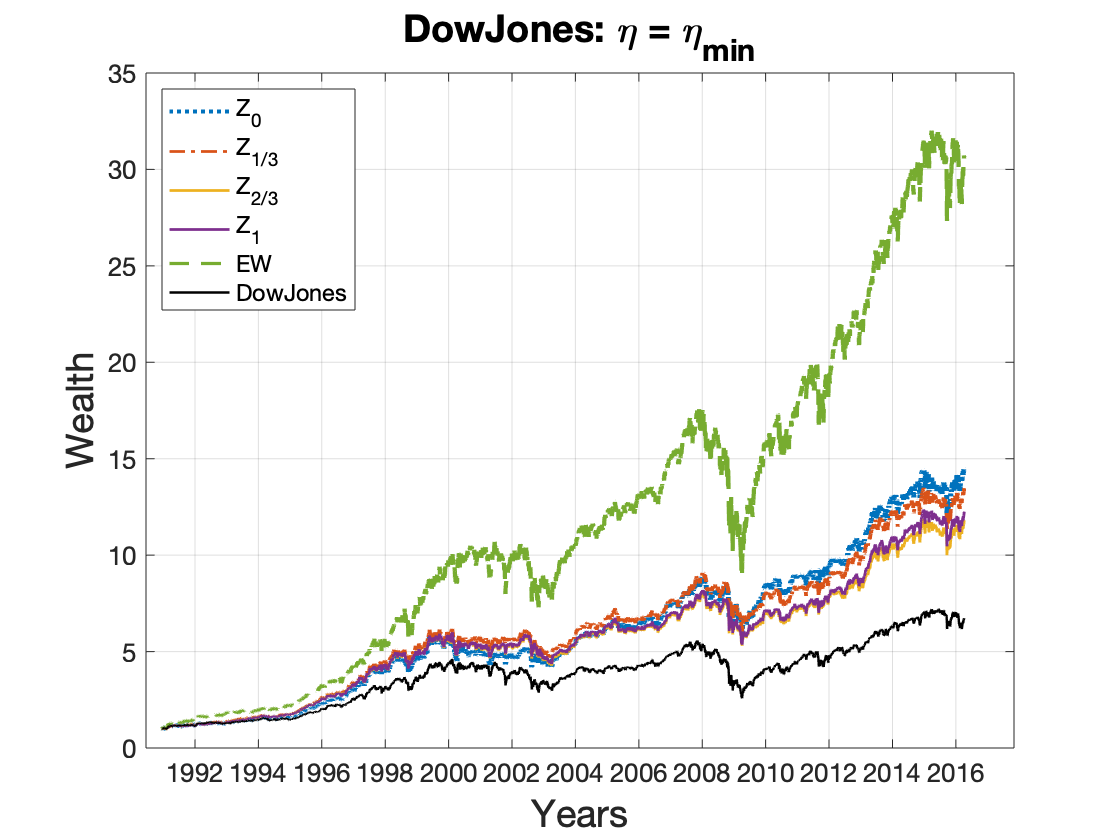}}
\hfill
\subfigure{\label{fig:CumulativeReturnsfor_eta1/4_DJ_daily_0.05}
\includegraphics[scale=0.19]{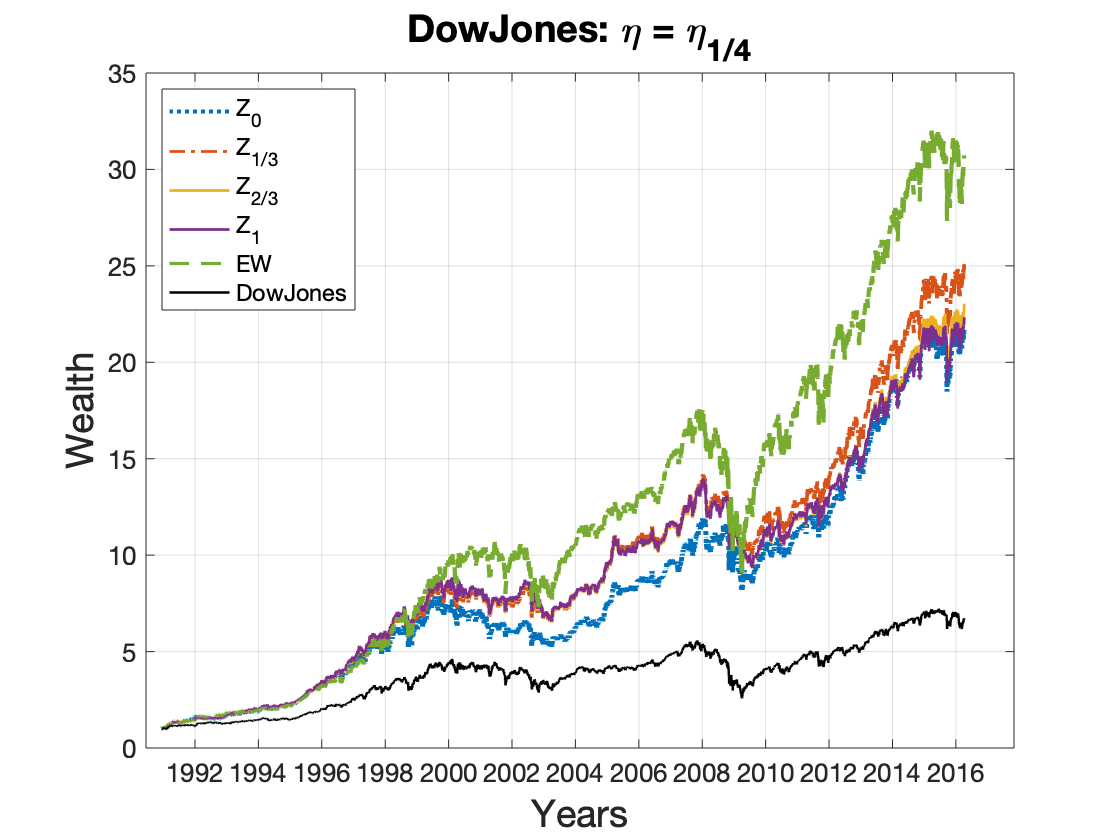}}
\vfill
\subfigure{\label{fig:CumulativeReturnsfor_eta1/2_DJ_daily_0.05}
\includegraphics[scale=0.19]{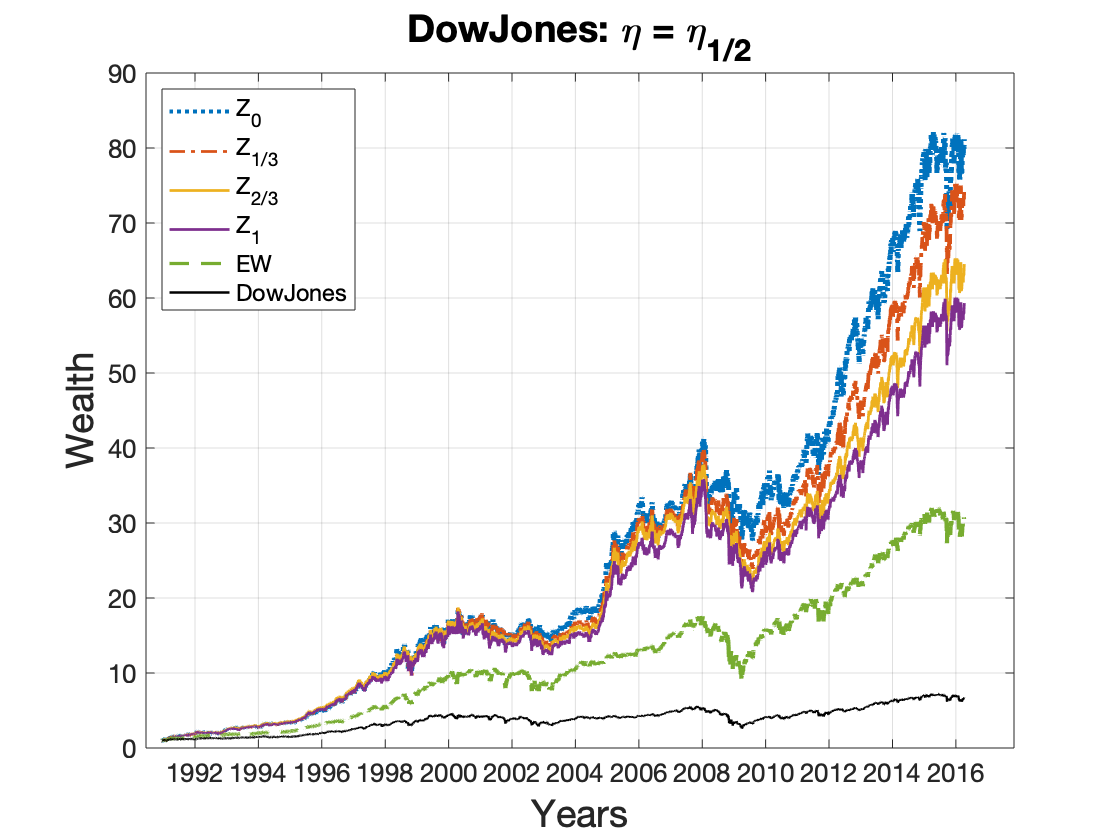}}
\hfill
\subfigure{\label{fig:CumulativeReturnsfor_eta3/4_DJ_daily_0.05}
\includegraphics[scale=0.19]{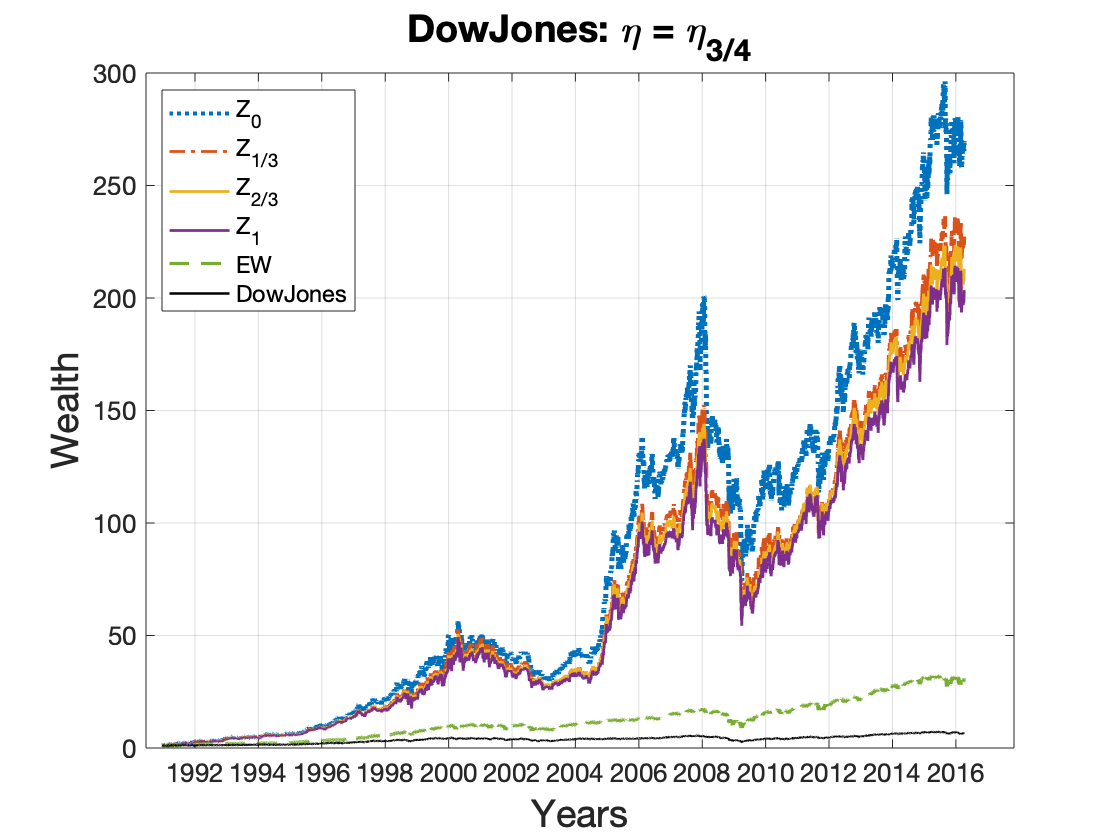}}
\caption{Cumulative out-of-sample portfolio returns using different levels of $\eta$ and $\varepsilon=5\%$ for the DowJones daily dataset}
\label{fig:CumulativeReturnsforfourlevelof targetreturnetawithepsilon=0.05_DJ_daily}
\end{figure}

\begin{figure}[h]
\centering
\subfigure{\label{fig:CumulativeReturnsfor_etamin_HangSeng_daily_0.01}
\includegraphics[scale=0.19]{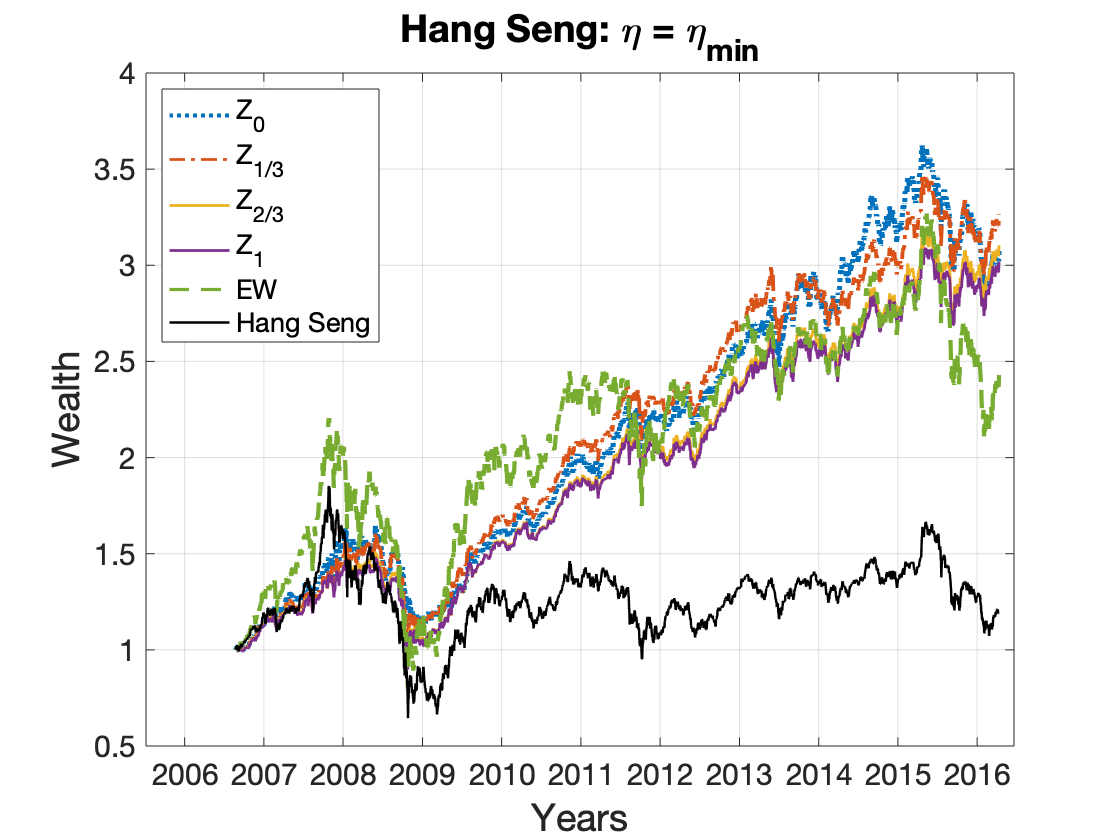}}
\hfill
\subfigure{\label{fig:CumulativeReturnsfor_eta1/4_HangSeng_daily_0.01}
\includegraphics[scale=0.19]{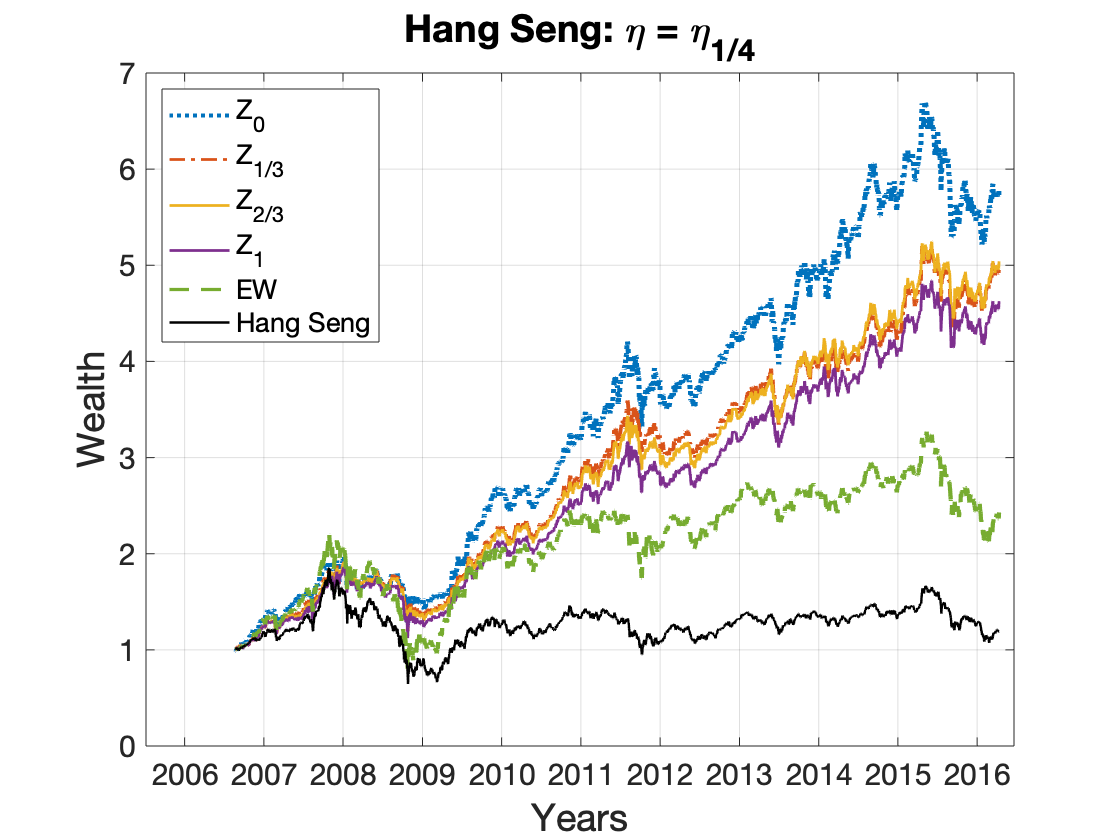}}
\vfill
\subfigure{\label{fig:CumulativeReturnsfor_eta1/2_HangSeng_daily_0.01}
\includegraphics[scale=0.19]{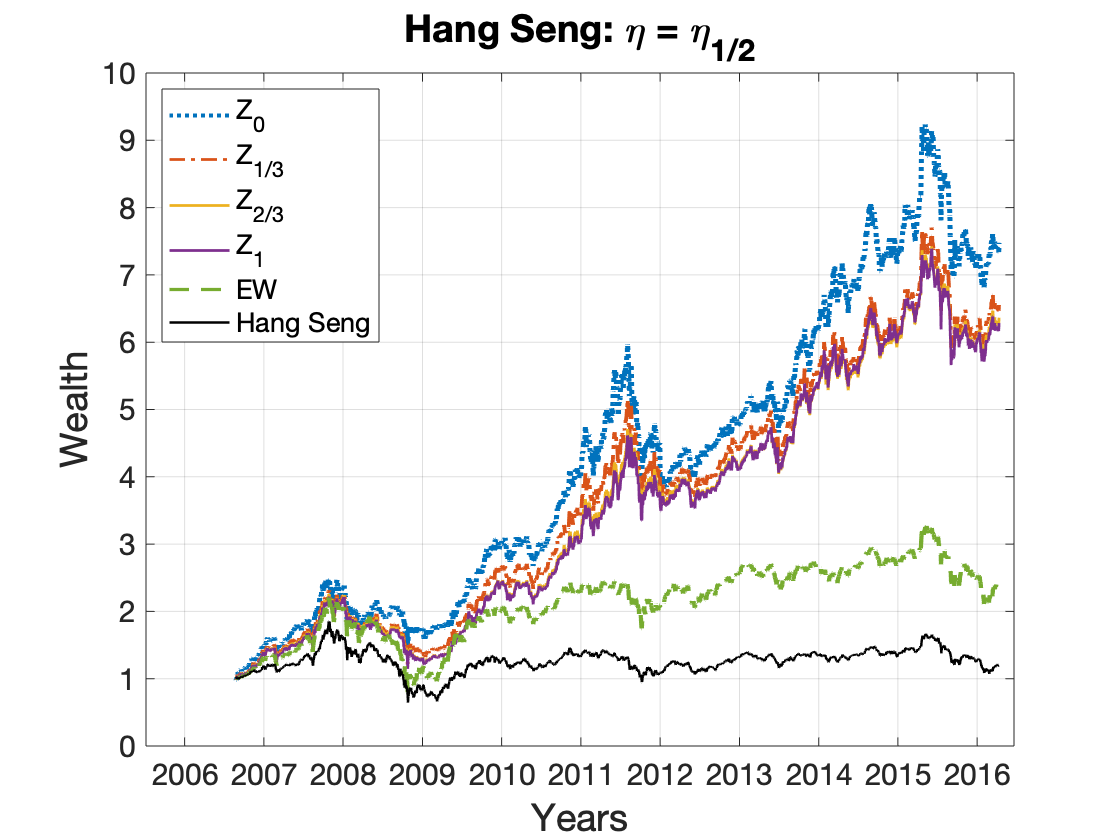}}
\hfill
\subfigure{\label{fig:CumulativeReturnsfor_eta3/4_HangSeng_daily_0.01}
\includegraphics[scale=0.19]{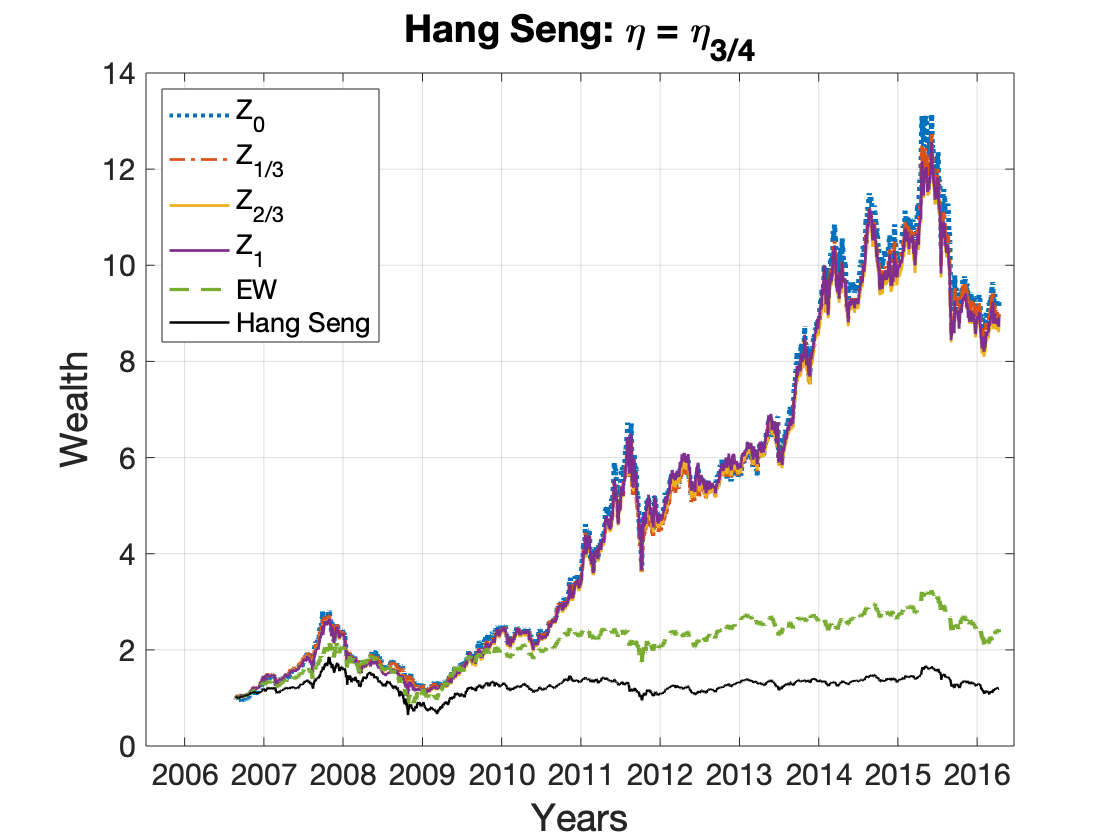}}
\caption{Cumulative out-of-sample portfolio returns using different levels of $\eta$ and $\varepsilon=1\%$ for the Hang Seng daily dataset}
\label{fig:CumulativeReturnsforfourlevelof targetreturnetawithepsilon=0.01_HangSeng_daily}
\end{figure}

\begin{figure}[h]
\centering
\subfigure{\label{fig:CumulativeReturnsfor_etamin_HangSeng_daily_0.05}
\includegraphics[scale=0.19]{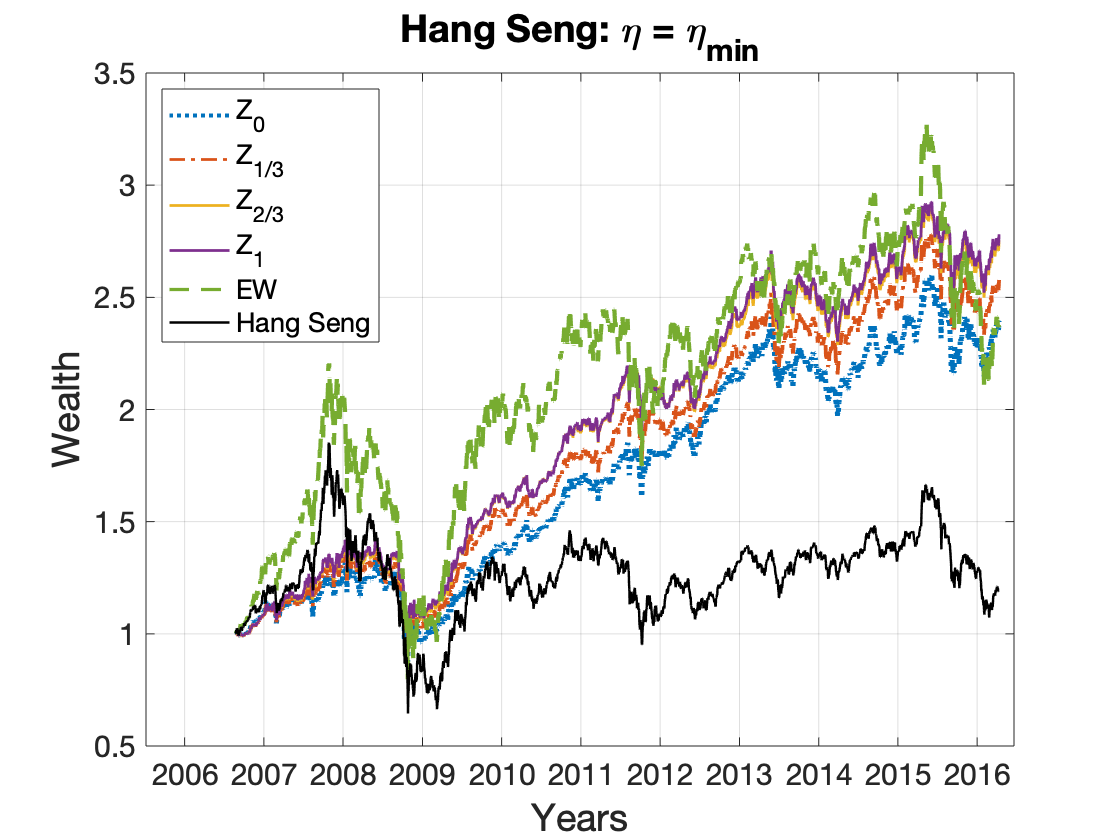}}
\hfill
\subfigure{\label{fig:CumulativeReturnsfor_eta1/4_HangSeng_daily_0.05}
\includegraphics[scale=0.19]{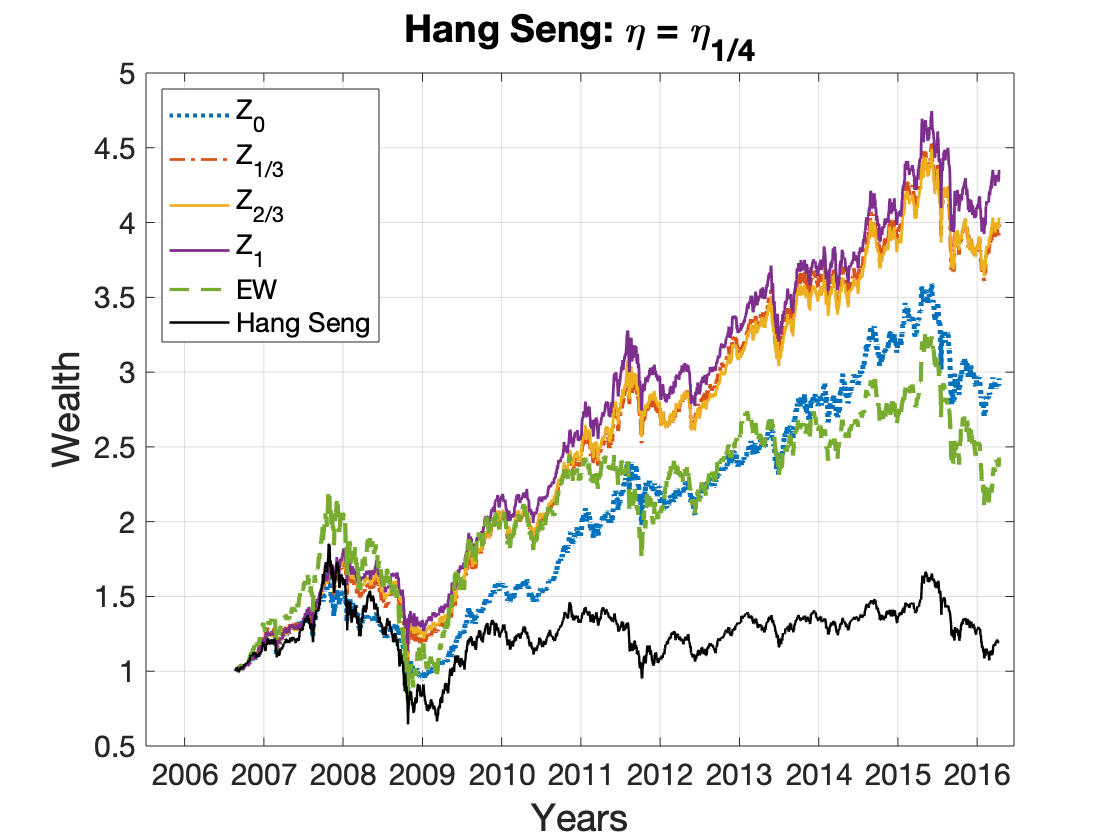}}
\vfill
\subfigure{\label{fig:CumulativeReturnsfor_eta1/2_HangSeng_daily_0.05}
\includegraphics[scale=0.19]{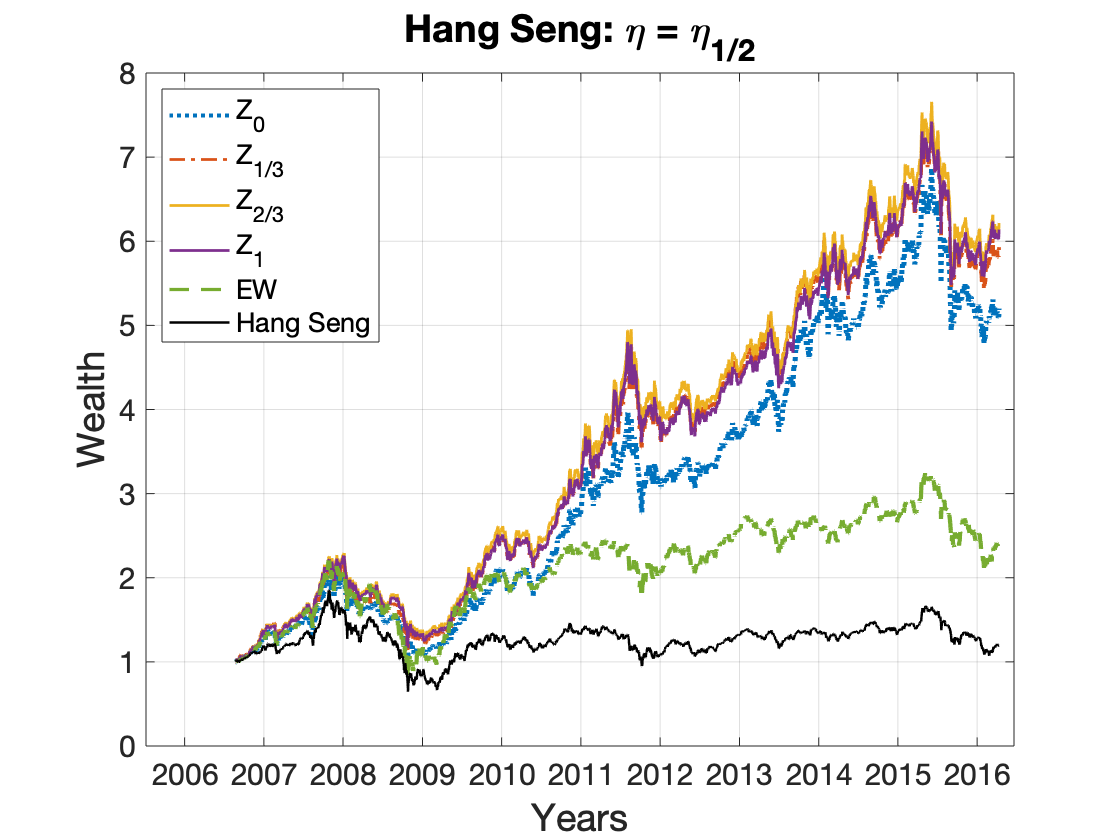}}
\hfill
\subfigure{\label{fig:CumulativeReturnsfor_eta3/4_HangSeng_daily_0.05}
\includegraphics[scale=0.19]{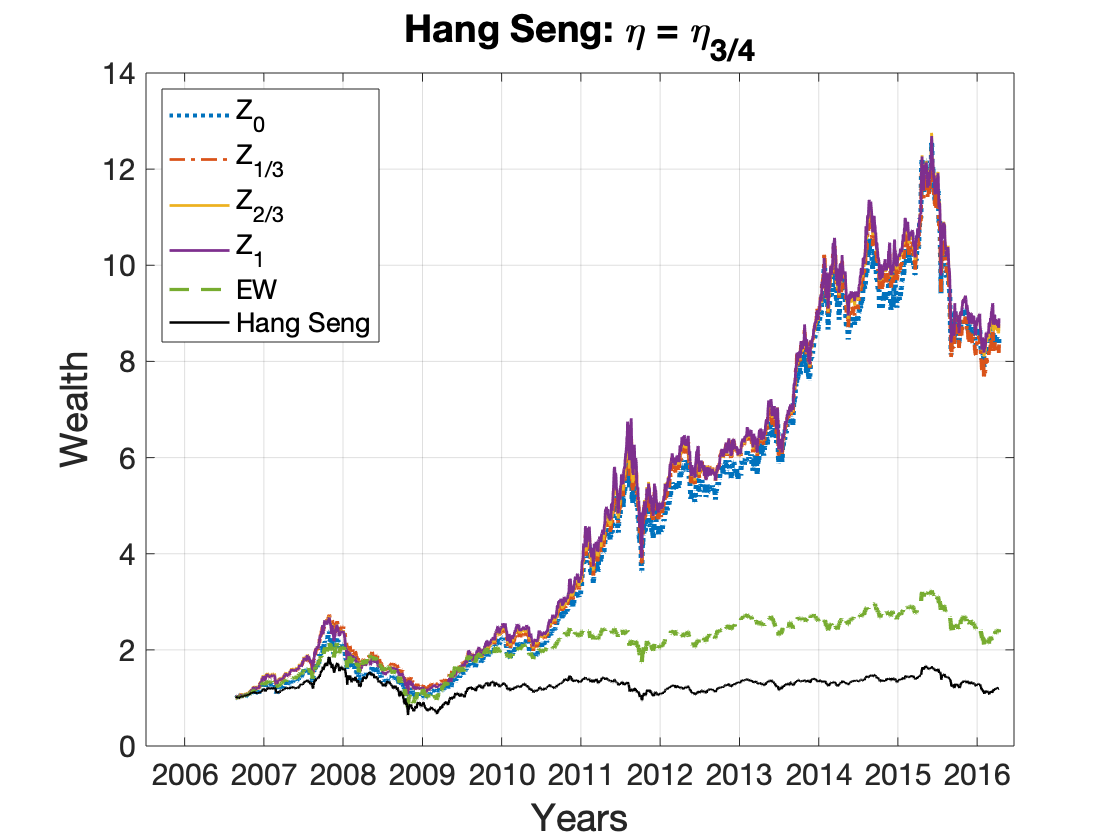}}
\caption{Cumulative out-of-sample portfolio returns using different levels of $\eta$ and $\varepsilon=5\%$ for the Hang Seng daily dataset}
\label{fig:CumulativeReturnsforfourlevelof targetreturnetawithepsilon=0.05_HangSeng_daily}
\end{figure}

For the sake of completeness, we report here additional tables containing
the computational results obtained by the 16 analyzed portfolio strategies
and by the benchmark EW portfolio
on DowJones weekly dataset with $\varepsilon=1\%$
(Tables \ref{tab:Weekly_DowJones_0.01}), on NASDAQ100 weekly dataset with $\varepsilon=1\%$ and
$\varepsilon=5\%$
(Tables \ref{tab:Weekly_NASDAQ100_0.01} and \ref{tab:Weekly_NASDAQ100_0.05}, respectively),
on FTSE100 weekly dataset with $\varepsilon=1\%$ and
$\varepsilon=5\%$
(Tables \ref{tab:Weekly_FTSE100_0.01} and \ref{tab:Weekly_FTSE100_0.05}, respectively),
on EuroStoxx 50 daily dataset with $\varepsilon=5\%$
(Table \ref{tab:Daily_EuroStoxx50_0.05}),
on DowJones daily dataset with $\varepsilon=1\%$ and
$\varepsilon=5\%$
(Tables \ref{tab:Daily_DowJones_0.01} and \ref{tab:Daily_DowJones_0.05}, respectively), and
on Hang Seng daily dataset with $\varepsilon=1\%$ and
$\varepsilon=5\%$
(Tables \ref{tab:Daily_HangSeng_0.01} and \ref{tab:Daily_HangSeng_0.05},
respectively).
Furthermore, we also report additional figures showing, for each portfolio strategy analyzed,
the cumulative out-of-sample portfolio returns
on DowJones weekly dataset with $\varepsilon=1\%$
(Figures \ref{fig:CumulativeReturnsforfourlevelof targetreturnetawithepsilon=0.01_DJ_weekly}, on NASDAQ100 weekly dataset with $\varepsilon=1\%$ and
$\varepsilon=5\%$
(Figures \ref{fig:CumulativeReturnsforfourlevelof targetreturnetawithepsilon=0.01_NASDAQ100_weekly} and \ref{fig:CumulativeReturnsforfourlevelof targetreturnetawithepsilon=0.05_NASDAQ100_weekly}, respectively),
on FTSE100 weekly dataset with $\varepsilon=1\%$ and
$\varepsilon=5\%$
(Figures \ref{fig:CumulativeReturnsforfourlevelof targetreturnetawithepsilon=0.01_FTSE100_weekly} and \ref{fig:CumulativeReturnsforfourlevelof targetreturnetawithepsilon=0.05_FTSE100_weekly}, respectively),
on EuroStoxx 50 daily dataset with $\varepsilon=5\%$
(Figure \ref{fig:CumulativeReturnsforfourlevelof targetreturnetawithepsilon=0.05_EuroStoxx50_daily}),
on DowJones daily dataset with $\varepsilon=1\%$ and
$\varepsilon=5\%$
(Figures \ref{fig:CumulativeReturnsforfourlevelof targetreturnetawithepsilon=0.01_DJ_daily} and \ref{fig:CumulativeReturnsforfourlevelof targetreturnetawithepsilon=0.05_DJ_daily}, respectively) and
on Hang Seng daily dataset with $\varepsilon=1\%$ and
$\varepsilon=5\%$
(Figures \ref{fig:CumulativeReturnsforfourlevelof targetreturnetawithepsilon=0.01_HangSeng_daily} and
\ref{fig:CumulativeReturnsforfourlevelof targetreturnetawithepsilon=0.05_HangSeng_daily}, respectively).
\newpage
\begin{table}[htbp]
	\centering
	\caption{Out-of-sample performance results for the DowJones weekly dataset
		with $\varepsilon=1\%$}
	\resizebox{\textwidth}{!}{
\begin{tabular}{|l|ccccccccccccccccc|}
			\cmidrule{3-18}    \multicolumn{1}{r}{} & \multicolumn{1}{r|}{} & \multicolumn{4}{c|}{\boldmath{$\eta_{min}$}} & \multicolumn{4}{c|}{\boldmath{$\eta_{1/4}$}} & \multicolumn{4}{c|}{\boldmath{$\eta_{1/2}$}} & \multicolumn{4}{c|}{\boldmath{$\eta_{3/4}$}} \\
			\cmidrule{2-18}    \multicolumn{1}{r|}{} & \multicolumn{1}{c|}{\textbf{EW}} & \multicolumn{1}{c|}{\boldmath{$z_0$}} & \multicolumn{1}{c|}{\boldmath{$z_{1/3}$}} & \multicolumn{1}{c|}{\boldmath{$z_{2/3}$}} & \multicolumn{1}{c|}{\boldmath{$z_1$}} & \multicolumn{1}{c|}{\boldmath{$z_0$}} & \multicolumn{1}{c|}{\boldmath{$z_{1/3}$}} & \multicolumn{1}{c|}{\textbf{$z_{2/3}$}} & \multicolumn{1}{c|}{\textbf{$z_1$}} & \multicolumn{1}{c|}{\boldmath{$z_0$}} & \multicolumn{1}{c|}{\boldmath{$z_{1/3}$}} & \multicolumn{1}{c|}{\textbf{$z_{2/3}$}} & \multicolumn{1}{c|}{\textbf{$z_1$}} & \multicolumn{1}{c|}{\boldmath{$z_0$}} & \multicolumn{1}{c|}{\boldmath{$z_{1/3}$}} & \multicolumn{1}{c|}{\textbf{$z_{2/3}$}} & \textbf{$z_1$} \\
\bottomrule
	\end{tabular}}%
	\label{tab:Weekly_DowJones_0.01}%
\end{table}%
\begin{table}[htbp]
	\centering
	\caption{Out-of-sample performance results for the NASDAQ100 weekly dataset
		with $\varepsilon=1\%$}
	\resizebox{\textwidth}{!}{\begin{tabular}{|l|ccccccccccccccccc|}
			\cmidrule{3-18}    \multicolumn{1}{r}{} & \multicolumn{1}{r|}{} & \multicolumn{4}{c|}{\boldmath{$\eta_{min}$}} & \multicolumn{4}{c|}{\boldmath{$\eta_{1/4}$}} & \multicolumn{4}{c|}{\boldmath{$\eta_{1/2}$}} & \multicolumn{4}{c|}{\boldmath{$\eta_{3/4}$}} \\
			\cmidrule{2-18}    \multicolumn{1}{r|}{} & \multicolumn{1}{c|}{\textbf{EW}} & \multicolumn{1}{c|}{\boldmath{$z_0$}} & \multicolumn{1}{c|}{\boldmath{$z_{1/3}$}} & \multicolumn{1}{c|}{\boldmath{$z_{2/3}$}} & \multicolumn{1}{c|}{\boldmath{$z_1$}} & \multicolumn{1}{c|}{\boldmath{$z_0$}} & \multicolumn{1}{c|}{\boldmath{$z_{1/3}$}} & \multicolumn{1}{c|}{\boldmath{$z_{2/3}$}} & \multicolumn{1}{c|}{\boldmath{$z_1$}} & \multicolumn{1}{c|}{\boldmath{$z_0$}} & \multicolumn{1}{c|}{\boldmath{$z_{1/3}$}} & \multicolumn{1}{c|}{\boldmath{$z_{2/3}$}} & \multicolumn{1}{c|}{\boldmath{$z_1$}} & \multicolumn{1}{c|}{\boldmath{$z_0$}} & \multicolumn{1}{c|}{\boldmath{$z_{1/3}$}} & \multicolumn{1}{c|}{\boldmath{$z_{2/3}$}} & \boldmath{$z_1$} \\
\bottomrule
	\end{tabular}}%
	\label{tab:Weekly_NASDAQ100_0.01}%
\end{table}%
%
\begin{table}[htbp]
	\centering
	\caption{Out-of-sample performance results for the NASDAQ100 weekly dataset
		with $\varepsilon=5\%$}
	\resizebox{\textwidth}{!}{\begin{tabular}{|l|ccccccccccccccccc|}
			\cmidrule{3-18}    \multicolumn{1}{r}{} & \multicolumn{1}{r|}{} & \multicolumn{4}{c|}{\boldmath{$\eta_{min}$}} & \multicolumn{4}{c|}{\boldmath{$\eta_{1/4}$}} & \multicolumn{4}{c|}{\boldmath{$\eta_{1/2}$}} & \multicolumn{4}{c|}{\boldmath{$\eta_{3/4}$}} \\
			\cmidrule{2-18}    \multicolumn{1}{r|}{} & \multicolumn{1}{c|}{\textbf{EW}} & \multicolumn{1}{c|}{\boldmath{$z_0$}} & \multicolumn{1}{c|}{\boldmath{$z_{1/3}$}} & \multicolumn{1}{c|}{\boldmath{$z_{2/3}$}} & \multicolumn{1}{c|}{\boldmath{$z_1$}} & \multicolumn{1}{c|}{\boldmath{$z_0$}} & \multicolumn{1}{c|}{\boldmath{$z_{1/3}$}} & \multicolumn{1}{c|}{\boldmath{$z_{2/3}$}} & \multicolumn{1}{c|}{\boldmath{$z_1$}} & \multicolumn{1}{c|}{\boldmath{$z_0$}} & \multicolumn{1}{c|}{\boldmath{$z_{1/3}$}} & \multicolumn{1}{c|}{\boldmath{$z_{2/3}$}} & \multicolumn{1}{c|}{\boldmath{$z_1$}} & \multicolumn{1}{c|}{\boldmath{$z_0$}} & \multicolumn{1}{c|}{\boldmath{$z_{1/3}$}} & \multicolumn{1}{c|}{\boldmath{$z_{2/3}$}} & \boldmath{$z_1$} \\
\bottomrule
	\end{tabular}}%
	\label{tab:Weekly_NASDAQ100_0.05}%
	\vspace{0.8cm}
\end{table}%
\newpage
%
\begin{table}[htbp]
	\centering
	\caption{Out-of-sample performance results for the FTSE100 weekly dataset
		with $\varepsilon=1\%$}
	\resizebox{\textwidth}{!}{\begin{tabular}{|l|ccccccccccccccccc|}
			\cmidrule{3-18}    \multicolumn{1}{r}{} & \multicolumn{1}{r|}{} & \multicolumn{4}{c|}{\boldmath{$\eta_{min}$}} & \multicolumn{4}{c|}{\boldmath{$\eta_{1/4}$}} & \multicolumn{4}{c|}{\boldmath{$\eta_{1/2}$}} & \multicolumn{4}{c|}{\boldmath{$\eta_{3/4}$}} \\
			\cmidrule{2-18}    \multicolumn{1}{r|}{} & \multicolumn{1}{c|}{\textbf{EW}} & \multicolumn{1}{c|}{\boldmath{$z_0$}} & \multicolumn{1}{c|}{\boldmath{$z_{1/3}$}} & \multicolumn{1}{c|}{\boldmath{$z_{2/3}$}} & \multicolumn{1}{c|}{\boldmath{$z_1$}} & \multicolumn{1}{c|}{\boldmath{$z_0$}} & \multicolumn{1}{c|}{\boldmath{$z_{1/3}$}} & \multicolumn{1}{c|}{\boldmath{$z_{2/3}$}} & \multicolumn{1}{c|}{\boldmath{$z_1$}} & \multicolumn{1}{c|}{\boldmath{$z_0$}} & \multicolumn{1}{c|}{\boldmath{$z_{1/3}$}} & \multicolumn{1}{c|}{\boldmath{$z_{2/3}$}} & \multicolumn{1}{c|}{\boldmath{$z_1$}} & \multicolumn{1}{c|}{\boldmath{$z_0$}} & \multicolumn{1}{c|}{\boldmath{$z_{1/3}$}} & \multicolumn{1}{c|}{\boldmath{$z_{2/3}$}} & \boldmath{$z_1$} \\
\bottomrule
	\end{tabular}}%
	\label{tab:Weekly_FTSE100_0.01}%
	\vspace{0.8cm}
\end{table}%
%
\begin{table}[htbp]
	\centering
	\caption{Out-of-sample performance results for the FTSE100 weekly dataset
		with $\varepsilon=5\%$}
	\resizebox{\textwidth}{!}{\begin{tabular}{|l|ccccccccccccccccc|}
			\cmidrule{3-18}    \multicolumn{1}{r}{} & \multicolumn{1}{r|}{} & \multicolumn{4}{c|}{\boldmath{$\eta_{min}$}} & \multicolumn{4}{c|}{\boldmath{$\eta_{1/4}$}} & \multicolumn{4}{c|}{\boldmath{$\eta_{1/2}$}} & \multicolumn{4}{c|}{\boldmath{$\eta_{3/4}$}} \\
			\cmidrule{2-18}    \multicolumn{1}{r|}{} & \multicolumn{1}{c|}{\textbf{EW}} & \multicolumn{1}{c|}{\boldmath{$z_0$}} & \multicolumn{1}{c|}{\boldmath{$z_{1/3}$}} & \multicolumn{1}{c|}{\boldmath{$z_{2/3}$}} & \multicolumn{1}{c|}{\boldmath{$z_1$}} & \multicolumn{1}{c|}{\boldmath{$z_0$}} & \multicolumn{1}{c|}{\boldmath{$z_{1/3}$}} & \multicolumn{1}{c|}{\boldmath{$z_{2/3}$}} & \multicolumn{1}{c|}{\boldmath{$z_1$}} & \multicolumn{1}{c|}{\boldmath{$z_0$}} & \multicolumn{1}{c|}{\boldmath{$z_{1/3}$}} & \multicolumn{1}{c|}{\boldmath{$z_{2/3}$}} & \multicolumn{1}{c|}{\boldmath{$z_1$}} & \multicolumn{1}{c|}{\boldmath{$z_0$}} & \multicolumn{1}{c|}{\boldmath{$z_{1/3}$}} & \multicolumn{1}{c|}{\boldmath{$z_{2/3}$}} & \boldmath{$z_1$} \\
\bottomrule
	\end{tabular}}%
	\label{tab:Weekly_FTSE100_0.05}%
\end{table}%
%
\begin{table}[htbp]
	\centering
	\caption{Out-of-sample performance results for the EuroStoxx 50 daily dataset
		with $\varepsilon=5\%$}
	\resizebox{\textwidth}{!}{\begin{tabular}{|l|ccccccccccccccccc|}
			\cmidrule{3-18}    \multicolumn{1}{r}{} & \multicolumn{1}{r|}{} & \multicolumn{4}{c|}{\boldmath{$\eta_{min}$}} & \multicolumn{4}{c|}{\boldmath{$\eta_{1/4}$}} & \multicolumn{4}{c|}{\boldmath{$\eta_{1/2}$}} & \multicolumn{4}{c|}{\boldmath{$\eta_{3/4}$}} \\
			\cmidrule{2-18}    \multicolumn{1}{r|}{} & \multicolumn{1}{c|}{\textbf{EW}} & \multicolumn{1}{c|}{\boldmath{$z_0$}} & \multicolumn{1}{c|}{\boldmath{$z_{1/3}$}} & \multicolumn{1}{c|}{\boldmath{$z_{2/3}$}} & \multicolumn{1}{c|}{\boldmath{$z_1$}} & \multicolumn{1}{c|}{\boldmath{$z_0$}} & \multicolumn{1}{c|}{\boldmath{$z_{1/3}$}} & \multicolumn{1}{c|}{\boldmath{$z_{2/3}$}} & \multicolumn{1}{c|}{\boldmath{$z_1$}} & \multicolumn{1}{c|}{\boldmath{$z_0$}} & \multicolumn{1}{c|}{\boldmath{$z_{1/3}$}} & \multicolumn{1}{c|}{\boldmath{$z_{2/3}$}} & \multicolumn{1}{c|}{\boldmath{$z_1$}} & \multicolumn{1}{c|}{\boldmath{$z_0$}} & \multicolumn{1}{c|}{\boldmath{$z_{1/3}$}} & \multicolumn{1}{c|}{\boldmath{$z_{2/3}$}} & \boldmath{$z_1$} \\
\bottomrule
	\end{tabular}}%
	\label{tab:Daily_EuroStoxx50_0.05}%
	\vspace{0.8cm}
\end{table}%
%
\begin{table}[htbp]
	\centering
	\caption{Out-of-sample performance results for the DowJones daily dataset
		with $\varepsilon=1\%$}
	\resizebox{\textwidth}{!}{\begin{tabular}{|l|ccccccccccccccccc|}
			\cmidrule{3-18}    \multicolumn{1}{r}{} & \multicolumn{1}{r|}{} & \multicolumn{4}{c|}{\boldmath{$\eta_{min}$}} & \multicolumn{4}{c|}{\boldmath{$\eta_{1/4}$}} & \multicolumn{4}{c|}{\boldmath{$\eta_{1/2}$}} & \multicolumn{4}{c|}{\boldmath{$\eta_{3/4}$}} \\
			\cmidrule{2-18}    \multicolumn{1}{r|}{} & \multicolumn{1}{c|}{\textbf{EW}} & \multicolumn{1}{c|}{\boldmath{$z_0$}} & \multicolumn{1}{c|}{\boldmath{$z_{1/3}$}} & \multicolumn{1}{c|}{\boldmath{$z_{2/3}$}} & \multicolumn{1}{c|}{\boldmath{$z_1$}} & \multicolumn{1}{c|}{\boldmath{$z_0$}} & \multicolumn{1}{c|}{\boldmath{$z_{1/3}$}} & \multicolumn{1}{c|}{\boldmath{$z_{2/3}$}} & \multicolumn{1}{c|}{\boldmath{$z_1$}} & \multicolumn{1}{c|}{\boldmath{$z_0$}} & \multicolumn{1}{c|}{\boldmath{$z_{1/3}$}} & \multicolumn{1}{c|}{\boldmath{$z_{2/3}$}} & \multicolumn{1}{c|}{\boldmath{$z_1$}} & \multicolumn{1}{c|}{\boldmath{$z_0$}} & \multicolumn{1}{c|}{\boldmath{$z_{1/3}$}} & \multicolumn{1}{c|}{\boldmath{$z_{2/3}$}} & \boldmath{$z_1$} \\
\bottomrule
	\end{tabular}}%
	\label{tab:Daily_DowJones_0.01}%
	\vspace{0.8cm}
\end{table}%
%
\begin{table}[htbp]
	\centering
	\caption{Out-of-sample performance results for the DowJones daily dataset
		with $\varepsilon=5\%$}
	\resizebox{\textwidth}{!}{\begin{tabular}{|l|ccccccccccccccccc|}
			\cmidrule{3-18}    \multicolumn{1}{r}{} & \multicolumn{1}{r|}{} & \multicolumn{4}{c|}{\boldmath{$\eta_{min}$}} & \multicolumn{4}{c|}{\boldmath{$\eta_{1/4}$}} & \multicolumn{4}{c|}{\boldmath{$\eta_{1/2}$}} & \multicolumn{4}{c|}{\boldmath{$\eta_{3/4}$}} \\
			\cmidrule{2-18}    \multicolumn{1}{r|}{} & \multicolumn{1}{c|}{\textbf{EW}} & \multicolumn{1}{c|}{\boldmath{$z_0$}} & \multicolumn{1}{c|}{\boldmath{$z_{1/3}$}} & \multicolumn{1}{c|}{\boldmath{$z_{2/3}$}} & \multicolumn{1}{c|}{\boldmath{$z_1$}} & \multicolumn{1}{c|}{\boldmath{$z_0$}} & \multicolumn{1}{c|}{\boldmath{$z_{1/3}$}} & \multicolumn{1}{c|}{\boldmath{$z_{2/3}$}} & \multicolumn{1}{c|}{\boldmath{$z_1$}} & \multicolumn{1}{c|}{\boldmath{$z_0$}} & \multicolumn{1}{c|}{\boldmath{$z_{1/3}$}} & \multicolumn{1}{c|}{\boldmath{$z_{2/3}$}} & \multicolumn{1}{c|}{\boldmath{$z_1$}} & \multicolumn{1}{c|}{\boldmath{$z_0$}} & \multicolumn{1}{c|}{\boldmath{$z_{1/3}$}} & \multicolumn{1}{c|}{\boldmath{$z_{2/3}$}} & \boldmath{$z_1$} \\
\bottomrule
	\end{tabular}}%
	\label{tab:Daily_DowJones_0.05}%
	\vspace{0.8cm}
\end{table}%
%
\begin{table}[htbp]
	\centering
	\caption{Out-of-sample performance results for the Hang Seng daily dataset
		with $\varepsilon=1\%$}
	\resizebox{\textwidth}{!}{\begin{tabular}{|l|ccccccccccccccccc|}
			\cmidrule{3-18}    \multicolumn{1}{r}{} & \multicolumn{1}{r|}{} & \multicolumn{4}{c|}{\boldmath{$\eta_{min}$}} & \multicolumn{4}{c|}{\boldmath{$\eta_{1/4}$}} & \multicolumn{4}{c|}{\boldmath{$\eta_{1/2}$}} & \multicolumn{4}{c|}{\boldmath{$\eta_{3/4}$}} \\
			\cmidrule{2-18}    \multicolumn{1}{r|}{} & \multicolumn{1}{c|}{\textbf{EW}} & \multicolumn{1}{c|}{\boldmath{$z_0$}} & \multicolumn{1}{c|}{\boldmath{$z_{1/3}$}} & \multicolumn{1}{c|}{\boldmath{$z_{2/3}$}} & \multicolumn{1}{c|}{\boldmath{$z_1$}} & \multicolumn{1}{c|}{\boldmath{$z_0$}} & \multicolumn{1}{c|}{\boldmath{$z_{1/3}$}} & \multicolumn{1}{c|}{\boldmath{$z_{2/3}$}} & \multicolumn{1}{c|}{\boldmath{$z_1$}} & \multicolumn{1}{c|}{\boldmath{$z_0$}} & \multicolumn{1}{c|}{\boldmath{$z_{1/3}$}} & \multicolumn{1}{c|}{\boldmath{$z_{2/3}$}} & \multicolumn{1}{c|}{\boldmath{$z_1$}} & \multicolumn{1}{c|}{\boldmath{$z_0$}} & \multicolumn{1}{c|}{\boldmath{$z_{1/3}$}} & \multicolumn{1}{c|}{\boldmath{$z_{2/3}$}} & \boldmath{$z_1$} \\
\bottomrule
	\end{tabular}}%
	\label{tab:Daily_HangSeng_0.01}%
	\vspace{0.8cm}
\end{table}%
%
\begin{table}[htbp]
	\centering
	\caption{Out-of-sample performance results for the Hang Seng daily dataset
		with $\varepsilon=5\%$}
	\resizebox{\textwidth}{!}{\begin{tabular}{|l|ccccccccccccccccc|}
			\cmidrule{3-18}    \multicolumn{1}{r}{} & \multicolumn{1}{r|}{} & \multicolumn{4}{c|}{\boldmath{$\eta_{min}$}} & \multicolumn{4}{c|}{\boldmath{$\eta_{1/4}$}} & \multicolumn{4}{c|}{\boldmath{$\eta_{1/2}$}} & \multicolumn{4}{c|}{\boldmath{$\eta_{3/4}$}} \\
			\cmidrule{2-18}    \multicolumn{1}{r|}{} & \multicolumn{1}{c|}{\textbf{EW}} & \multicolumn{1}{c|}{\boldmath{$z_0$}} & \multicolumn{1}{c|}{\boldmath{$z_{1/3}$}} & \multicolumn{1}{c|}{\boldmath{$z_{2/3}$}} & \multicolumn{1}{c|}{\boldmath{$z_1$}} & \multicolumn{1}{c|}{\boldmath{$z_0$}} & \multicolumn{1}{c|}{\boldmath{$z_{1/3}$}} & \multicolumn{1}{c|}{\boldmath{$z_{2/3}$}} & \multicolumn{1}{c|}{\boldmath{$z_1$}} & \multicolumn{1}{c|}{\boldmath{$z_0$}} & \multicolumn{1}{c|}{\boldmath{$z_{1/3}$}} & \multicolumn{1}{c|}{\boldmath{$z_{2/3}$}} & \multicolumn{1}{c|}{\boldmath{$z_1$}} & \multicolumn{1}{c|}{\boldmath{$z_0$}} & \multicolumn{1}{c|}{\boldmath{$z_{1/3}$}} & \multicolumn{1}{c|}{\boldmath{$z_{2/3}$}} & \boldmath{$z_1$} \\
\bottomrule
	\end{tabular}}%
	\label{tab:Daily_HangSeng_0.05}%
	\vspace{0.8cm}
\end{table}%
%
%
%
%
%
%
%
%
%

\end{document}